\documentclass[12pt,double]{article}

\usepackage{times}
\usepackage{graphicx} 
\usepackage{amsmath}
\usepackage{amsthm}
\usepackage{amssymb}
\usepackage{enumerate}
\usepackage{verbatim}
\usepackage{float}
\usepackage[margin=0.8in]{geometry}
\usepackage{multicol,ragged2e,setspace}
\usepackage{bm}
\usepackage[round]{natbib}

\usepackage{subfigure}

\usepackage{array}
\newcolumntype{H}{>{\setbox0=\hbox\bgroup}c<{\egroup}@{}}

\newcommand{\E}{\mathrm{E}}

\usepackage{siunitx}

\begin{document} 
\title{\textbf{Improved Generalized Raking Estimators to Address Dependent Covariate and Failure-Time Outcome Error}}
\author{}

\date{\vspace{-5ex}}

\maketitle

\begin{center}
Eric J. Oh$^{\footnotemark 1}$, Bryan E. Shepherd$^{2}$, Thomas Lumley$^{3}$, Pamela A. Shaw$^{1}$  \\
\vspace{0.1in}
$^1$University of Pennsylvania, Perelman School of Medicine \\
Department of Biostatistics, Epidemiology, and Informatics \\
\vspace{0.1in}
$^{2}$Vanderbilt University School of Medicine \\
Department of Biostatistics \\
\vspace{0.1in}
$^{3}$University of Auckland \\
Department of Statistics
\end{center}

\footnotetext[1]{Corresponding author: ericoh@pennmedicine.upenn.edu}

\doublespacing

\begin{abstract}
{Biomedical studies that use electronic health records (EHR) data for inference are often subject to bias due to measurement error. The measurement error present in EHR data is typically complex, consisting of errors of unknown functional form in covariates and the outcome, which can be dependent. To address the bias resulting from such errors, generalized raking has recently been proposed as a robust method that yields consistent estimates without the need to model the error structure. We provide rationale for why these previously proposed raking estimators can be expected to be inefficient in failure-time outcome settings involving misclassification of the event indicator. We propose raking estimators that utilize multiple imputation, to impute either the target variables or auxiliary variables, to improve the efficiency. We also consider outcome-dependent sampling designs and investigate their impact on the efficiency of the raking estimators, either with or without multiple imputation. We present an extensive numerical study to examine the performance of the proposed estimators across various measurement error settings. We then apply the proposed methods to our motivating setting, in which we seek to analyze HIV outcomes in an observational cohort with electronic health records data from the Vanderbilt Comprehensive Care Clinic.} 
\end{abstract}

\section{Introduction}

Modern biomedical studies are increasingly using non-traditional data sources such as electronic health records (EHR), which are not primarily collected for research purposes. These data sources have enormous potential to advance research of population-level health outcomes due to their large sample sizes and low cost compared to prospectively collected data (\citealp{beresniak2016cost,hillestad2005can,jensen2012mining,van2014opportunities}). EHR data, however, have also been shown to be vulnerable to measurement error (\citealp{botsis2010secondary,weiskopf2013methods,floyd2012use,kiragga2011quality,duda2012measuring}). If such errors are not accounted for in the data analysis, estimated effects of interest can be biased, which in turn can mislead researchers and potentially harm patients.

The measurement error found in EHR data can be complex, consisting of errors in both an outcome and covariates of interest, which in turn can be dependent. This complexity stems from the fact that variables of interest are often not directly observed in EHR data; instead, they need to be derived from other existing variables in the data. For example, HIV/AIDS studies might be interested in evaluating the association between a lab value at the date of antiretroviral therapy (ART) initiation and the time from ART initiation to some event of interest. Both the exposure and outcome in the above example depend on the ART initiation date; thus, if the initiation date is incorrect, the outcome and covariate in the analysis will both contain measurement error that is dependent (in addition to potential misclassification of the event). 

Covariate measurement error, particularly classical measurement error or extensions of it, has been well studied in the literature and methods to correct the bias resulting from such error have been well developed (\citealp{carroll2006measurement}). Although less attention has been given to errors in an outcome of interest, there has been some recent work looking at errors in binary outcomes (\citealp{magder1997logistic, edwards2013accounting, le2016evaluating}), discrete time-to-event outcomes (\citealp{meier2003discrete, magaret2008incorporating, hunsberger2010analysis}), and to a lesser extent, continuous time-to-event outcomes (\citealp{oh2018considerations,gravel2018validation}). There has been even less work to understand the impact of errors in both covariates and a time-to-event outcome, but it has recently been shown that ignoring such errors can cause severe bias in estimates of effects of interest (\citealp{oh2019raking,boe2020approximate,giganti2020tdmi}). 

In some cases, errors can be handled by retrospectively reviewing records and correcting all data points; however in most scenarios this will be too time-consuming and expensive to feasibly carry out. Instead, one can use a two-phase design, which involves reviewing and correcting only a subset of the records, to obtain consistent estimates of effects of interest. There have been some methods proposed recently that employ this framework to incorporate the large error-prone data with the smaller validated data to improve statistical inference, including regression calibration (\citealp{oh2019raking,boe2020approximate}), multiple imputation (\citealp{giganti2020tdmi}), and generalized raking (\citealp{oh2019raking}). Generalized raking in particular has been shown to be robust to the structure of the measurement error, which can be quite complex for EHR data (\citealp{oh2019raking, han2019combining}). Specifically, generalized raking estimators use the error-prone data as auxiliary variables to improve the efficiency of the analysis of the validated data without having to model the error structure, making them appealing for EHR settings where the true structure is likely unknown. Thus, we focus on the generalized raking methods in this manuscript. 

In the measurement error setting, an error-prone version of the target variable is generally available on all subjects at phase one, which can be used to construct auxiliary variables for raking. While generalized raking estimators are robust, their statistical efficiency is dependent on the quality of the raking variables. Specifically, the efficiency of raking estimators depends on the (linear) correlation between the auxiliary variables and the target variable (\citealp{deville1992calibration}). We show that for a  time-to-event outcome, where the event indicator is subject to misclassification, this linear correlation is generally low and results in inefficient estimates. In this manuscript, we propose generalized raking estimators that utilize multiple imputation to construct improved auxiliary variables using imputed values of either the error-prone data or direct imputation of the auxiliary variables themselves to improve the linear correlation and ultimately, the efficiency of the raking estimator. 

Our contributions in this manuscript are twofold. First, we develop generalized raking estimators that utilize multiple imputation to construct improved auxiliary variables in the presence of event indicator misclassification. Second, we evaluate the performance of various sampling designs with respect to their impact on the efficiency of the standard or proposed raking estimators. The rest of the paper proceeds as follows. We present our time-to-event outcome model and measurement error framework, and we introduce generalized raking estimators in Section~\ref{chpt3:model_setup}. Section~\ref{chpt3:better_auxiliary} discusses how the auxiliary variables relate to the efficiency of raking estimators and the need for their improvement in time-to-event settings with event indicator misclassification. Section~\ref{chpt3:raking_MI} develops the proposed generalized raking estimators using multiple imputation. Section~\ref{chpt3:simulations} compares the relative performance of the proposed estimators with simulation studies for various parameter settings and study designs. In Section~\ref{chpt3:dataexample}, we apply our methods to evaluate HIV outcomes in an HIV cohort with error-prone EHR data. We conclude with a discussion in Section~\ref{chpt3:discussion}.  

\section{Model setup and design framework}\label{chpt3:model_setup}

This section introduces the design and estimation framework, including the time-to-event outcome model, measurement error framework, and generalized raking methods used to estimate parameters of interest.

\subsection{Time-to-Event outcome model}

Let $T_i$ and $C_i$,  be the failure time and right censoring time, respectively, for subjects $i=1,\ldots,N$ on a finite follow-up time interval, $[0,\tau]$. Define $U_i=\textrm{min}(T_i,C_i)$ and the corresponding failure indicator $\Delta_i = I(T_i \leq C_i)$. Let $Y_i(t) = I(U_i \geq t)$ and $N_i(t)=I(U_i \leq t, \Delta_i=1)$ denote the at-risk indicator and counting process for observed events, respectively. Let $X_i$ be a \emph{p}-dimensional vector of continuous covariates that are measured with error and $Z_i$ a \emph{q}-dimensional vector of precisely measured discrete and/or continuous covariates that may be correlated with $X_i$. We assume $C_i$ is independent of $T_i$ given $(X_i,Z_i)$ and that $(T_i,C_i,X_i,Z_i)$ are i.i.d. 

In this paper, we consider estimating the parameters of a Cox proportional hazards model. Let the hazard rate for subject $i$ at time $t$ be given by $\lambda_i(t) = \lambda_0(t)\exp(\beta_X' X_i + \beta_Z' Z_i)$, where $\lambda_0(t)$ is an unspecified baseline hazard function. Then to estimate $\beta = \left(\beta_X, \beta_Z\right)$, we solve the partial likelihood score equation
\begin{equation}\label{cox_score}
\sum_{i=1}^N \int_0^{\tau} \left\{\left\{X_i,Z_i\right\}' - \frac{\sum_{j=1}^N Y_j(t)\left\{X_j,Z_j\right\}'\exp(\beta_X' X_j + \beta_Z' Z_j)}{\sum_{j=1}^N Y_j(t)\exp(\beta_X' X_j + \beta_Z' Z_j)}\right\} dN_i(t) = 0
\end{equation}

\subsubsection{Error framework}

Instead of observing $(X,Z,U,\Delta)$, we observe $(X^{\star},Z,U^{\star},\Delta^{\star})$, where $X^{\star}$, $U^{\star}$, and $\Delta^{\star}$ are the error-prone versions of  $X$, $U$, and $\Delta$, respectively. We do not impose any assumptions on the structure of the measurement error except that the error must have finite variance. In addition, we allow any of the errors to be correlated.  

\subsection{Two-phase design}

We consider a retrospective two-phase design where at phase one, a set of possibly error-prone covariates and outcome information is collected on a large group of subjects. At phase two, the large cohort is augmented by selecting a subset of the subjects ($n < N$) to be validated, i.e., to have error-free covariates and outcome information measured. As a result, the phase two data is often referred to as the validation subset. Since the validation subset is selected retrospectively, the sampling probabilities are known. This type of sampling strategy accommodates both fixed subsample sizes (e.g. simple random sampling) as well as more complex designs with random subsample sizes (e.g. case-cohort). Specifically, let $R_i$ be the indicator for whether subject $i=1,\ldots,N$ is selected to be in the validation subset with known sampling probability $0 < \pi_i \leq 1$. Then the observed data is given by $(X_i^{\star},Z_i,U^{\star}_i,\Delta_i^{\star})$ for $R_i=0$ and $(X_i^{\star},X_i,Z_i,U^{\star}_i,U_i,\Delta_i^{\star},\Delta_i)$ for $R_i=1$.

\subsection{Generalized Raking}\label{raking_description}

To estimate parameters in the two-phase design framework, we use generalized raking, a design-based estimator that combines the error-prone phase one data with the error-free phase two data to obtain efficient estimates that take advantage of all the measured data. Let $\beta_0$ denote the parameter defined by the population estimating equations $\sum_{i=1}^N \psi_i(\beta_0) = 0$. One classical estimator for two-phase designs is the Horvitz-Thompson (HT) estimator, $\hat{\beta}_{\textrm{HT}}$, which is defined as the solution to $\sum_{i=1}^N \frac{R_i}{\pi_i}\psi_i(\beta) = 0$. Under suitable regularity conditions, $\hat{\beta}_{\textrm{HT}}$ is a consistent estimator of $\beta_0$; however, it has been shown to be inefficient due to not using all of the available data at phase one (\citealp{robins1994estimation}). Let $A_i$ denote a $p+q$-dimensional vector of auxiliary variables that are available for all $N$ phase one subjects and correlated with the phase two data. Then generalized raking estimators modify the HT estimator design weights to new weights that incorporate the auxiliary variables such that $\sum_{i=1}^N A_i$, the known population total of auxiliary variables, is exactly estimated by the phase 2 subset. However, the new weights are constructed so that they are as close as possible to the HT weights while still satisfying the constraint. Specifically, for some distance measure $d(.,.)$, the objective can be written
\begin{equation*}
\textrm{minimize} \hspace{0.1in} \sum_{i=1}^N R_i d\left(\frac{g_i}{\pi_i}, \frac{1}{\pi_i}\right) \hspace{0.1in} \textrm{subject to} \hspace{0.1in} \sum_{i=1}^N A_i = \sum_{i=1}^N R_i \frac{g_i}{\pi_i} A_i,
\end{equation*}
where $\frac{g_i}{\pi_i}$ are the raking weights that can be solved for using Lagrange multipliers (\citealp{deville1992calibration}). Note that the constraints above are known as the calibration equations. Therefore, the generalized raking estimator is defined by the solution to 
\begin{equation}\label{raking_ee}
\sum_{i=1}^N R_i\frac{g_i}{\pi_i}\psi_i(\beta) = 0.
\end{equation}
Under suitable regularity conditions, the solution to (\ref{raking_ee}) has been shown to be a $\sqrt{N}$ consistent and asymptotically normal estimator of $\beta_0$ (\citealp{saegusa2013weighted}). When $\beta_0$ are the regression parameters from a correctly specified Cox proportional hazards model, $\psi_i(\beta)=\psi(X_i,Z_i,U_i,\Delta_i;\beta)$ is the Cox partial score equation (\ref{cox_score}) and the distance measure $d(a,b)=a\log(a/b)-a+b$ is used. Let $\lambda$ denote a $p+q$-dimensional vector of Lagrange multipliers. Then solving the constrained minimization problem yields $g_i=\exp\left(\hat{\lambda}'A_i\right)$, where $\hat{\lambda} = \hat{B}^{-1} \left(\sum_{i=1}^N \frac{R_i}{\pi_i} A_i - \sum_{i=1}^N A_i\right) + O_p(n^{-1})$ and $\hat{B} = \sum_{i=1}^N \frac{R_i}{\pi_i} A_i' A_i$ (\citealp{deville1992calibration}).
 
\section{Construction of Better Auxiliary Variables}\label{chpt3:better_auxiliary}

To quantify the gain in efficiency of raking estimators compared to the HT estimator, it is useful to consider the calibration equations, which constrain the raking weights to exactly estimate the known population total of the auxiliary variables. \citet{deville1992calibration} argued that ``weights that perform well for the auxiliary variable also should perform well for the study variable'' to provide support for such a construction. Note that study variable in this context represents the variable that is only observed on the phase two sample. Furthermore, there is an implicit assumption underlying this argument; namely that there exists a linear relationship between the variable of interest and the auxiliary variables of the form $S_i = \gamma_0 + \gamma_1 A_i + \epsilon_i$, where $S_i$ and $A_i$ are the variable of interest and auxiliary variables, respectively, and $\epsilon_i$ is random error. Thus, the efficiency gain of raking estimators depends directly on the (linear) correlation between the variable of interest and auxiliary variables. For more details, see \citet{lumley2011connections}. The true relationship between $S_i$ and $A_i$ determines how to best use the auxiliary variables, which we hope to capture with the working model. If the true relationship between the study variable and auxiliary variables is nonlinear, standard generalized raking could be quite inefficient.

Assessing whether a linear working model is appropriate requires precise definitions for the variable of interest and auxiliary variables. In the setting of estimating regression parameters, many common estimators can be written as a population mean of influence function (or efficient influence function for semiparametric models) terms, $\tilde{\ell}_0(X_i,Z_i,U_i,\Delta_i)$, using their asymptotically linear expansion. Thus, $\tilde{\ell}_0(X_i,Z_i,U_i,\Delta_i)$ is considered to be the variable of interest and the auxiliary variables should be constructed to be highly correlated with the influence function contributions. The optimal auxiliary variable was shown by \citet{breslow2009improved} to be $\mathrm{E}(\tilde{\ell}_0(X_i,Z_i,U_i,\Delta_i)|V)$, where $V = (X^{\star},Z,U^{\star},\Delta^{\star})$, which is unavailable in practice. \citet{oh2019raking}, however, proposed an approximation, $\tilde{\ell}_0(X_i^{\star},Z_i,U_i^{\star},\Delta_i^{\star})$, as the auxiliary variable, motivated by settings involving correlated measurement error in covariates and a censored event-time only.

\begin{figure}[H]
    \caption{Plots of the true influence function $\tilde{\ell}_0(X_i,Z_i,U_i,\Delta_i)$ against the error-prone version $\tilde{\ell}_0^{\star}$ with the variables subject to measurement error noted in the graph subtitle. For example, (a) displays $\tilde{\ell}_0(X_i,Z_i,U_i,\Delta_i)$ against $\tilde{\ell}_0(X^{\star}_i,Z_i,U_i,\Delta_i)$. Univariate and normally distributed $X$ and $Z$ were generated. Survival times were generated from an exponential distribution with rate $\lambda_0 \exp(\beta_X X + \beta_Z Z)$, where $\lambda_0 = 0.1$, $\beta_X=\log(1.5)$, and $\beta_Z=\log(0.5)$, with $90\%$ independent censoring. The error was generated as $X^{\star}=0.2+X-0.1Z-0.4\Delta+0.25U+\epsilon$, $U^{\star}=U + \sigma_{\nu}\cdot 3 - 0.2X - 1.05 Z + \nu$, and $\Delta^{\star}=\textrm{Bernoulli}\left(\textrm{expit}\left(-1.1+3 \Delta - 0.3 X - 0.2 U + 0.1 Z\right)\right)$, where $(\epsilon,\nu)$ were normally distributed with $(\mu_{\epsilon},\mu_{\nu})=(0,0)$, variances $(\sigma^2_\epsilon,\sigma^2_\nu)=(0.5,0.5)$, and $\rho_{\epsilon,\nu}=0.5$.}
    \label{IF_plots}
    \centering
    \subfigure[$X$]{\includegraphics[width=0.4\linewidth]{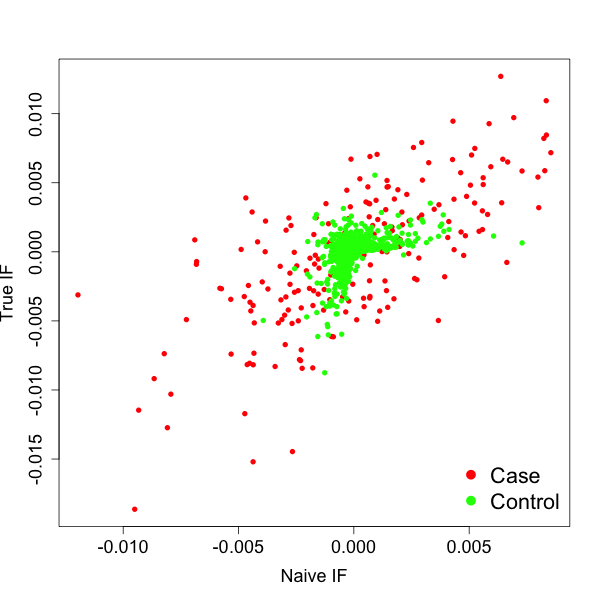}}
    \subfigure[$U$]{\includegraphics[width=0.4\linewidth]{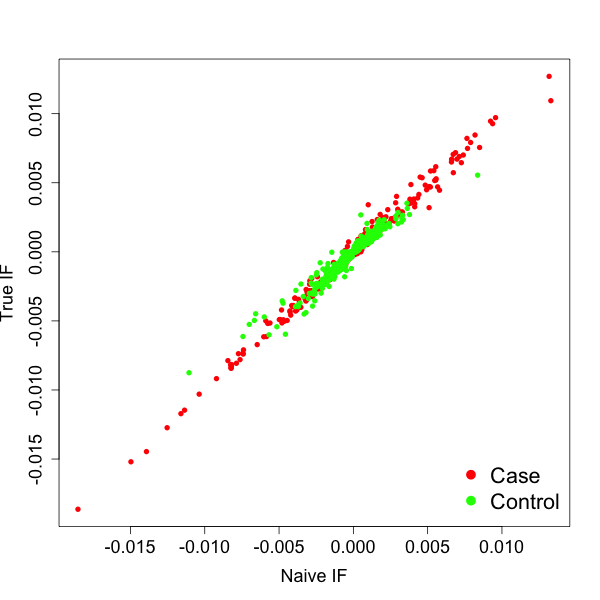}}
    \subfigure[$\Delta$]{\includegraphics[width=0.4\linewidth]{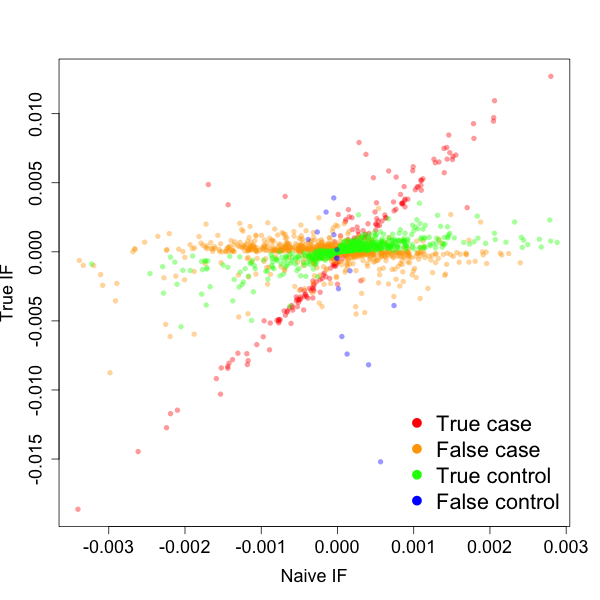}}
    \subfigure[$\Delta$, $U$, $X$]{\includegraphics[width=0.4\linewidth]{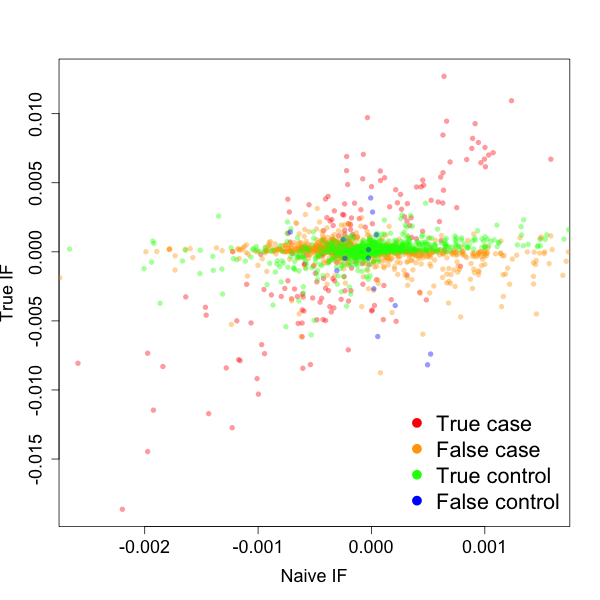}}
\end{figure}

Thus, the linear working model underlying the estimator from \citet{oh2019raking} is given by \\ $\tilde{\ell}_{0}(X_i,Z_i,U_i,\Delta_i) = \gamma_0 + \gamma_1 \tilde{\ell}_{0}(X_i^{\star},Z_i,U_i^{\star},\Delta_i^{\star}) + \epsilon_i$. To assess whether the linear fit is appropriate, we plot $\tilde{\ell}_{0}(X_i,Z_i,U_i,\Delta_i)$ against $\tilde{\ell}_{0}(X_i^{\star},Z_i,U_i^{\star},\Delta_i^{\star})$ from simulated data for various measurement error scenarios. Specifically, we plot empirical approximations of $\tilde{\ell}_0$ using delta-beta residuals (see \citet{oh2019raking} for more detail on their calculation) for settings with covariate error, time-to-event error, and misclassification only, as well as combinations of all three in Figure~\ref{IF_plots}. The plots of $\tilde{\ell}_{0}(X_i,Z_i,U_i,\Delta_i)$ against $\tilde{\ell}_{0}(X_i^{\star},Z_i,U_i^{\star},\Delta_i^{\star})$ for additive errors in the time-to-event or covariate show that the assumption of a linear relationship is mostly justified, albeit with some heteroscedasticity. However, when there is misclassification of the event indicator, a linear working model appears to be a very poor fit, and including additional errors in variables as in Figure 1d worsens the fit. 

\subsection{Model-calibration}

\citet{wu2001model} proposed an alternative calibration method to handle settings where the true relationship between the variable of interest and the auxiliary variables may be nonlinear. Specifically, they assume the relationship between $S_i$ and $A_i$ can be characterized by the first and second moments, $\mathrm{E}(S_i|A_i)=\mu(A_i;\theta)$ and $\mathrm{Var}(S_i|A_i)=v_i^2\sigma^2$, where $\mu$ is a known function of $A_i$ and $\theta$, $v$ is a known function of $A_i$ or $\mu$, and $(\theta,\sigma^2)$ are unknown parameters. Then using the validation subset, one obtains fitted values of $\mu(x_i;\theta)$, $\mu(x_i;\hat{\theta})$, and performs the raking procedure using them as auxiliary variables. Specifically, the generalized raking objective can be written as
\begin{equation}\label{wu_sitter_raking}
\textrm{minimize} \hspace{0.1in} \sum_{i=1}^N R_i d\left(\frac{g_i}{\pi_i}, \frac{1}{\pi_i}\right) \hspace{0.1in} \textrm{subject to} \hspace{0.1in} \sum_{i=1}^N \mu(x_i;\hat{\theta}) = \sum_{i=1}^N R_i \frac{g_i}{\pi_i} \mu(x_i;\hat{\theta})
\end{equation}
\citet{wu2001model} showed that this method yields more efficient estimates than the traditional raking estimator but still retains all of its statistical properties for a true nonlinear relationship between the variable of interest and auxiliary variables. Inspired by the model-calibration approach, we propose a data imputation approach that imputes the true $\Delta$ to obtain an auxiliary variable that has higher linear correlation with $\tilde{\ell}_{0}(X_i,Z_i,U_i,\Delta_i)$ than $\tilde{\ell}_0(X_i^{\star},Z_i,U_i^{\star},\Delta_i^{\star})$ does. Additionally, we propose a novel application of the \citet{wu2001model} approach that directly imputes $\tilde{\ell}_{0}(X_i,Z_i,U_i,\Delta_i)$ based on a (potentially nonlinear) working model.

\section{Proposed Multiple Imputation Methods for Generalized Raking}\label{chpt3:raking_MI}

In this section, we propose methods to improve the efficiency of the generalized raking estimators under measurement error settings involving event indicator misclassification. Our methods use multiple imputation to impute the event indicator and then constructs new auxiliary variables using the imputed values to solve the raking estimating equation. For settings involving errors beyond just misclassification (e.g. additional time-to-event and/or covariate error), we propose a method using the fully conditional specification multiple imputation procedure that additionally imputes the other error-prone variables iteratively. These methods are related to those of \citet{han2016combining}, who proposed combining an empirical likelihood approach with multiple imputation to construct multiply robust estimators that are consistent if one of the sampling models or data generating models are correctly specified. Our approach differs in that we assume known phase two sampling probabilities possibly specified using a complex sampling design and study specific efficiency issues for time-to-event data. We additionally consider directly imputing the true population influence functions via a working model to use as auxiliary variables as a novel application of \citet{wu2001model}. Lastly, we consider various study designs, including outcome-dependent sampling designs, for the selection of the validation subset in the two-phase design framework and discuss their varying impact on the efficiency of the proposed methods.

Note that due to raking being a design-based method, it will yield consistent estimates of the parameter that would be estimated with error-free data on the full cohort. The proposed methods all focus on adjusting the working model of the population influence functions to construct auxiliary variables closer to the optimal auxiliary variable. If the working model is misspecified, or does not capture the true relationship well, the proposed estimators still yield consistent and asympotically normal estimates (\citealp{breslow2009improved}). If however, the working model is correct, the estimators will yield the most efficient design-consistent estimator (\citealp{han2016combining}). 

\subsection{Multiple Imputation for the Event Indicator}\label{MI_delta}

Traditional multiple imputation in missing data settings (\citealp{rubin2004multiple}) involves developing statistical models for the distributions of the variables subject to missingness conditional on the fully observed variables. The missing variables are sampled $M$ times from their distribution to generate $M$ imputations of the missing data. The original data is augmented with the imputations, yielding $M$ complete imputed datasets. Each of the $M$ imputed datasets are then used to separately estimate the parameters of interest and the average of the $M$ estimates is the multiple imputation estimator. The variance of the estimates can be calculated using Rubin's rules (\citealp{barnard1999small}) or the estimators proposed by \citet{robins2000inference}.

Multiple imputation for generalized raking follows similarly, with the exception that the $M$ imputed datasets are first used to construct auxiliary variables for the influence functions for the target parameters. 

First, we posit an imputation model for $\Delta$, $f(\Delta|\Delta^{\star},X^{\star},U^{\star},Z;\eta)$, with parameter vector $\eta$, and specify a non-informative prior distribution, $f(\eta)$. We then fit the imputation model using the validation subset, generate the posterior distribution for $\eta$, and then sample $M$ times from this posterior distribution to obtain $\eta_{\star}^{(1)},\ldots,\eta_{\star}^{(M)}$. The parameter draws are used to sample $\hat{\Delta}_i^{(m)} \sim f(\Delta|\Delta_i^{\star},X_i^{\star},U_i^{\star},Z_i;\eta_{\star}^{(m)})$ for all $N$ phase one subjects and $m=1,\ldots,M$. $\hat{\Delta}^{(1)},\ldots,\hat{\Delta}^{(M)}$ are then augmented with the phase one data to yield $M$ complete imputed datasets. Then for $m=1,\ldots,M$, the estimating equation $\sum_{i=1}^N \psi(X_i^{\star}, Z_i, U^{\star}_i, \hat{\Delta}_i^{(m)};\beta)=0$
is solved to obtain $\hat{\beta}^{(m)}$. For each subject $i=1,\ldots,N$, the auxiliary variable $\hat{A}_i$ is defined as
\begin{equation*}
    \hat{A}_i= \frac{1}{M}\sum_{m=1}^M \tilde{\ell}_0(X^{\star}_i,Z_i,U^{\star}_i,\hat{\Delta}^{(m)}_i;\hat{\beta}^{(m)}),
\end{equation*}
where $\tilde{\ell}_0(X_i^{\star}, Z_i, U_i^{\star}, \hat{\Delta}_i^{(m)})$ is the influence function for the estimating equation from the $m$-th imputation and can be empirically approximated as
\begin{align*}
\tilde{\ell}_0(X_i^{\star}, Z_i, U_i^{\star}, \hat{\Delta}_i^{(m)}) & \approx  \hat{\Delta}_i^{(m)}\left\{\left\{X_i^{\star},Z_i\right\}' - \frac{S^{(1)\star}(\beta,t)}{S^{(0)\star}(\beta,t))}\right\} \\
&- \sum_{i=1}^n \int_0^{\tau} \frac{\exp(\beta_X' X_i^{\star} + \beta_Z' Z_i)}{S^{(0)\star}(\beta,t)} \left\{\left\{X_i^{\star},Z_i\right\}' - \frac{S^{(1)\star}(\beta,t)}{S^{(0)\star}(\beta,t))}\right\} d\hat{N}_i(t),
\end{align*}
where $S^{(r)\star}(\beta,t) = n^{-1}\sum_{j=1}^n Y^{\star}_j(t)\left\{X_j^{\star},Z_j\right\}^{'\otimes r}\exp(\beta_X' X_j^{\star} + \beta_Z' Z_j)$ ($a^{\otimes 1}$ is the vector $a$ and $a^{\otimes 0}$ is the scalar 1), $Y^{\star}_j(t) = I(U_j^{\star} \geq t)$, and $\hat{N}_i(t)=I(U_i^{\star} \leq t, \hat{\Delta}_i^{(m)}=1)$.

Finally, to obtain estimates of the parameter of interest, we solve the raking estimating equation with adjusted weights calculated using $\hat{A}_i$ as auxiliary variables in (\ref{raking_ee}).

\subsection{Fully Conditional Specification Multiple Imputation}\label{MI_all}

If there exists measurement error in variables beyond just the event indicator (e.g. additional time-to-event and/or covariate error), it is possible to gain efficiency by additionally imputing all error-prone variables iteratively using the fully conditional specification multiple imputation (FCSMI) method (\citealp{van2007multiple}). FCSMI involves specifying univariate models for the conditional distribution of each of the variables observed only at phase two given all phase one variables. Each missing variable is repeatedly imputed using the specified models and conditioning on the most recent imputations of the other variables. We explicate the FCSMI method for generalized raking in the presence of misclassification, covariate error, and time-to-event error. The method assumes a working model for the censored time-to-event that takes the form $U^{\star}=U+R(\Delta,X,Z)$, where $R(\Delta,X,Z)$ is an arbitrary function of $\Delta$, $X$, and $Z$. Note that if the working error model is misspecified, the raking estimator will still be consistent, albeit with some loss of efficiency. 

First, we posit imputation models for $\Delta$, $X$, and $R$, as well as non-informative prior distributions for their parameter vectors $\eta$, $\theta$, and $\omega$, respectively, to generate posterior distributions. We then draw parameters from their posteriors as follows: $\eta_{\star}^{(0)} \sim f(\Delta|\Delta^{\star},X^{\star},U^{\star},Z;\eta_V)f(\eta_V)$, $\theta_{\star}^{(0)} \sim f(X|\Delta^{\star},X^{\star},U^{\star},Z;\theta_V)f(\theta_V)$, and $\omega_{\star}^{(0)} \sim f(R|\Delta^{\star},X^{\star},Z;\omega_V)f(\omega_V)$. Then $\Delta$, $X$, and $U$ are imputed for all $N$ phase one subjects by sampling from the imputation models using the initial parameter draws: $\hat{\Delta}^{(0)} \sim f(\Delta|\Delta^{\star},X^{\star},U^{\star},Z;\eta_{\star}^{(0)})$, $\hat{X}^{(0)} \sim f(X|\Delta^{\star},X^{\star},U^{\star},Z;\theta_{\star}^{(0)})$, and $\hat{U}^{(0)} = U^{\star}-\hat{R}^{(0)}$ where $\hat{R}^{(0)} \sim f(R|\Delta^{\star},X^{\star},Z;\omega_{\star}^{(0)})$. Then for iteration $l=1,\ldots,L$, the algorithm proceeds as follows
\begin{align*}
\eta_{\star}^{(l)} &\sim f(\Delta|\Delta^{\star},\hat{X}^{(l-1)},\hat{U}^{(l-1)},Z;\eta)f(\eta) \\
\hat{\Delta}^{(l)} &\sim f(\Delta|\Delta^{\star},\hat{X}^{(l-1)},\hat{U}^{(l-1)},Z;\eta_{\star}^{(l)}) \\
\theta_{\star}^{(l)} &\sim f(X|\hat{\Delta}^{(l)},X^{\star},\hat{U}^{(l-1)},Z;\theta)f(\theta) \\
\hat{X}^{(l)} &\sim f(X|\hat{\Delta}^{(l)},X^{\star},\hat{U}^{(l-1)},Z;\theta_{\star}^{(l)}) \\
\omega_{\star}^{(l)} &\sim f(R|\hat{\Delta}^{(l)},\hat{X}^{(l)},Z;\omega)f(\omega) \\ 
\hat{U}^{(l)}=U^{\star}-\hat{R}^{(l)} \hspace{0.1in} &\textrm{where} \hspace{0.1in} \hat{R}^{(l)} \sim f(R|\hat{\Delta}^{(l)},\hat{X}^{(l)},Z;\omega_{\star}^{(l)})
\end{align*}
The algorithm continues sampling and imputing $\hat{\Delta}$, $\hat{X}$, and $\hat{U}$ for $L$ iterations, after which it is assumed a stationary distribution has been reached. The above steps are repeated for $M$ iterations, where $\hat{\Delta}^{(L)}$, $\hat{X}^{(L)}$, and $\hat{U}^{(L)}$ are taken to be the imputed values of $\Delta$, $X$, and $U$, respectively, for each $m=1,\ldots,M$. $\hat{\Delta}^{(m)}$, $\hat{X}^{(m)}$, and $\hat{U}^{(m)}$ are then augmented with the phase one data to yield $M$ complete imputed datasets. Then for $m=1,\ldots,M$, the estimating equation $\sum_{i=1}^N \psi(\hat{X}_i^{(m)}, Z_i, \hat{U}_i^{(m)}, \hat{\Delta}_i^{(m)};\beta)=0$ is solved to obtain $\hat{\beta}^{(m)}$. Then the auxiliary variable for each subject, $\hat{A}_i$, is defined as
\begin{equation*}
    \hat{A}_i= \frac{1}{M}\sum_{m=1}^M \tilde{\ell}_0(\hat{X}_i^{(m)}, Z_i, \hat{U}_i^{(m)}, \hat{\Delta}_i^{(m)};\hat{\beta}^{(m)}),
\end{equation*}
and $\tilde{\ell}_0(\hat{X}_i^{(m)}, Z_i, \hat{U}_i^{(m)}, \hat{\Delta}_i^{(m)})$ can be empirically approximated as
\begin{align*}
\tilde{\ell}_0(\hat{X}_i^{(m)}, Z_i, \hat{U}_i^{(m)}, \hat{\Delta}_i^{(m)}) & \approx  \hat{\Delta}_i^{(m)}\left\{\left\{\hat{X}_i^{(m)},Z_i\right\}' - \frac{\hat{S}^{(1)}(\beta,t)}{\hat{S}^{(0)}(\beta,t))}\right\} \\
&- \sum_{i=1}^n \int_0^{\tau} \frac{\exp(\beta_X' \hat{X}_i^{(m)} + \beta_Z' Z_i)}{\hat{S}^{(0)}(\beta,t)} \left\{\left\{\hat{X}_i^{(m)},Z_i\right\}' - \frac{\hat{S}^{(1)}(\beta,t)}{\hat{S}^{(0)}(\beta,t))}\right\} d\hat{N}_i(t),
\end{align*}
where $\hat{S}^{(r)}(\beta,t) = n^{-1}\sum_{j=1}^n \hat{Y}_j(t)\left\{\hat{X}_j^{(m)},Z_j\right\}^{'\otimes r}\exp(\beta_X' \hat{X}_j^{(m)} + \beta_Z' Z_j)$ ($a^{\otimes 1}$ is the vector $a$ and $a^{\otimes 0}$ is the scalar 1), $\hat{Y}_j(t) = I(\hat{U}_j^{(m)} \geq t)$, and $\hat{N}_i(t)=I(\hat{U}_i^{(m)} \leq t, \hat{\Delta}_i^{(m)}=1)$.

Lastly, to obtain estimates of the parameter of interest, we solve the raking estimating equation with adjusted weights calculated using $\hat{A}_i$ as auxiliary variables in (\ref{raking_ee}). 

\subsection{Model-calibration multiple imputation}\label{wusittermethod}

We propose a multiple imputation application of the \citet{wu2001model} model-calibration approach by specifying a working model for the population influence function and using the fitted values as auxiliary variables for raking in repeated iterations. First, we impute the error-prone variable(s) using MI or FCSMI as described in Sections~\ref{MI_delta} and \ref{MI_all}. For the purposes of exposition, assume that FCSMI is used to impute $\Delta$, $X$, and $U$ to obtain $\hat{\Delta}^{(m)}$, $\hat{X}^{(m)}$, and $\hat{U}^{(m)}$. We posit a working model
\begin{equation*}
\E(\tilde{\ell}_{0}(X_i,Z_i,U_i,\Delta_i)|\tilde{\ell}_{0}(\hat{X}_i^{(m)},Z_i,\hat{U}_i^{(m)},\hat{\Delta}_i^{(m)})) = \mu(\tilde{\ell}_{0}(\hat{X}_i^{(m)},Z_i,\hat{U}_i^{(m)},\hat{\Delta}_i^{(m)});\gamma^{(m)}),
\end{equation*}
where $\tilde{\ell}_{0}(\hat{X}_i^{(m)},Z_i,\hat{U}_i^{(m)},\hat{\Delta}_i^{(m)})$ is constructed using the empirical approximation given in Section~\ref{MI_all}. Here, $\mu$ can capture nonlinear relationships and the model is fit on the validation subset to obtain $\hat{\gamma}^{(m)}$. The above steps are repeated $m=1,\ldots,M$ iterations to obtain $\hat{\gamma}^{(1)},\ldots,\hat{\gamma}^{(M)}$. The auxiliary variable for each subject, $\hat{A}_i$, is then defined as
\begin{equation*}
    \hat{A}_i= \frac{1}{M}\sum_{m=1}^M \mu(\tilde{\ell}_0(\hat{X}_i^{(m)}, Z_i, \hat{U}_i^{(m)}, \hat{\Delta}_i^{(m)});\hat{\gamma}^{(m)})
\end{equation*}
Finally, estimates of the parameter of interest are obtained by solving the raking estimating equation with adjusted weights calculated using $\hat{A}_i$ as auxiliary variables in (\ref{raking_ee}).

\subsection{Sampling Design Considerations}

In validation study settings, such as those considered in this manuscript, researchers can define the phase two sampling probabilities as functions of the phase one data to select more informative subjects for increased efficiency. For example, researchers may want to oversample cases in rare-event settings or oversample subjects at underrepresented levels of informative covariates. Although generalized raking can easily accommodate such designs, the interplay between sampling designs and raking has not been well studied. We consider the effects of three different sampling designs on the efficiency of raking estimates: simple random sampling (SRS), case-control (CC), and covariate stratified case-control (SCC).  

\section{Simulation Study}\label{chpt3:simulations}

In this section, we study the finite sample performance of the proposed raking estimators utilizing multiple imputation in the presence of event indicator misclassification. We compare these estimators to the raking estimator that constructs auxiliary variables using the naive error-prone data (GRN), the HT estimator, and the true estimator, i.e., the Cox proportional hazards model fit with the error-free data for all subjects. We considered three different measurement error scenarios where different variables are observed with error: 1) $(X,Z,U,\Delta^{\star})$, 2) $(X,Z,U^{\star},\Delta^{\star})$, and 3) $(X^{\star},Z,U^{\star},\Delta^{\star})$. For each error scenario, we considered the proposed raking estimator utilizing MI to impute the event indicator only, referred to as Generalized Raking Multiple Imputation (GRMI) hereafter. For error scenarios 2 and 3, which include errors in other variables besides the event indicator, we additionally considered the proposed raking estimator utilizing FCSMI to impute all error-prone variables iteratively, referred to as Generalized Raking Fully Conditional Specification Multiple Imputation (GRFCSMI) hereafter. We refer to these estimators as encompassing the data imputation approach. For all three error scenarios, we also considered the corresponding model-calibration multiple imputation methods described in Section~\ref{wusittermethod}, which we similarly refer to as encompassing the influence function (IF) imputation approach. We present $\%$ biases, average model standard errors (ASE), empirical standard errors (ESE), relative efficiency (RE) calculated with respect to the HT ESE, mean squared errors (MSE), and 95$\%$ coverage probabilities (CP) for varying values of the log hazard ratio $\beta_X$, $\%$ censoring, cohort and validation subset sizes, and validation subset sampling designs. We additionally present type 1 error results for $\beta_X=0$ and $\alpha = 0.05$. All standard errors were calculated using sandwich variance estimators.

\subsection{Simulation set-up}\label{sim_setup}

All simulations were run 2000 times using $\textsf{R}$ version 3.6.2 (\citealp{R_ref}). Cohort and validation subset sizes of $\{N,n\}=\{2000, 400\}$ and $\{N,n\}=\{10000, 2000\}$ were considered. Univariate $X$ and $Z$ were considered and were generated as a bivariate normal distribution with means $(\mu_X,\mu_Z)=(0,2)$, variances $(\sigma^2_X,\sigma^2_Z)=(1,1)$, and $\rho_{X,Z}=0.5$. The true log hazard ratios were set to be $\beta_X \in \{\log(1.5),\log(3)\}$ and $\beta_Z=\log(0.5)$. The true survival time $T$ was generated from an exponential distribution with rate equal to $\lambda_0 \exp(\beta_X X + \beta_Z Z)$, where $\lambda_0 = 0.1$. Censoring times were simulated for each $\beta_X$ and $\beta_Z$ to yield $50\%$, $75\%$, and $90\%$ censoring rates. Specifically, they were generated from Uniform distributions of varying lengths to mimic studies of different lengths. 

The error-prone data were generated as follows:
\begin{enumerate}
\item Scenario 1: $(X, Z, U, \Delta^{\star})$, where
\begin{equation*}
\Delta^{\star}=\textrm{Bernoulli}\left(\textrm{expit}\left(-1.1+3 \Delta - 0.3 X - 0.2 U + 0.1 Z\right)\right)
\end{equation*}
\item Scenario 2: $(X, Z, U^{\star}, \Delta^{\star})$, where
\begin{align*}
\Delta^{\star}&=\textrm{Bernoulli}\left(\textrm{expit}\left(-1.1+3 \Delta - 0.3 X - 0.2 U + 0.1 Z\right)\right) \\
U^{\star} &= U +R = U + \sigma_{\nu}\cdot 3 - 0.2X - 1.05 Z + \nu
\end{align*}
\item Scenario 3: $(X^{\star}, Z, U^{\star}, \Delta^{\star})$, where
\begin{align*}
\Delta^{\star}&=\textrm{Bernoulli}\left(\textrm{expit}\left(-1.1+3 \Delta - 0.3 X - 0.2 U + 0.1 Z\right)\right) \\
U^{\star} &= U + R = U + \sigma_{\nu}\cdot 3 - 0.2X - 1.05 Z + \nu \\
X^{\star} &= 0.2 + X - 0.1 Z -0.4 \Delta + 0.25 U + \epsilon
\end{align*}
\end{enumerate}

Note that the choice of the intercept term in the event time error model is such that the error-prone time is a valid event time (i.e., greater than zero) with high probability. The few censored event times that were less than 0 were reflected across 0 to generate valid outcomes. For scenario 3, the error terms $(\epsilon,\nu)$ were generated from a bivariate normal distribution with means $(\mu_{\epsilon},\mu_{\nu})=(0,0)$, variances $(\sigma^2_\epsilon,\sigma^2_\nu)=(0.5,0.5)$, and $\rho_{\epsilon,\nu}=0.5$. $\nu$ was generated from a univariate normal distribution for scenario 2 with the same mean and variance as in scenario 3. Supplementary Materials Table~\ref{chpt3:misclass_srs_metrics} presents the sensitivity, specificity, positive predictive value, and negative predictive value for the misclassified event indicator across all error scenarios.

For the working imputation models, we fit logistic regression models for $\Delta$ and linear regression models for $X$ and $R$. Under the error generating process considered in this section, analytical expressions for the true imputation models do not exist. Therefore, we considered two types of working imputation models: those including only main effects and those additionally adding all possible interaction effects to potentially specify an imputation model closer to the truth. Specifically, the imputations models including only main effects (referred to as Generalized Raking Multiple Imputation Simple (GRMIS) and Generalized Raking Fully Conditional Specification Multiple Imputation Simple (GRFCSMIS) hereafter) were specified as follows:
\begin{enumerate}
\item Scenario 1: $(X,Z,U,\Delta^{\star})$
\begin{equation*}
\textrm{logit}(P(\Delta=1)|\Delta^{\star},X,U,Z)=\eta_0+\eta_1 \Delta^{\star} + \eta_2 X + \eta_3 U + \eta_4 Z 
\end{equation*}
\item Scenario 2: $(X,Z,U^{\star},\Delta^{\star})$
\begin{align*}
\textrm{logit}(P(\Delta=1)|\Delta^{\star},X,U^{\star},Z)&=\eta_0+\eta_1 \Delta^{\star} + \eta_2 X + \eta_3 U^{\star} + \eta_4 Z\\
\E(R|\Delta^{\star},X,Z)&=\omega_0+\omega_1 \Delta^{\star} + \omega_2 X + \omega_3 Z 
\end{align*}
\item Scenario 3: $(X^{\star},Z,U^{\star},\Delta^{\star})$
\begin{align*}
\textrm{logit}(P(\Delta=1)|\Delta^{\star},X^{\star},U^{\star},Z)&=\eta_0+\eta_1 \Delta^{\star} + \eta_2 X^{\star} + \eta_3 U^{\star} + \eta_4 Z\\
\E(R|\Delta^{\star},X^{\star},Z)&=\omega_0+\omega_1 \Delta^{\star} + \omega_2 X^{\star} + \omega_3 Z\\
\E(X|\Delta^{\star},X^{\star},U^{\star},Z)&=\theta_0 + \theta_1 \Delta^{\star} + \theta_2 X^{\star} + \theta_3 U^{\star} + \theta_4 Z
\end{align*}
\end{enumerate}
The imputation models containing interaction terms (referred to as Generalized Raking Multiple Imputation Complex (GRMIC) and Generalized Raking Fully Conditional Specification Multiple Imputation Complex (GRFCSMIC) hereafter) include the same predictors as above as well as all possible interaction terms. For each error scenario and all parameter settings, the number of imputation iterations was set to 50 and the FCSMI estimators performed 500 iterative updates to the imputed variables per imputation iteration. Appendix B provides further detail on the implementation of the multiple imputation procedures. For the IF imputation approach, linear regression models were fit for the working models of the true influence function for each covariate. For example, the following model was fit for error scenario 1:
\begin{align*}
\E(\tilde{\ell}_{0}|\hat{\tilde{\ell}}_{0}) &= \gamma_0 + \gamma_1 \hat{\tilde{\ell}}_{0} + \gamma_2 \hat{\Delta} + \gamma_3 U + \gamma_4 X + \gamma_5 Z  \\
& +\gamma_6\left(\hat{\tilde{\ell}}_{0}\times\hat{\Delta}\right) + \gamma_7\left(\hat{\tilde{\ell}}_{0}\times U\right) + \gamma_8\left(\hat{\tilde{\ell}}_{0}\times X\right) + \gamma_9\left(\hat{\tilde{\ell}}_{0}\times Z\right).
\end{align*}
For error scenarios 2 and 3, the same models were fit except $U$ and $X$ were replaced by $\hat{U}$ and $\hat{X}$.

We considered validation subsets selected via simple random sampling for all three error scenarios. For the rare-event setting of $90\%$ censoring in error scenarios 2 and 3, we additionally compared the performance of the estimators using validation subsets selected via case-control sampling and stratified case-control sampling. For these sampling design comparisons, we considered $\{N,n\}=\{4000,800\}$ and generated the error-prone event indicator according to the model described in Supplementary Materials Table~\ref{chpt3:misclass_design_metrics}. The covariate and time-to-event error were generated using the same previous models. To perform case-control sampling, all error-prone cases were selected and a simple random sample of error-prone controls were selected to yield a nearly one-to-one ratio of error-prone cases to controls. To perform stratified case-control sampling, we stratified the continuous covariate $X$ (or $X^{\star}$ for settings involving covariate error) into four discrete categories by setting cutpoints at the $20^{th}$, $50^{th}$, and $80^{th}$ percentiles. We then selected an equal number of subjects from each of the eight strata defined by the combinations of the error-prone case status and the covariate strata (i.e., the balanced sampling design proposed by \citet{breslow1999design}) . Note that for CC and SCC sampling, the data imputation models and influence function working models for the IF imputation approach were inverse-probability weighted to account for the sampling design of the validation subsets. For the proposed raking estimators utilizing MI or FCSMI for data imputation only, the imputation models were not weighted as we included all stratification variables in the models (\citealp{cochran2007sampling}) and we noticed no empirical differences between including weights or not.

\subsection{Simulation results}

In the scenarios considered, all of the considered estimators were nearly unbiased for all settings, as expected, with the exception of a few specific rare-event settings with $\{N,n\}=\{2000,400\}$ and simple random sampling, due to relatively few true events (40 on average) in the validation subset. Since the proposed estimators construct improved auxiliary variables to increase efficiency compared to GRN, we focus on the ESE, RE (with respect to the HT estimator), MSE, and CP and how these performance measures differed across settings.  

Table~\ref{data_imp_err_1} presents the relative performance under error scenario 1 for estimating $\beta_X \in \{\log(1.5),\log(3)\}$ using the data imputation approach for $\{N,n\}=\{2000, 400\}$, $\{50\%,75\%,90\%\}$ censoring, and simple random sampling of the validation subset. GRN had increased efficiency compared to HT with the RE ranging from $1.24$ for $50\%$ censoring to $1.06$ for $90\%$ censoring. However, GRMIS and GRMIC both had higher RE than GRN for nearly all parameter settings, ranging from $1.41$ for $50\%$ censoring to $1.16$ for $90\%$ censoring. GRMIS and GRMIC had comparable REs, lower MSE than HT and GRN, and CPs near $95\%$ for all parameter settings.

Supplementary Materials Table~\ref{data_imp_err_2} presents the relative performance under error scenario 2 for estimating $\beta_X \in \{\log(1.5),\log(3)\}$ using the data imputation approach for $\{N,n\}=\{2000, 400\}$, $\{50\%,75\%,90\%\}$ censoring, and simple random sampling of the validation subset. GRN again had increased efficiency compared to HT with the RE ranging from $1.21$ to $1.07$. GRMIS, GRMIC, GRFCSMIS, and GRFCSMIC, however, all had higher RE than GRN for all parameter settings, ranging from $1.43$ to $1.14$ for GRMI and $1.45$ to $1.14$ for GRFCSMI. Comparing GRMIS to GRMIC and GRFCSMIS to GRFCSMIC, we observed nearly no difference in efficiency. Comparing GRMI to GRFCSMI, GRFCSMI had higher or equal RE for nearly all settings, although the difference was sometimes small. In addition, GRMI and GRFCSMI had lower MSE than HT and GRN and CPs by $5-6\%$ for all settings.

Table~\ref{data_imp_err_3} presents the relative performance under error scenario 3 for estimating $\beta_X \in \{\log(1.5),\log(3)\}$ using the data imputation approach for $\{N,n\}=\{2000, 400\}$, $\{50\%,75\%,90\%\}$ censoring, and simple random sampling of the validation subset. In this more complex error scenario, GRN had a small improvement in efficiency over HT, with its RE peaking around $1.05$ across all settings. GRMIS and GRMIC similarly showed minor efficiency improvements compared to HT with its RE ranging from $1$ to $1.06$. However, GRFCSMIS and GRFCSMIC had appreciable gains in efficiency, with RE ranging from $1.12$ to $1.25$ for all settings except for $90\%$ censoring, where the RE was less than $1.1$. These efficiency gains suggest that, in the presence of covariate measurement error that depends on the outcome, multiply imputing all error-prone variables was advantageous over only imputing the misclassified event indicator. Overall, GRFCSMI had lower MSE than all other estimators (albeit with some bias for $90\%$ censoring) and CPs that ranged from $94-95\%$ for all settings.

Results for $\{N,n\}=\{10000, 2000\}$, keeping all other parameters the same as Table~\ref{data_imp_err_1}, Supplementary Materials Table~\ref{data_imp_err_2}, and Table~\ref{data_imp_err_3}, are presented in Supplementary Materials Tables~\ref{data_imp_large_cohort_err_1}, \ref{data_imp_large_cohort_err_2}, and \ref{data_imp_large_cohort_err_3}, respectively. The conclusions for these large cohort settings were similar to those with $\{N,n\}=\{2000, 400\}$. For error scenario 1, GRMI provided appreciable efficiency gain over GRN. For error scenario 2, both GRMI and GRFCSMI provided comparable and significant efficiency gain over GRN. For error scenario 3, only GRFCSMI yielded appreciable efficiency gain over GRN and both GRMI and GRFCSMI were nearly unbiased even with $90\%$ censoring.

We present the type 1 error results under error scenario 3 for estimating $\beta_X=0$ using the data imputation approach for $\{N,n\}=\{10000, 2000\}$, $\{50\%,75\%,90\%\}$ censoring, and simple random sampling of the validation subset in Supplementary Materials Table~\ref{data_imp_type1_err}. For the $50\%$ and $75\%$ censoring levels, the type 1 error of the proposed GRMI and GRFCSMI estimators ranged from $0.052$ to $0.064$. For the $90\%$ censoring setting, the number of cases in the phase two data was very small at 40, and the type 1 error ranged from $0.068$ to $0.072$ for the proposed methods. However, we note that the type 1 error could likely be improved by using the bootstrap to calculate standard errors instead of the sandwich variance estimators (see \citet{oh2019raking} for more detail).

Results for the IF imputation approach under error scenario 3 for $\{N,n\}=\{2000, 400\}$, keeping all other parameters the same as Table~\ref{data_imp_err_3}, are presented in Table~\ref{IF_imp_err_3}. We note that the RE of the proposed estimators cannot be directly compared to those from the data imputation tables due to the HT ESE varying slightly. Overall, the conclusions for this approach were very similar to those of the data imputation approach. We observed that GRFCSMI was more efficient (by RE) and had lower MSE than all other estimators, albeit with some bias. Comparing the IF imputation estimators to the data imputation estimators, the ESE was very similar across all settings; this suggests that in the relatively simple error settings considered, the data imputation improved most of the auxiliary variable nonlinearity issues. Similar tables for error scenarios 1 and 2 are presented in Supplementary Materials Tables~\ref{IF_imp_err_1} and \ref{IF_imp_err_2} and similar conclusions were reached. Results for the IF imputation approach for $\{N,n\}=\{10000, 2000\}$, keeping all other parameters the same as Supplementary Tables~\ref{data_imp_large_cohort_err_1}, \ref{data_imp_large_cohort_err_2}, and \ref{data_imp_large_cohort_err_3}, are presented in Supplementary Materials Tables~\ref{IF_imp_large_cohort_err_1}, \ref{IF_imp_large_cohort_err_2}, and \ref{IF_imp_large_cohort_err_3}, respectively. The efficiency conclusions were similar to those observed under $\{N,n\}=\{2000, 400\}$, with the larger sample sizes again removing any observed bias. 

Table~\ref{data_imp_design_compare_err_3} presents the relative performance under error scenario 3 for estimating $\beta_X$ using the data imputation approach comparing simple random sampling to case-control and stratified case-control sampling where $\{N,n\}=\{4000, 800\}$ and censoring was $90\%$. GRFCSMI had increased efficiency compared to HT and GRN for nearly all designs whereas GRMI did not; however, the absolute gain in efficiency varied by sampling design. The RE for GRFCSMI was higher for SRS than for CC and SCC, ranging from $1.10$ to $1.15$ for SRS compared to $0.99$ to $1.11$ for CC and SCC. Although the RE for the proposed estimators was lower for the CC and SCC designs than for SRS, the actual standard errors (ESE and ASE) themselves were lower under these outcome-dependent designs. HT was quite inefficient under SRS, leading to a greater gain in efficiency for GRFCSMI; in contrast, HT under SCC was often nearly as efficient as GRFCSMI under SRS. For instance, the ESE of HT for $\beta_X=\log(3)$ and SCC is $0.126$, compared to the ESE of $0.128$ for GRFCSMIC for SRS. Similar conclusions were observed for error scenario 2 in Supplementary Materials Table~\ref{data_imp_design_compare_err_2}, with all other parameters the same as Table~\ref{data_imp_design_compare_err_3}, except both GRMI and GRFCSMI had slightly increased efficiency compared to HT and GRN for all designs. The RE for GRMI and GRFCSMI ranged from $1.21$ to $1.26$ for SRS; for CC and SCC designs, however, the RE ranged from $1.09$ to $1.15$. The RE for GRFCSMI was higher for SRS than for CC and SCC, ranging from $1.10$ to $1.15$ for SRS compared to $0.99$ to $1.11$ for CC and SCC. Thus, we observed less overall efficiency gain in the outcome-dependent sampling designs for the proposed methods but still constructed more efficient estimators generally. Results for the IF imputation approach, keeping all other parameters the same as Supplementary Materials Table~\ref{data_imp_design_compare_err_2} and Table~\ref{data_imp_design_compare_err_3}, are presented in Supplementary Materials Tables~\ref{IF_imp_design_compare_err_2} and~\ref{IF_imp_design_compare_err_3}, respectively. The conclusions follow very similarly to those of the data imputation approach. 

We considered the relative performance of our proposed methods under error scenario 3 where the misclassification generation process additionally included interaction terms (shown in Supplementary Materials Table~\ref{chpt3:misclass_interact_metrics}). Results for estimating $\beta_X$ using the data imputation and IF imputation approaches are shown in Supplementary Materials Tables~\ref{data_imp_misclass_interact_err_3} and \ref{IF_imp_misclass_interact_err_3}, respectively, with $\{N,n\}=\{2000, 400\}$ and simple random sampling of the validation subset. While the conclusions regarding the comparisons of GRMI and GRFCSMI to GRN were very similar to previous tables under error scenario 3, the efficiency gains of GRFCSMI were much larger than under the more simple misclassification scenarios. Overall, the RE ranged from $1.03$ to $1.34$ and the reduction in MSE compared to that of GRN was appreciable across all settings. These results suggest that our methods yield larger efficiency gains with increased nonlinearity. In addition, we observed greater efficiency gains for GRFCSMIC compared to GRFCSMIS, especially for $75\%$ and $90\%$ censoring where the positive predictive value (PPV) was very low. This high censoring and low PPV setting is common for EHR studies and thus suggests that more complex multiple imputation models to model potential nonlinearity would be helpful. The same set of results for error scenarios 1 and 2, namely with added interaction terms into the error models, were also generated (not presented) and we observed even greater efficiency gains for both GRMI and GRFCSMI with the more complex imputation approaches.

\section{VCCC Data example}\label{chpt3:dataexample}

In this section, we applied the proposed raking methods to electronic health records data on 4797 patients from the Vanderbilt Comprehensive Care Clinic (VCCC), a large HIV clinic. Health care providers at the clinic routinely collect and electronically record data on patients, including demographics, laboratory measurements, pharmacy dispensations, opportunistic infections, and vital status. A recent project at the VCCC performed a full chart review for all records to validate important clinical variables, including antiretroviral dispensations and AIDS-defining events (ADEs). Due to the comprehensive chart reviews, two full datasets were available; the first, which we refer to as the unvalidated data, contains the values for all patients prior to chart review and the second, which we refer to as the validated data, contains the true values after chart review. Additional details on the study design and data validation are in \citet{giganti2020tdmi}.

In this example, we were interested in estimating the association between the covariates CD4 cell count and age at the time of antiretroviral therapy (ART) and the outcome of time from the start of ART to the first ADE. As is common for studies based on EHR data, the outcome and covariates used in the analysis were derived variables. Specifically, CD4 cell count and age at the time of ART were extracted from tables of laboratory measurements and demographics, respectively, by matching the test date or visit date to the ART start date. In addition, the time from ART start to first ADE is extracted by finding the date of first opportunistic infection and the ART start date and calculating the time elapsed. A comparison of the unvalidated data to the validated data revealed errors in the ART start date in about $41\%$ of subjects, which led to downstream errors in the covariates and outcome of the statistical analysis. In addition, the ADE event was very rare with $93.8\%$ censoring and was subject to appreciable misclassification at $11\%$, suggesting that raking estimators that ignore the misclassification will be inefficient. The misclassification yielded sensitivity, specificity, positive predictive value, and negative predictive value of $0.879$, $0.892$, $0.351$, and $0.991$, respectively. The exact eligibility criteria used for the analysis and degree of measurement error in the covariates and outcome are given in Appendix I.

For this analysis, we considered the validated data to be the ``truth'' and defined the hazard ratio (HR) estimates calculated using the entire validated dataset to be the true, gold-standard estimates. The naive estimator that calculates the HRs using the entire unvalidated dataset was also considered, along with the HT estimator, the GRN estimator proposed by \citet{oh2019raking}, and the proposed raking estimators using multiple imputation (GRMI and GRFCSMI) for both the data imputation and IF imputation approaches. Although we had a fully validated dataset, we retrospectively sampled 100 different validation subsets as if we did not have validated data for all records in order to examine the estimators' performance. Due to the rare-event setting, we considered two different validation subset sampling designs: CC and SCC. Two variants of SCC were considered: 1) stratified case-control balanced (SCCB), which is described in Section~\ref{sim_setup}, and 2) stratified case-control Neyman allocation (SCCN), where the number of subjects sampled in each strata is proportional to the product of the phase one stratum size and the within-stratum (error-prone) influence function standard deviation. In addition, we considered two different validation subset sizes, 340 and 680, representing roughly $21\%$ and $43\%$ of the cohort respectively. For CC, all 248 error-prone cases were selected along with a random sample of 92 (or 432)  error-prone controls. For SCCB and SCCN, CD4 count was stratified at cutpoints of 100, 200, and 400 to create four discrete covariate groups for sampling. These cutpoints were selected to strategically oversample more informative subjects. Specifically, given that CD4 count is an important indicator of HIV severity, someone with CD4 count below $200$ $\textrm{cells}/\textrm{mm}^3$ is considered to be at high risk of getting an ADE. Thus, we selected cutpoints at $100$ and $200$ $\textrm{cells}/\textrm{mm}^3$ to oversample subjects clinically defined as high risk for an ADE to try to select more true cases and increase efficiency. For each sampling design, the same imputation models (both with and without interaction terms) and influence function working models were fit as described in the simulation section for error scenario 3 with CD4 cell count and age at ART start corresponding to $X^\star$ and $Z$, respectively.    

The median of the 100 HRs and the median of their corresponding $95\%$ confidence interval widths for the proposed methods using the data imputation approach are presented in Table~\ref{data_imp_vccc_table}. For each subset size and sampling design, the naive estimator had significant bias (calculated with respect to the true estimator) for both covariates ($31.3\%$ for CD4 and $31.1\%$ for age). In contrast, HT and all of the raking estimators yielded nearly unbiased estimates of the true estimates for both covariates. In addition, GRN had narrower $95\%$ confidence interval (CI) widths than that of HT for all sampling designs. For a subset size of 340, GRMI and GRFCSMI both had narrower CI widths than those of GRN for all sampling designs. However, the degree of efficiency gain differed by sampling design; namely, we observed a larger increase in efficiency (around a $5\%$ decrease in CI width) from GRMI and GRFCSMI under CC sampling compared to SCCB or SCCN (at most a $3\%$ decrease in CI width). GRMI and GRFCSMI under CC sampling had the narrowest median CI widths among all estimators for the 340 subset size. When the validation size was 680, the efficiency gain from GRMI and GRFCSMI over GRN was comparable across sampling designs and the median widths of the confidence intervals were similar. The more modest efficiency gains from GRMI and GRFCSMI over GRN compared to those observed in the simulations can likely be attributed to relatively poor imputation models. The small number of cases at phase one and low PPV of the error-prone event indicator made imputation models difficult to build due to the the validation subset containing an extremely small number of true cases. Across the 100 sampled validation subsets, the average ROC-AUC for the imputed event indicator ranged from $0.652$ to $0.670$ across all sampling designs, suggesting that the imputations of the event indicator were poor. Interestingly, GRMI had comparable, if not narrower, confidence interval widths than GRFCSMI across sampling designs and subset sizes. This is likely due to the fact that the amount of covariate error present was very small, which corresponds to error scenario 2 in the simulations where GRMI and GRFCSMI had comparable efficiency. Supplementary Materials Table~\ref{IF_imp_vccc_table} presents the median HRs and $95\%$ confidence interval widths across the 100 validation subsets for the IF imputation approach. The conclusions about the comparisons of the naive, HT, and GRN estimators are very similar to those of the data imputation approach. For both subset sizes, GRMI and GRFCSMI under CC and SCCB were less efficient than GRN, except for GRMIC under SCCB for the 340 subset size. GRMI and GRFCSMI under SCCN had slightly better performance, with narrower CI widths for the 340 subset size but not the 680 subset size. The lack of efficiency gains observed for the IF imputation approach can be attributed to the very poor influence function imputation working models. Across the 100 sampled validation subsets, the average R-squared for the CD4 influence function working models ranged from $0.099$ to $0.194$, indicating a lack of predictive accuracy. In small samples, such low correlation between the target and auxiliary variables can limit the improvement over the HT estimator, indicating the need to carefully examine the performance of the imputation working models, especially under complex error scenarios. In the rare event setting, validation sampling strategies that target missed true cases, such as by stratifying on risk factors that may be less prone to error, will also help efficiency.

\section{Discussion}\label{chpt3:discussion}

The increasing availability of EHR data collected on large patient populations has allowed researchers to study possible associations between a wide array of risk factors and health outcomes rapidly and cost-effectively. However, estimating such associations without bias requires precisely measured data on the variables of interest, an assumption that is often not met with EHR data due to errors in derived variables, error-prone record entry, or other error mechanisms. To address such bias, \citet{oh2019raking} proposed validating a subset of records and applying generalized raking estimators, including GRN studied in this manuscript. However, we demonstrated in this manuscript that GRN, which builds the raking variables from the error-prone data, is inefficient in the presence of event indicator misclassification. In addition, we proposed two classes of generalized raking estimators utilizing multiple imputation to estimate the optimal auxiliary variable, one that yields the optimal efficiency. Both MI approaches yield estimates of the expected value of the influence function for the target parameter based on the error-free data. The data imputation estimators impute either the event-indicator or all error-prone variables (if applicable) to construct auxiliary variables with increased degree of linearity with the true population influence functions. The IF imputation estimators take the data imputations and then fit a (potentially flexible, nonlinear) working model of the true population influence functions to construct auxiliary variables. These raking estimators are very appealing for the analysis of EHR data because their validity is not sensitive to the true measurement error structure nor do they require correct specification of the imputation or influence function working models, all of which are generally unknown for such large observational data. These features do, however, affect their efficiency and thus constructing estimators with increased efficiency has been the main focus of this manuscript. 

Overall, the proposed raking estimators using multiple imputation performed well, yielding nearly unbiased estimators, the highest RE, and the lowest MSE across all simulation settings. For settings involving misclassification only or misclassification and event-time error, both GRMI and GRFCSMI had large efficiency gains compared to GRN for all parameter settings. For the most complex error setting involving errors in the covariates, event-time, and event indicator, GRFCSMI had appreciable efficiency gains compared to GRN and GRMI for all parameter settings, which increased when nonlinear error functions were simulated. For all error scenarios, we observed more appreciable efficiency gains under $50\%$ and $75\%$ censoring compared to $90\%$ censoring. It is of note that these simulations involved error settings with very low sensitivity or PPV to mimic real EHR analysis scenarios. In simulations with higher sensitivity or PPV (not presented), larger efficiency gains were realized for GRMI and GRFCSMI, with RE greater than $1.5$. The data analysis, which involved an event with over $90\%$ censoring and very low PPV, resulted in similar conclusions. Nevertheless, we observed that GRMI and GRFCSMI yielded around a $5\%$ reduction in CI widths for both covariates, an appreciable gain in a data poor setting. In addition, we considered outcome-dependent sampling designs to select the validation subset to increase efficiency in rare event settings where the number of cases is small. Specifically, we evaluated case-control and stratified case-control sampling designs and found that while the gain in efficiency for GRMI or GRFCSMI over GRN is smaller compared to the efficiency gain under SRS, the overall standard errors are lower, yielding the most efficient estimates across all designs. While good imputation models are difficult to construct in rare events settings, one can still obtain more precise estimates overall by selecting more informative subjects to be validated at phase two. 

Another possible estimation approach for the considered settings is the direct multiple imputation estimator, which uses MI to impute the error-prone variables and plug into the Cox model to obtain estimates without the use of raking. \citet{giganti2020tdmi} considered this approach using discrete failure time models but noted challenges with correctly specifying the imputation model. While the MI estimator will be more efficient than raking estimators if the regression and imputation models are correctly specified, \citet{han2019combining} showed that in the nearly-true model framework of \citet{lumley2017robustness}, even slight misspecification of the models result in bias and worse MSE than raking. This robustness makes raking a very appealing approach in practical settings where the true models are generally unknown.

The two-phase design framework considered in this manuscript is a specific missing data setting where the data are missing by design. This missing data lens allows us to consider the augmented inverse probability weighted (AIPW) estimators proposed by \citet{robins1994estimation}, who showed that the class of AIPW estimators contains all regular asymptotically linear estimators consistent for the design-based parameter of interest. There is a close relationship between AIPW and raking estimators, in that the class of AIPW estimators contains the raking estimators, but the raking estimators include all of the best AIPW estimators (\citealp{lumley2011connections}). Thus, raking estimators are asymptotically efficient among design-based estimators and provide simple, easy to compute AIPW estimators. In particular, the raking estimators utilizing multiple imputation proposed in this manuscript yield practical methods to approximate the optimal AIPW estimator in settings involving complex measurement error that is often seen in EHR data. In addition, these estimators are consistent without requiring correct specification of the imputation or working models; however, they yield the most efficient design-based estimator if the models are correctly specified.

In this work we proposed a novel estimation method to improve raking estimators and showed additional efficiency could be gained by pairing these estimators with an efficient two-phase sampling design. While this manuscript considered outcome-dependent sampling designs to improve efficiency in rare-event settings, we believe that more theoretical and empirical work studying efficient sampling designs and their effects on efficiency for failure time outcomes is needed. In particular, constructing multi-phase sampling designs would be a fruitful avenue for future work. See \citet{mcisaac2015adaptive}, \citet{chen2020optimal}, and \citet{han2020twophase} for some initial work. These authors considered designs where a pilot sample could initially be selected from the cohort to obtain information on the validated data that can be used to guide follow-up sampling waves. We believe more work is needed to understand how best to take advantage of such strategies for the continuous failure time setting. 

\section{Acknowledgements}
We would like to thank Timothy Sterling, MD and the co-investigators of the Vanderbilt Comprehensive Care Clinic (VCCC) for use of their data. This work was supported in part by the U.S. National Institutes of Health (NIH) [R01-AI131771, P30-AI110527, R01-AI093234, U01-AI069923, and U01-AI069918] and the Patient Centered Outcomes Research Institute (PCORI) Award [R-1609-36207]. The statements in this manuscript are solely the responsibility of the authors and do not necessarily represent the views of NIH or PCORI.

\bibliographystyle{plainnat}
\bibliography{improving_raking}

\newpage 

\begin{table}[H]
\sisetup{
input-decimal-markers={.},
round-mode=places,
detect-all,
round-integer-to-decimal
}
\caption{Simulation results for estimating $\beta_x$ using the data imputation approach for error scenario 1 (error only in event indicator) with $N=2000$, $n=400$, and simple random sampling. The $\%$ bias, empirical standard error (ESE), relative efficiency (RE), average standard error (ASE), mean squared error, and coverage probabilities (CP) are presented for 2000 simulated datasets.}
\label{data_imp_err_1}
\centering 
\scalebox{0.80}{
\begin{tabular}{
c
c
c
c
*6{S[table-format=1.3,round-precision=3]}
}
$\beta_z$ & \% Cens & $\beta_x$ & Method & {\% Bias} & {ESE}  & {RE}   & {ASE}  & {MSE}  & {CP} \\ \hline
log(0.5)  & 50      & log(1.5)  & True   & -0.03595    & 0.039644 & 2.289193 & 0.039422 & 0.001572 & 0.956  \\
          &         &           & HT     & 1.228958    & 0.090753 & 1        & 0.087874 & 0.008261 & 0.949  \\
          &         &           & GRN    & 1.40684     & 0.07401  & 1.226214 & 0.072528 & 0.00551  & 0.95   \\
          &         &           & GRMIS  & -0.08937    & 0.065693 & 1.381469 & 0.06386  & 0.004316 & 0.948  \\
          &         &           & GRMIC  & -0.42527    & 0.064598 & 1.404892 & 0.063389 & 0.004176 & 0.946  \\
          &         &           &        &             &          &          &          &          &        \\
          &         & log(3)    & True   & 0.041168    & 0.041582 & 2.453957 & 0.04415  & 0.001729 & 0.948  \\
          &         &           & HT     & 0.631089    & 0.10204  & 1        & 0.097775 & 0.01046  & 0.939  \\
          &         &           & GRN    & 0.282312    & 0.082568 & 1.235824 & 0.080447 & 0.006827 & 0.942  \\
          &         &           & GRMIS  & 0.108883    & 0.072166 & 1.413959 & 0.069818 & 0.005209 & 0.948  \\
          &         &           & GRMIC  & 0.007399    & 0.072275 & 1.411835 & 0.069152 & 0.005224 & 0.948  \\
          &         &           &        &             &          &          &          &          &        \\
          & 75      & log(1.5)  & True   & 0.119394    & 0.051672 & 2.266392 & 0.053276 & 0.00267  & 0.954  \\
          &         &           & HT     & 0.781363    & 0.117109 & 1        & 0.118644 & 0.013725 & 0.952  \\
          &         &           & GRN    & 0.916624    & 0.097339 & 1.203106 & 0.096548 & 0.009489 & 0.945  \\
          &         &           & GRMIS  & 0.188371    & 0.093537 & 1.252017 & 0.091773 & 0.00875  & 0.944  \\
          &         &           & GRMIC  & 0.096736    & 0.096302 & 1.216058 & 0.090939 & 0.009274 & 0.94   \\
          &         &           &        &             &          &          &          &          &        \\
          &         & log(3)    & True   & -0.01311    & 0.06088  & 2.241353 & 0.059211 & 0.003706 & 0.949  \\
          &         &           & HT     & 1.034735    & 0.136454 & 1        & 0.131041 & 0.018749 & 0.938  \\
          &         &           & GRN    & 0.386125    & 0.119288 & 1.143905 & 0.113786 & 0.014248 & 0.934  \\
          &         &           & GRMIS  & 0.197862    & 0.102954 & 1.325384 & 0.102518 & 0.010604 & 0.943  \\
          &         &           & GRMIC  & 0.040924    & 0.101157 & 1.348933 & 0.101394 & 0.010233 & 0.944  \\
          &         &           &        &             &          &          &          &          &        \\
          & 90      & log(1.5)  & True   & 0.0138      & 0.084364 & 2.222885 & 0.083155 & 0.007117 & 0.947  \\
          &         &           & HT     & 1.805251    & 0.187531 & 1        & 0.184444 & 0.035222 & 0.943  \\
          &         &           & GRN    & 0.30929     & 0.167181 & 1.121725 & 0.165789 & 0.027951 & 0.94   \\
          &         &           & GRMIS  & 0.192308    & 0.161702 & 1.159732 & 0.160033 & 0.026148 & 0.944  \\
          &         &           & GRMIC  & -0.55691    & 0.159657 & 1.174587 & 0.158312 & 0.025495 & 0.936  \\
          &         &           &        &             &          &          &          &          &        \\
          &         & log(3)    & True   & -0.04654    & 0.088525 & 2.315872 & 0.089229 & 0.007837 & 0.95   \\
          &         &           & HT     & 1.160558    & 0.205013 & 1        & 0.197598 & 0.042193 & 0.938  \\
          &         &           & GRN    & 0.945284    & 0.194363 & 1.054793 & 0.187058 & 0.037885 & 0.941  \\
          &         &           & GRMIS  & 0.26163     & 0.175215 & 1.17007  & 0.16969  & 0.030708 & 0.94   \\
          &         &           & GRMIC  & -0.31527    & 0.17402  & 1.178102 & 0.169034 & 0.030295 & 0.939 
\end{tabular}
}
\end{table}

\begin{table}[H]
\sisetup{
input-decimal-markers={.},
round-mode=places,
detect-all,
round-integer-to-decimal
}
\caption{Simulation results for estimating $\beta_x$ using the data imputation approach for error scenario 3 (errors in event indicator, failure time, and X) with $N=2000$, $n=400$, and simple random sampling. The $\%$ bias, empirical standard error (ESE), relative efficiency (RE), average standard error (ASE), mean squared error, and coverage probabilities (CP) are presented for 2000 simulated datasets.}
\label{data_imp_err_3}
\centering 
\scalebox{0.80}{
\begin{tabular}{
c
c
c
c
*6{S[table-format=1.3,round-precision=3]}
}
$\beta_z$ & \% Cens & $\beta_x$ & Method & {\% Bias}  & {ESE}      & {RE}       & {ASE}      & {MSE}      & {CP}    \\ \hline
log(0.5)  & 50      & log(1.5)  & True   & 0.07661  & 0.039569 & 2.328964 & 0.039418 & 0.001566 & 0.95  \\
          &         &           & HT     & 1.342474 & 0.092155 & 1        & 0.088213 & 0.008522 & 0.937 \\
          &         &           & GRN    & 2.100967 & 0.093898 & 0.98143  & 0.087678 & 0.008889 & 0.928 \\
          &         &           & GRMIS  & 1.308134 & 0.092762 & 0.993454 & 0.088003 & 0.008633 & 0.935 \\
          &         &           & GRMIC  & 1.276032 & 0.092487 & 0.996408 & 0.088048 & 0.008581 & 0.935 \\
          &         &           & GRFCSMIS & 0.798683 & 0.075276 & 1.224217 & 0.074605 & 0.005677 & 0.948 \\
          &         &           & GRFCSMIC & 0.407051 & 0.073879 & 1.247364 & 0.074138 & 0.005461 & 0.942 \\
          &         &           &        &          &          &          &          &          &       \\
          &         & log(3)    & True   & -0.00837 & 0.041674 & 2.491381 & 0.04412  & 0.001737 & 0.951 \\
          &         &           & HT     & 0.777852 & 0.103825 & 1        & 0.097835 & 0.010853 & 0.944 \\
          &         &           & GRN    & 1.177403 & 0.101197 & 1.02597  & 0.097568 & 0.010408 & 0.943 \\
          &         &           & GRMIS  & 0.846726 & 0.103247 & 1.005597 & 0.097632 & 0.010747 & 0.944 \\
          &         &           & GRMIC  & 0.816276 & 0.103057 & 1.007453 & 0.097623 & 0.010701 & 0.945 \\
          &         &           & GRFCSMIS & 0.60425  & 0.088082 & 1.178736 & 0.087678 & 0.007802 & 0.939 \\
          &         &           & GRFCSMIC & 0.333361 & 0.088859 & 1.168426 & 0.087836 & 0.007909 & 0.938 \\
          &         &           &        &          &          &          &          &          &       \\
          & 75      & log(1.5)  & True   & -0.11172 & 0.050646 & 2.445003 & 0.053272 & 0.002565 & 0.946 \\
          &         &           & HT     & 2.494616 & 0.12383  & 1        & 0.119095 & 0.015436 & 0.945 \\
          &         &           & GRN    & 3.469044 & 0.121951 & 1.015403 & 0.116576 & 0.01507  & 0.944 \\
          &         &           & GRMIS  & 3.493033 & 0.123515 & 1.002552 & 0.117963 & 0.015456 & 0.94  \\
          &         &           & GRMIC  & 3.755253 & 0.123088 & 1.006027 & 0.117855 & 0.015383 & 0.938 \\
          &         &           & GRFCSMIS & 1.830123 & 0.107038 & 1.156879 & 0.103193 & 0.011512 & 0.946 \\
          &         &           & GRFCSMIC & 1.848374 & 0.106441 & 1.163367 & 0.102455 & 0.011386 & 0.947 \\
          &         &           &        &          &          &          &          &          &       \\
          &         & log(3)    & True   & -0.01819 & 0.05804  & 2.37266  & 0.05929  & 0.003369 & 0.948 \\
          &         &           & HT     & 0.939605 & 0.13771  & 1        & 0.13192  & 0.019071 & 0.95  \\
          &         &           & GRN    & 1.291495 & 0.133879 & 1.028617 & 0.129447 & 0.018125 & 0.947 \\
          &         &           & GRMIS  & 1.13204  & 0.134928 & 1.020621 & 0.130678 & 0.01836  & 0.948 \\
          &         &           & GRMIC  & 1.21211  & 0.137343 & 1.002675 & 0.130285 & 0.01904  & 0.947 \\
          &         &           & GRFCSMIS & 0.749482 & 0.123367 & 1.116261 & 0.119853 & 0.015287 & 0.946 \\
          &         &           & GRFCSMIC & 0.725826 & 0.120588 & 1.14199  & 0.119671 & 0.014605 & 0.944 \\
          &         &           &        &          &          &          &          &          &       \\
          & 90      & log(1.5)  & True   & 0.0138   & 0.084364 & 2.227607 & 0.083155 & 0.007117 & 0.947 \\
          &         &           & HT     & 2.839981 & 0.18793  & 1        & 0.184457 & 0.03545  & 0.944 \\
          &         &           & GRN    & 4.005694 & 0.180168 & 1.043079 & 0.178185 & 0.032724 & 0.94  \\
          &         &           & GRMIS  & 4.361114 & 0.177508 & 1.05871  & 0.17808  & 0.031822 & 0.937 \\
          &         &           & GRMIC  & 4.460343 & 0.178246 & 1.054326 & 0.176558 & 0.032099 & 0.936 \\
          &         &           & GRFCSMIS & 1.373064 & 0.176884 & 1.062444 & 0.170686 & 0.031319 & 0.943 \\
          &         &           & GRFCSMIC & 2.936905 & 0.173456 & 1.08344  & 0.169147 & 0.030229 & 0.938 \\
          &         &           &        &          &          &          &          &          &       \\
          &         & log(3)    & True   & -0.04654 & 0.088525 & 2.300257 & 0.089229 & 0.007837 & 0.95  \\
          &         &           & HT     & 0.99248  & 0.203631 & 1        & 0.198896 & 0.041584 & 0.945 \\
          &         &           & GRN    & 1.644558 & 0.192597 & 1.05729  & 0.193718 & 0.03742  & 0.942 \\
          &         &           & GRMIS  & 1.503862 & 0.196311 & 1.037287 & 0.192159 & 0.038811 & 0.945 \\
          &         &           & GRMIC  & 1.581827 & 0.19941  & 1.021165 & 0.19132  & 0.040067 & 0.943 \\
          &         &           & GRFCSMIS & 1.162629 & 0.195142 & 1.043502 & 0.189566 & 0.038243 & 0.947 \\
          &         &           & GRFCSMIC & 1.150689 & 0.196691 & 1.035282 & 0.188478 & 0.038847 & 0.946
\end{tabular}
}
\end{table}

\begin{table}[H]
\sisetup{
input-decimal-markers={.},
round-mode=places,
detect-all,
round-integer-to-decimal
}
\caption{Simulation results for estimating $\beta_x$ using the IF imputation approach for error scenario 3 (errors in event indicator, failure time, and X) with $N=2000$, $n=400$, and simple random sampling. The $\%$ bias, empirical standard error (ESE), relative efficiency (RE), average standard error (ASE), mean squared error, and coverage probabilities (CP) are presented for 2000 simulated datasets.}
\label{IF_imp_err_3}
\centering 
\scalebox{0.80}{
\begin{tabular}{
c
c
c
c
*6{S[table-format=1.3,round-precision=3]}
}
$\beta_z$ & \% Cens & $\beta_x$ & Method & {\% Bias}  & {ESE}      & {RE}       & {ASE}      & {MSE}      & {CP}    \\ \hline
log(0.5)  & 50      & log(1.5)  & True   & -0.03595 & 0.039644 & 2.357942 & 0.039422 & 0.001572 & 0.956 \\
          &         &           & HT     & 0.968647 & 0.093478 & 1        & 0.087968 & 0.008754 & 0.944 \\
          &         &           & GRN    & 2.48905  & 0.092254 & 1.013273 & 0.087589 & 0.008613 & 0.942 \\
          &         &           & GRMIS  & 1.27812  & 0.095618 & 0.977622 & 0.082415 & 0.00917  & 0.904 \\
          &         &           & GRMIC  & 0.702605 & 0.094581 & 0.988339 & 0.082057 & 0.008954 & 0.912 \\
          &         &           & GRFCSMIS & 1.176668 & 0.076746 & 1.218027 & 0.072894 & 0.005913 & 0.932 \\
          &         &           & GRFCSMIC & 0.766525 & 0.076361 & 1.224153 & 0.072638 & 0.005841 & 0.938 \\
          &         &           &        &          &          &          &          &          &       \\
          &         & log(3)    & True   & 0.041168 & 0.041582 & 2.490834 & 0.04415  & 0.001729 & 0.948 \\
          &         &           & HT     & 0.313211 & 0.103573 & 1        & 0.097851 & 0.010739 & 0.942 \\
          &         &           & GRN    & 0.725322 & 0.104082 & 0.995114 & 0.097532 & 0.010897 & 0.945 \\
          &         &           & GRMIS  & 1.421894 & 0.102001 & 1.015417 & 0.091883 & 0.010648 & 0.924 \\
          &         &           & GRMIC  & 1.487215 & 0.102937 & 1.006184 & 0.091601 & 0.010863 & 0.926 \\
          &         &           & GRFCSMIS & 0.262352 & 0.096256 & 1.076016 & 0.08654  & 0.009274 & 0.934 \\
          &         &           & GRFCSMIC & 0.102202 & 0.095132 & 1.088739 & 0.08656  & 0.009051 & 0.934 \\
          &         &           &        &          &          &          &          &          &       \\
          & 75      & log(1.5)  & True   & 0.119394 & 0.051672 & 2.316876 & 0.053276 & 0.00267  & 0.954 \\
          &         &           & HT     & 1.004049 & 0.119718 & 1        & 0.118566 & 0.014349 & 0.948 \\
          &         &           & GRN    & 1.661829 & 0.119968 & 0.997919 & 0.116507 & 0.014438 & 0.945 \\
          &         &           & GRMIS  & 4.68564  & 0.122646 & 0.976125 & 0.107653 & 0.015403 & 0.92  \\
          &         &           & GRMIC  & 5.039218 & 0.121839 & 0.98259  & 0.107163 & 0.015262 & 0.916 \\
          &         &           & GRFCSMIS & 1.012435 & 0.104425 & 1.146449 & 0.100447 & 0.010921 & 0.948 \\
          &         &           & GRFCSMIC & 1.16514  & 0.108355 & 1.104869 & 0.099865 & 0.011763 & 0.946 \\
          &         &           &        &          &          &          &          &          &       \\
          &         & log(3)    & True   & -0.01311 & 0.06088  & 2.250031 & 0.059211 & 0.003706 & 0.949 \\
          &         &           & HT     & 0.836351 & 0.136982 & 1        & 0.131293 & 0.018849 & 0.952 \\
          &         &           & GRN    & 1.114833 & 0.133936 & 1.022745 & 0.12923  & 0.018089 & 0.952 \\
          &         &           & GRMIS  & 1.098573 & 0.134396 & 1.019243 & 0.119741 & 0.018208 & 0.931 \\
          &         &           & GRMIC  & 1.354594 & 0.135155 & 1.013522 & 0.119708 & 0.018488 & 0.93  \\
          &         &           & GRFCSMIS & -0.52327 & 0.128106 & 1.069285 & 0.115569 & 0.016444 & 0.928 \\
          &         &           & GRFCSMIC & -0.46431 & 0.127535 & 1.074077 & 0.115312 & 0.016291 & 0.934 \\
          &         &           &        &          &          &          &          &          &       \\
          & 90      & log(1.5)  & True   & 0.0138   & 0.084364 & 2.251745 & 0.083155 & 0.007117 & 0.947 \\
          &         &           & HT     & 1.897751 & 0.189966 & 1        & 0.183082 & 0.036146 & 0.94  \\
          &         &           & GRN    & 1.897914 & 0.183304 & 1.036344 & 0.176042 & 0.03366  & 0.942 \\
          &         &           & GRMIS  & 8.193088 & 0.198884 & 0.955159 & 0.163381 & 0.040658 & 0.902 \\
          &         &           & GRMIC  & 8.29543  & 0.195141 & 0.97348  & 0.162322 & 0.039211 & 0.894 \\
          &         &           & GRFCSMIS & 4.745953 & 0.177903 & 1.067808 & 0.159259 & 0.03202  & 0.918 \\
          &         &           & GRFCSMIC & 3.798847 & 0.181029 & 1.049366 & 0.157469 & 0.033009 & 0.908 \\
          &         &           &        &          &          &          &          &          &       \\
          &         & log(3)    & True   & -0.04654 & 0.088525 & 2.348938 & 0.089229 & 0.007837 & 0.95  \\
          &         &           & HT     & 0.928622 & 0.207941 & 1        & 0.196985 & 0.043343 & 0.939 \\
          &         &           & GRN    & 1.061707 & 0.203441 & 1.022115 & 0.192655 & 0.041524 & 0.943 \\
          &         &           & GRMIS  & 4.095097 & 0.206598 & 1.006498 & 0.181218 & 0.044707 & 0.913 \\
          &         &           & GRMIC  & 3.94024  & 0.205562 & 1.011573 & 0.180065 & 0.044129 & 0.91  \\
          &         &           & GRFCSMIS & 1.614577 & 0.194645 & 1.068304 & 0.175683 & 0.038201 & 0.906 \\
          &         &           & GRFCSMIC & 1.290465 & 0.198216 & 1.049058 & 0.174164 & 0.039491 & 0.904
\end{tabular}
}
\end{table}

\begin{table}[H]
\sisetup{
input-decimal-markers={.},
round-mode=places,
detect-all,
round-integer-to-decimal
}
\caption{Simulation results for estimating $\beta_x$ using the data imputation approach for error scenario 3 (errors in event indicator, failure time, and X) with $N=4000$, $n=800$ comparing simple random sampling (SRS), case-control sampling (CC), and stratified case-control sampling (SCC). The $\%$ bias, empirical standard error (ESE), relative efficiency (RE), average standard error (ASE), mean squared error, and coverage probabilities (CP) are presented for 2000 simulated datasets.}
\label{data_imp_design_compare_err_3}
\centering 
\scalebox{0.80}{
\begin{tabular}{
c
c
c
c
c
*6{S[table-format=1.3,round-precision=3]}
}
$\beta_z$ & \% Cens & $\beta_x$ & Design & Method & {\% Bias}  & {ESE}      & {RE}       & {ASE}      & {MSE}      & {CP}    \\ \hline
log(0.5)  & 90      & log(1.5)  & SRS    & True   & -0.19575 & 0.056825 & 2.306052 & 0.058701 & 0.00323  & 0.953 \\
          &         &           &        & HT     & 1.348122 & 0.131041 & 1        & 0.130666 & 0.017202 & 0.943 \\
          &         &           &        & GRN    & 0.599619 & 0.123075 & 1.064728 & 0.120298 & 0.015153 & 0.942 \\
          &         &           &        & GRMIS  & 1.352229 & 0.120274 & 1.089521 & 0.121238 & 0.014496 & 0.942 \\
          &         &           &        & GRMIC  & 1.064436 & 0.12481  & 1.049923 & 0.120657 & 0.015596 & 0.938 \\
          &         &           &        & GRFCSMIS & 0.372602 & 0.116359 & 1.126176 & 0.115015 & 0.013542 & 0.938 \\
          &         &           &        & GRFCSMIC & 0.262224 & 0.118828 & 1.102777 & 0.114345 & 0.014121 & 0.936 \\
          &         &           &        &        &          &          &          &          &          &       \\
          &         &           & CC     & True   & -0.19575 & 0.056825 & 2.307166 & 0.058701 & 0.00323  & 0.953 \\
          &         &           &        & HT     & 1.278795 & 0.131104 & 1        & 0.121309 & 0.017215 & 0.938 \\
          &         &           &        & GRN    & 1.295066 & 0.128054 & 1.023824 & 0.120734 & 0.016425 & 0.943 \\
          &         &           &        & GRMIS  & 1.925384 & 0.129153 & 1.015113 & 0.122768 & 0.016741 & 0.942 \\
          &         &           &        & GRMIC  & 1.665403 & 0.12981  & 1.009974 & 0.123029 & 0.016896 & 0.94  \\
          &         &           &        & GRFCSMIS & 1.221831 & 0.119855 & 1.093861 & 0.11281  & 0.01439  & 0.938 \\
          &         &           &        & GRFCSMIC & 0.804186 & 0.117846 & 1.112503 & 0.112475 & 0.013898 & 0.938 \\
          &         &           &        &        &          &          &          &          &          &       \\
          &         &           & SCC    & True   & -0.19575 & 0.056825 & 1.941845 & 0.058701 & 0.00323  & 0.953 \\
          &         &           &        & HT     & -0.6459  & 0.110345 & 1        & 0.110845 & 0.012183 & 0.957 \\
          &         &           &        & GRN    & -0.09306 & 0.109473 & 1.00797  & 0.110642 & 0.011984 & 0.952 \\
          &         &           &        & GRMIS  & 0.081196 & 0.110714 & 0.996668 & 0.111346 & 0.012258 & 0.954 \\
          &         &           &        & GRMIC  & -0.02695 & 0.108453 & 1.017446 & 0.111431 & 0.011762 & 0.954 \\
          &         &           &        & GRFCSMIS & -0.16322 & 0.101909 & 1.082777 & 0.105767 & 0.010386 & 0.954 \\
          &         &           &        & GRFCSMIC & -0.10748 & 0.100555 & 1.097364 & 0.105699 & 0.010111 & 0.952 \\
          &         &           &        &        &          &          &          &          &          &       \\
          &         & log(3)    & SRS    & True   & 0.1293   & 0.064842 & 2.25486  & 0.06303  & 0.004206 & 0.954 \\
          &         &           &        & HT     & 0.974558 & 0.146209 & 1        & 0.140603 & 0.021492 & 0.948 \\
          &         &           &        & GRN    & 0.744679 & 0.129418 & 1.129747 & 0.130516 & 0.016816 & 0.94  \\
          &         &           &        & GRMIS  & 0.713557 & 0.131614 & 1.110893 & 0.131276 & 0.017384 & 0.942 \\
          &         &           &        & GRMIC  & 0.650456 & 0.131029 & 1.115852 & 0.131065 & 0.01722  & 0.94  \\
          &         &           &        & GRFCSMIS & 0.627308 & 0.127227 & 1.149195 & 0.127457 & 0.016234 & 0.942 \\
          &         &           &        & GRFCSMIC & 0.60765  & 0.128461 & 1.138158 & 0.126735 & 0.016547 & 0.944 \\
          &         &           &        &        &          &          &          &          &          &       \\
          &         &           & CC     & True   & 0.1293   & 0.064842 & 2.208732 & 0.06303  & 0.004206 & 0.954 \\
          &         &           &        & HT     & 1.422661 & 0.143218 & 1        & 0.130477 & 0.020756 & 0.928 \\
          &         &           &        & GRN    & 1.646294 & 0.141186 & 1.014393 & 0.129232 & 0.020261 & 0.927 \\
          &         &           &        & GRMIS  & 1.614425 & 0.1409   & 1.016452 & 0.130462 & 0.020167 & 0.931 \\
          &         &           &        & GRMIC  & 1.506875 & 0.139858 & 1.024024 & 0.130487 & 0.019834 & 0.926 \\
          &         &           &        & GRFCSMIS & 1.395031 & 0.13998  & 1.023132 & 0.124715 & 0.019829 & 0.925 \\
          &         &           &        & GRFCSMIC & 1.32011  & 0.137537 & 1.041307 & 0.124594 & 0.019127 & 0.922 \\
          &         &           &        &        &          &          &          &          &          &       \\
          &         &           & SCC    & True   & 0.1293   & 0.064842 & 1.938671 & 0.06303  & 0.004206 & 0.954 \\
          &         &           &        & HT     & 0.82001  & 0.125707 & 1        & 0.123465 & 0.015883 & 0.938 \\
          &         &           &        & GRN    & 0.693561 & 0.126412 & 0.99442  & 0.122793 & 0.016038 & 0.94  \\
          &         &           &        & GRMIS  & 0.733702 & 0.126538 & 0.99343  & 0.123577 & 0.016077 & 0.94  \\
          &         &           &        & GRMIC  & 0.70857  & 0.127711 & 0.984303 & 0.123601 & 0.016371 & 0.936 \\
          &         &           &        & GRFCSMIS & 0.771774 & 0.127503 & 0.985911 & 0.119766 & 0.016329 & 0.944 \\
          &         &           &        & GRFCSMIC & 0.614896 & 0.124678 & 1.008254 & 0.119554 & 0.01559  & 0.946
\end{tabular}
}
\end{table}

\begin{table}[H]
\sisetup{
input-decimal-markers={.},
round-mode=places,
detect-all,
round-integer-to-decimal
}
\caption{The median hazard ratios (HR) and their corresponding $95\%$ confidence interval widths calculated using the data imputation method from 100 different sampled validation subsets for a 100 cell/$\textrm{mm}^3$ increase in CD4 count at ART initiation and 10-year increase in age at CD4 count measurement.}
\label{data_imp_vccc_table}
\centering 
\scalebox{0.75}{
\begin{tabular}{
c
c
c
*4{S[table-format=1.3,round-precision=3]}
}
Subset size & Sampling & Method   & {CD4 HR} & {CD4 CI width} & {Age HR} & {Age CI width} \\ \hline
340         & CC       & True     & 0.693  & 0.19         & 0.829  & 0.361        \\
            &          & Naive    & 0.91   & 0.125        & 1.087  & 0.275        \\
            &          & HT       & 0.669  & 0.313        & 0.829  & 0.579        \\
            &          & GRN      & 0.674  & 0.274        & 0.819  & 0.465        \\
            &          & GRMIS    & 0.679  & 0.26         & 0.824  & 0.44         \\
            &          & GRMIC    & 0.678  & 0.264        & 0.83   & 0.438        \\
            &          & GRFCSMIS & 0.675  & 0.265        & 0.824  & 0.444        \\
            &          & GRFCSMIC & 0.677  & 0.261        & 0.824  & 0.44         \\
            &          &          &        &              &        &              \\
            & SCCB     & True     & 0.693  & 0.19         & 0.829  & 0.361        \\
            &          & Naive    & 0.91   & 0.125        & 1.087  & 0.275        \\
            &          & HT       & 0.686  & 0.283        & 0.823  & 0.573        \\
            &          & GRN      & 0.687  & 0.28         & 0.82   & 0.494        \\
            &          & GRMIS    & 0.689  & 0.272        & 0.835  & 0.496        \\
            &          & GRMIC    & 0.689  & 0.278        & 0.826  & 0.491        \\
            &          & GRFCSMIS & 0.687  & 0.275        & 0.839  & 0.498        \\
            &          & GRFCSMIC & 0.689  & 0.276        & 0.814  & 0.495        \\
            &          &          &        &              &        &              \\
            & SCCN     & True     & 0.693  & 0.19         & 0.829  & 0.361        \\
            &          & Naive    & 0.91   & 0.125        & 1.087  & 0.275        \\
            &          & HT       & 0.69   & 0.308        & 0.779  & 0.665        \\
            &          & GRN      & 0.688  & 0.308        & 0.807  & 0.599        \\
            &          & GRMIS    & 0.684  & 0.303        & 0.813  & 0.608        \\
            &          & GRMIC    & 0.684  & 0.299        & 0.807  & 0.596        \\
            &          & GRFCSMIS & 0.687  & 0.302        & 0.818  & 0.614        \\
            &          & GRFCSMIC & 0.69   & 0.297        & 0.803  & 0.598        \\
            &          &          &        &              &        &              \\
680         & CC       & True     & 0.693  & 0.19         & 0.829  & 0.361        \\
            &          & Naive    & 0.91   & 0.125        & 1.087  & 0.275        \\
            &          & HT       & 0.692  & 0.237        & 0.826  & 0.412        \\
            &          & GRN      & 0.693  & 0.23         & 0.825  & 0.385        \\
            &          & GRMIS    & 0.693  & 0.228        & 0.826  & 0.38         \\
            &          & GRMIC    & 0.697  & 0.228        & 0.826  & 0.382        \\
            &          & GRFCSMIS & 0.693  & 0.228        & 0.826  & 0.383        \\
            &          & GRFCSMIC & 0.696  & 0.229        & 0.821  & 0.382        \\
            &          &          &        &              &        &              \\
            & SCCB     & True     & 0.693  & 0.190        & 0.829  & 0.361        \\
            &          & Naive    & 0.910  & 0.125        & 1.087  & 0.275        \\
            &          & HT       & 0.695  & 0.234        & 0.837  & 0.416        \\
            &          & GRN      & 0.695  & 0.233        & 0.830  & 0.395        \\
            &          & GRMIS    & 0.693  & 0.232        & 0.829  & 0.393        \\
            &          & GRMIC    & 0.697  & 0.233        & 0.831  & 0.393        \\
            &          & GRFCSMIS & 0.693  & 0.231        & 0.826  & 0.393        \\
            &          & GRFCSMIC & 0.694  & 0.232        & 0.832  & 0.394        \\
            &          &          &        &              &        &              \\
            & SCCN     & True     & 0.693  & 0.19         & 0.829  & 0.361        \\
            &          & Naive    & 0.91   & 0.125        & 1.087  & 0.275        \\
            &          & HT       & 0.69   & 0.229        & 0.826  & 0.43         \\
            &          & GRN      & 0.689  & 0.228        & 0.821  & 0.406        \\
            &          & GRMIS    & 0.689  & 0.226        & 0.823  & 0.404        \\
            &          & GRMIC    & 0.689  & 0.228        & 0.825  & 0.401        \\
            &          & GRFCSMIS & 0.689  & 0.226        & 0.822  & 0.403        \\
            &          & GRFCSMIC & 0.689  & 0.228        & 0.821  & 0.406       
\end{tabular}
}
\end{table}

\pagebreak
\newpage

\begin{center}
\textbf{\Large Improved Generalized Raking Estimators Supplementary Materials}
\end{center}

\begin{center}
Eric J. Oh$^{\star 1}$, Bryan E. Shepherd$^{2}$, Thomas Lumley$^{3}$, Pamela A. Shaw$^{1}$  \\
\vspace{0.1in}
$^1$University of Pennsylvania, Perelman School of Medicine \\
Department of Biostatistics, Epidemiology, and Informatics \\
\vspace{0.1in}
$^{2}$Vanderbilt University School of Medicine \\
Department of Biostatistics \\
\vspace{0.1in}
$^{3}$University of Auckland \\
Department of Statistics
\end{center}

\setcounter{equation}{0}
\setcounter{figure}{0}
\setcounter{table}{0}
\makeatletter
\renewcommand{\theequation}{S\arabic{equation}}
\renewcommand{\thefigure}{S\arabic{figure}}
\renewcommand{\bibnumfmt}[1]{[S#1]}
\renewcommand{\citenumfont}[1]{S#1}

\section*{Appendix A: Misclassification metrics for simulations}\label{chpt3:misclass_metrics}

\begin{table}[H]
\caption{The sensitivity (Sens), specificity (Spec), positive predictive value (PPV), and negative predictive value (NPV) for the event indicator generated for error scenarios 1, 2, and 3 in the simple random sampling simulations.}
\label{chpt3:misclass_srs_metrics}
\centering
\begin{tabular}{ccccccc}
$\beta_z$   & $\%$ Cens & $\beta_x$ & Sens     & Spec     & PPV      & NPV      \\ \hline
$\log(0.5)$ & 50        & log(1.5)  & 0.465 & 0.947 & 0.878 & 0.684 \\
            &           & log(3)    & 0.479 & 0.948 & 0.893 & 0.669 \\
            &           &           &          &          &          &          \\
            & 75        & log(1.5)  & 0.672 & 0.905 & 0.693 & 0.897 \\
            &           & log(3)    & 0.705 & 0.889 & 0.659 & 0.908 \\
            &           &           &          &          &          &          \\
            & 90        & log(1.5)  & 0.822 & 0.820  & 0.330  & 0.977 \\
            &           & log(3)    & 0.819 & 0.796 & 0.294 & 0.977
\end{tabular}
\end{table}

\begin{table}[H]
\caption{Misclassification generation process for the sampling design comparison simulations. The sensitivity (Sens), specificity (Spec), positive predictive value (PPV), and negative predictive value (NPV) for the event indicator are presented.}
\label{chpt3:misclass_design_metrics}
\centering 
\begin{tabular}{cccccccc}
$\beta_z$   & $\%$ Cens & $\beta_x$ & $\Delta^{\star}$                                                  & Sens  & Spec  & PPV   & NPV   \\ \hline
$\log(0.5)$ & 90        & log(1.5)  & \begin{tabular}[c]{@{}c@{}}$\textrm{Bernoulli}(\textrm{expit}(-1+4*\Delta+$\\ $0.5*X-0.5*U-0.5*Z))$\end{tabular}   & 0.718 & 0.961 & 0.665 & 0.969 \\
            &           & log(3)    & \begin{tabular}[c]{@{}c@{}}$\textrm{Bernoulli}(\textrm{expit}(-1.5+4*\Delta+$\\ $0.5*X-0.5*U-0.5*Z))$\end{tabular} & 0.715 & 0.970 & 0.710 & 0.971
\end{tabular}
\end{table}

\section*{Appendix B: Multiple Imputation Details}\label{chpt3:MI_details}

We explicate the multiple imputation implementation details below for imputation models without interaction terms. Define
\begin{itemize}
\item $V_i=(1_i,\Delta_i^{\star},X_i^{\star},U_i^{\star},Z_i)'$
\item $V_{-U,i}=(1_i,\Delta_i^{\star},X_i^{\star},Z_i)'$
\item $V^{(l)}_{\Delta,i} = (1_i,\Delta^{\star}_i,\hat{X}^{(l-1)}_i,\hat{U}^{(l-1)}_i,Z_i)$
\item $V^{(l)}_{X,i} = (1_i,\hat{\Delta}^{(l)}_i,X^{\star}_i,\hat{U}^{(l-1)}_i,Z_i)$ 
\item $V^{(l)}_{U,i} = (1_i,\hat{\Delta}^{(l)}_i,\hat{X}^{(l)}_i,Z_i)$,
\end{itemize}
and let the lower case versions denote their observed counterparts. MI with interaction terms follows exactly the same except the terms defined above contain all possible interaction terms. 

\subsection*{Multiple imputation for $\Delta$ only}
\begin{enumerate}
    \item Fit the logistic regression model $\textrm{logit}(P(\Delta_i=1)|V_i=v_i)=v_i'\eta$ using the validation subset to obtain $\hat{\eta}$. This corresponds to characterizing a posterior distribution for $\eta$ given the phase two data under a non-informative prior distribution.
    \item For $m=1,\ldots,M$ iterations:
    \item Generate $\eta_{\star}^{(m)} \sim \textrm{N}(\hat{\eta},\tau^2_{\Delta,\star}(V'V)^{-1})$, where $\tau^2_{\Delta,\star} \sim \hat{\tau}^2_{\Delta}\frac{n-p_\eta}{\chi^2_{n-p_\eta}}$, $\hat{\tau}^2_{\Delta}$ is the squared sum of the working residuals from the logistic regression model, and $p_\eta$ is the dimension of $\eta$.
    \item Sample and impute $\hat{\Delta}_i^{(m)} \sim \textrm{Bernoulli}(\textrm{expit}(v'_i\eta_{\star}^{(m)}))$ for all phase one subjects.
    \item Stop after $M$ iterations
\end{enumerate}

\subsection*{Fully conditional specification multiple imputation}
\begin{enumerate}
    \item Fit the logistic regression model $\textrm{logit}(P(\Delta_i=1)|V_i=v_i)=v_i'\eta_V$ and linear regression models $\mathrm{E}(X_i|V_i=v_i) = v_i'\theta_V$ and $\mathrm{E}(R_i|V_{-U,i}=v_{-U,i})=v'_{-U,i}\omega_V$ using the validation subset to obtain $\hat{\eta}_V$, $\hat{\theta}_V$, and $\hat{\omega}_V$.
    \item For $m=1,\ldots,M$ iterations:
    \item Generate $\eta_{\star}^{(0)} \sim \textrm{N}(\hat{\eta}_V,\tau^2_{\Delta,V,\star}(V'V)^{-1})$, $\theta_{\star}^{(0)} \sim \textrm{N}(\hat{\theta}_V,\tau^2_{X,V,\star}(V'V)^{-1})$, and $\omega_{\star}^{(0)} \sim \textrm{N}(\hat{\omega}_V,\tau^2_{U,V,\star}(V'_{-U}V_{-U})^{-1})$ where $\tau^2_{\Delta,V,\star} \sim \hat{\tau}^2_{\Delta,V}\frac{n-p_{\eta_V}}{\chi^2_{n-p_{\eta_V}}}$, $\tau^2_{X,V,\star} \sim \hat{\tau}^2_{X,V}\frac{n-p_{\theta_V}}{\chi^2_{n-p_{\theta_V}}}$, $\tau^2_{U,V,\star} \sim \hat{\tau}^2_{U,V}\frac{n-p_{\omega_V}}{\chi^2_{n-p_{\omega_V}}}$, $\hat{\tau}^2_{\Delta,V}$, $\hat{\tau}^2_{X,V}$, and $\hat{\tau}^2_{U,V}$ are the squared sum of working residuals/residual sum of squares from their respective regression models, and $p_{\eta_V}$, $p_{\theta_V}$, and $p_{\omega_V}$ are the dimensions of their respective parameters. 
    \item Sample and impute $\hat{\Delta}_i^{(0)} \sim \textrm{Bernoulli}(\textrm{expit}(v'_i\eta_{\star}^{(0)}))$ and $\hat{X}_i^{(0)} \sim \textrm{N}(v'_i\theta_{\star}^{(0)}, \tau^2_{X,V,\star})$ for all phase one subjects. Sample $\hat{R}_i^{(0)} \sim \textrm{N}(v'_{-U,i}\omega_{\star}^{(0)},\tau^2_{U,V,\star})$ and impute $\hat{U}^{(0)}_i = U^{\star}_i-\hat{R}^{(0)}_i$ for all phase one subjects.
    \item For $l=1,\ldots,L$ iterations:
    \item Fit the logistic regression model $\textrm{logit}(P(\Delta_i=1)|V^{(l)}_{\Delta,i}=v^{(l)}_{\Delta,i})=v'^{(l)}_{\Delta,i}\eta$ on the validation subset to obtain $\hat{\eta}^{(l)}$.
    \item Generate $\eta_{\star}^{(l)} \sim \textrm{N}(\hat{\eta}^{(l)},\tau^2_{\Delta,\star}(V_{\Delta}^{(l)'}V_{\Delta}^{(l)})^{-1})$ where $\tau^2_{\Delta,\star} \sim \hat{\tau}^2_{\Delta}\frac{n-p_{\eta}}{\chi^2_{n-p_{\eta}}}$.
    \item Sample and impute $\hat{\Delta}_i^{(l)} \sim \textrm{Bernoulli}(\textrm{expit}(v^{(l)'}_{\Delta,i}\eta_{\star}^{(l)}))$ for all phase one subjects.
    \item Fit the linear regression model $\mathrm{E}(X_i|V^{(l)}_{X,i}=v^{(l)}_{X,i}) = v'^{(l)}_{X,i}\theta$ on the validation subset to obtain $\hat{\theta}^{(l)}$.
    \item Generate $\theta_{\star}^{(l)} \sim \textrm{N}(\hat{\theta}^{(l)},\tau^2_{X,\star}(V_{X}^{(l)'}V_{X}^{(l)})^{-1})$ where $\tau^2_{X,\star} \sim \hat{\tau}^2_{X}\frac{n-p_{\theta}}{\chi^2_{n-p_{\theta}}}$.
    \item Sample and impute $\hat{X}^{(l)}_i \sim \textrm{N}(v^{(l)'}_{X,i}\theta_{\star}^{(l)}, \tau^2_{X,\star})$ for all phase one subjects.
    \item Fit the linear regression model $\mathrm{E}(R_i|V^{(l)}_{U,i}=v^{(l)}_{U,i}) = v'^{(l)}_{U,i}\omega$ on the validation subset to obtain $\hat{\omega}^{(l)}$.
    \item Generate $\omega_{\star}^{(l)} \sim \textrm{N}(\hat{\omega}^{(l)},\tau^2_{U,\star}(V_{U}^{(l)'}V_{U}^{(l)})^{-1})$ where $\tau^2_{U,\star} \sim \hat{\tau}^2_{U}\frac{n-p_{\omega}}{\chi^2_{n-p_{\omega}}}$.
    \item Sample $\hat{R}^{(l)}_i \sim \textrm{N}(v^{(l)'}_{U,i}\omega_{\star}^{(l)},\tau^2_{U,\star})$ and impute $\hat{U}^{(l)}_i = U^{\star}_i - \hat{R}^{(l)}_i$ for all phase one subjects.
    \item Stop after $L$ iterations
    \item Stop after $M$ iterations
\end{enumerate}

\section*{Appendix C: Error scenario 2 - data imputation results}

\begin{table}[H]
\sisetup{
input-decimal-markers={.},
round-mode=places,
detect-all,
round-integer-to-decimal
}
\caption{Simulation results for estimating $\beta_x$ using the data imputation approach for error scenario 2 (errors in event indicator and failure time) with $N=2000$, $n=400$, and simple random sampling. The $\%$ bias, empirical standard error (ESE), relative efficiency (RE), average standard error (ASE), mean squared error, and coverage probabilities (CP) are presented for 2000 simulated datasets.}
\label{data_imp_err_2}
\centering 
\scalebox{0.80}{
\begin{tabular}{
c
c
c
c
*6{S[table-format=1.3,round-precision=3]}
}
$\beta_z$ & \% Cens & $\beta_x$ & Method & {\% Bias}  & {ESE}      & {RE}       & {ASE}      & {MSE}      & {CP}    \\ \hline
log(0.5)  & 50      & log(1.5)  & True   & 0.07661  & 0.039569 & 2.322461 & 0.039418 & 0.001566 & 0.95  \\
          &         &           & HT     & 0.862615 & 0.091897 & 1        & 0.087962 & 0.008457 & 0.937 \\
          &         &           & GRN    & 0.858763 & 0.076007 & 1.209062 & 0.072554 & 0.005789 & 0.946 \\
          &         &           & GRMIS  & 1.031894 & 0.06415  & 1.432527 & 0.063852 & 0.004133 & 0.948 \\
          &         &           & GRMIC  & 0.855469 & 0.064549 & 1.423681 & 0.063311 & 0.004179 & 0.948 \\
          &         &           & GRFCSMIS & 0.85294  & 0.063187 & 1.454371 & 0.063714 & 0.004005 & 0.95  \\
          &         &           & GRFCSMIC & 0.876251 & 0.064668 & 1.421072 & 0.063193 & 0.004195 & 0.947 \\
          &         &           &        &          &          &          &          &          &       \\
          &         & log(3)    & True   & -0.00837 & 0.041674 & 2.365635 & 0.04412  & 0.001737 & 0.951 \\
          &         &           & HT     & 0.44906  & 0.098585 & 1        & 0.097536 & 0.009743 & 0.942 \\
          &         &           & GRN    & 0.21076  & 0.081199 & 1.214119 & 0.080367 & 0.006599 & 0.944 \\
          &         &           & GRMIS  & 0.017758 & 0.070751 & 1.393417 & 0.070419 & 0.005006 & 0.944 \\
          &         &           & GRMIC  & 0.054635 & 0.070446 & 1.399443 & 0.069713 & 0.004963 & 0.946 \\
          &         &           & GRFCSMIS & 0.097153 & 0.068036 & 1.449007 & 0.070358 & 0.00463  & 0.948 \\
          &         &           & GRFCSMIC & -0.0385  & 0.0696   & 1.416459 & 0.069488 & 0.004844 & 0.944 \\
          &         &           &        &          &          &          &          &          &       \\
          & 75      & log(1.5)  & True   & -0.11172 & 0.050646 & 2.511236 & 0.053272 & 0.002565 & 0.946 \\
          &         &           & HT     & 1.592927 & 0.127184 & 1        & 0.118928 & 0.016218 & 0.938 \\
          &         &           & GRN    & 0.404828 & 0.099182 & 1.282334 & 0.096779 & 0.00984  & 0.946 \\
          &         &           & GRMIS  & 0.15642  & 0.091196 & 1.394624 & 0.091856 & 0.008317 & 0.941 \\
          &         &           & GRMIC  & -0.60126 & 0.093375 & 1.362087 & 0.090987 & 0.008725 & 0.945 \\
          &         &           & GRFCSMIS & -0.27014 & 0.091629 & 1.388037 & 0.091734 & 0.008397 & 0.942 \\
          &         &           & GRFCSMIC & -0.9402  & 0.090891 & 1.399307 & 0.090986 & 0.008276 & 0.945 \\
          &         &           &        &          &          &          &          &          &       \\
          &         & log(3)    & True   & -0.01819 & 0.05804  & 2.371545 & 0.05929  & 0.003369 & 0.948 \\
          &         &           & HT     & 0.563709 & 0.137646 & 1        & 0.131856 & 0.018985 & 0.938 \\
          &         &           & GRN    & 0.513363 & 0.122738 & 1.121461 & 0.114183 & 0.015096 & 0.938 \\
          &         &           & GRMIS  & 0.130314 & 0.111008 & 1.239958 & 0.103685 & 0.012325 & 0.946 \\
          &         &           & GRMIC  & -0.18315 & 0.110439 & 1.246354 & 0.10261  & 0.012201 & 0.941 \\
          &         &           & GRFCSMIS & -0.0051  & 0.109314 & 1.259179 & 0.103718 & 0.01195  & 0.942 \\
          &         &           & GRFCSMIC & -0.31075 & 0.109638 & 1.255454 & 0.102901 & 0.012032 & 0.939 \\
          &         &           &        &          &          &          &          &          &       \\
          & 90      & log(1.5)  & True   & 0.0138   & 0.084364 & 2.231975 & 0.083155 & 0.007117 & 0.947 \\
          &         &           & HT     & 1.89313  & 0.188298 & 1        & 0.184107 & 0.035515 & 0.944 \\
          &         &           & GRN    & 0.747537 & 0.167743 & 1.122541 & 0.166189 & 0.028147 & 0.94  \\
          &         &           & GRMIS  & 0.392568 & 0.160833 & 1.170769 & 0.159171 & 0.02587  & 0.929 \\
          &         &           & GRMIC  & 0.50219  & 0.162774 & 1.156806 & 0.156793 & 0.0265   & 0.928 \\
          &         &           & GRFCSMIS & 0.250184 & 0.160999 & 1.169562 & 0.159623 & 0.025922 & 0.933 \\
          &         &           & GRFCSMIC & 0.860638 & 0.163186 & 1.153889 & 0.157477 & 0.026642 & 0.93  \\
          &         &           &        &          &          &          &          &          &       \\
          &         & log(3)    & True   & -0.04654 & 0.088525 & 2.286886 & 0.089229 & 0.007837 & 0.95  \\
          &         &           & HT     & 1.420837 & 0.202447 & 1        & 0.199373 & 0.041229 & 0.944 \\
          &         &           & GRN    & 1.282639 & 0.188611 & 1.073361 & 0.188717 & 0.035773 & 0.946 \\
          &         &           & GRMIS  & 0.11986  & 0.177052 & 1.143433 & 0.175632 & 0.031349 & 0.944 \\
          &         &           & GRMIC  & -0.27151 & 0.175575 & 1.153052 & 0.173819 & 0.030836 & 0.938 \\
          &         &           & GRFCSMIS & 0.852398 & 0.177915 & 1.137887 & 0.176882 & 0.031741 & 0.945 \\
          &         &           & GRFCSMIC & -0.01923 & 0.176458 & 1.147285 & 0.173343 & 0.031137 & 0.935
\end{tabular}
}
\end{table}

\section*{Appendix D: Large cohort simulation results}\label{chpt3:large_cohort_results}

\begin{table}[H]
\sisetup{
input-decimal-markers={.},
round-mode=places,
detect-all,
round-integer-to-decimal
}
\caption{Simulation results for estimating $\beta_x$ using the data imputation approach for error scenario 1 (error only in event indicator) with $N=10000$, $n=2000$, and simple random sampling. The $\%$ bias, empirical standard error (ESE), relative efficiency (RE), average standard error (ASE), mean squared error, and coverage probabilities (CP) are presented for 2000 simulated datasets.}
\label{data_imp_large_cohort_err_1}
\centering 
\scalebox{0.80}{
\begin{tabular}{
c
c
c
c
*6{S[table-format=1.3,round-precision=3]}
}
$\beta_z$ & \% Cens & $\beta_x$ & Method & {\% Bias} & {ESE}  & {RE}   & {ASE}  & {MSE}  & {CP} \\ \hline
log(0.5)  & 50      & log(1.5)  & True   & 0.054808    & 0.017284 & 2.362416 & 0.017613 & 0.000299 & 0.951  \\
          &         &           & HT     & -0.02749    & 0.040833 & 1        & 0.039395 & 0.001667 & 0.958  \\
          &         &           & GRN    & -0.17062    & 0.032145 & 1.270254 & 0.032467 & 0.001034 & 0.946  \\
          &         &           & GRMIS  & 0.11896     & 0.028221 & 1.446867 & 0.02839  & 0.000797 & 0.952  \\
          &         &           & GRMIC  & 0.171039    & 0.028357 & 1.43994  & 0.02831  & 0.000805 & 0.95   \\
          &         &           &        &             &          &          &          &          &        \\
          &         & log(3)    & True   & -0.05218    & 0.020492 & 2.155264 & 0.019701 & 0.00042  & 0.942  \\
          &         &           & HT     & -0.06065    & 0.044165 & 1        & 0.043853 & 0.001951 & 0.946  \\
          &         &           & GRN    & -0.09902    & 0.038639 & 1.143009 & 0.035966 & 0.001494 & 0.945  \\
          &         &           & GRMIS  & -0.08065    & 0.031253 & 1.413158 & 0.030816 & 0.000978 & 0.946  \\
          &         &           & GRMIC  & -0.13329    & 0.030914 & 1.428659 & 0.030735 & 0.000958 & 0.948  \\
          &         &           &        &             &          &          &          &          &        \\
          & 75      & log(1.5)  & True   & -0.34616    & 0.023745 & 2.238788 & 0.023804 & 0.000566 & 0.956  \\
          &         &           & HT     & -0.29482    & 0.053159 & 1        & 0.053226 & 0.002827 & 0.944  \\
          &         &           & GRN    & -0.23871    & 0.043041 & 1.235098 & 0.043158 & 0.001853 & 0.943  \\
          &         &           & GRMIS  & -0.30717    & 0.042088 & 1.263042 & 0.041024 & 0.001773 & 0.954  \\
          &         &           & GRMIC  & -0.11051    & 0.043443 & 1.223657 & 0.040882 & 0.001888 & 0.949  \\
          &         &           &        &             &          &          &          &          &        \\
          &         & log(3)    & True   & -0.03541    & 0.027097 & 2.085792 & 0.026437 & 0.000734 & 0.942  \\
          &         &           & HT     & 0.006602    & 0.056519 & 1        & 0.058911 & 0.003194 & 0.948  \\
          &         &           & GRN    & -0.03041    & 0.052345 & 1.079745 & 0.050762 & 0.00274  & 0.948  \\
          &         &           & GRMIS  & -0.14913    & 0.045914 & 1.230966 & 0.045108 & 0.002111 & 0.95   \\
          &         &           & GRMIC  & -0.17122    & 0.045346 & 1.246394 & 0.044947 & 0.00206  & 0.948  \\
          &         &           &        &             &          &          &          &          &        \\
          & 90      & log(1.5)  & True   & 0.300231    & 0.038368 & 2.189938 & 0.037121 & 0.001474 & 0.946  \\
          &         &           & HT     & -0.27875    & 0.084024 & 1        & 0.082716 & 0.007061 & 0.95   \\
          &         &           & GRN    & -0.15046    & 0.076347 & 1.100555 & 0.074226 & 0.005829 & 0.951  \\
          &         &           & GRMIS  & -0.12485    & 0.07321  & 1.147715 & 0.070996 & 0.00536  & 0.946  \\
          &         &           & GRMIC  & 0.114554    & 0.072765 & 1.154728 & 0.070949 & 0.005295 & 0.946  \\
          &         &           &        &             &          &          &          &          &        \\
          &         & log(3)    & True   & -0.10845    & 0.040031 & 2.20517  & 0.039781 & 0.001604 & 0.948  \\
          &         &           & HT     & -0.03262    & 0.088274 & 1        & 0.08865  & 0.007792 & 0.952  \\
          &         &           & GRN    & -0.06333    & 0.085414 & 1.033488 & 0.083241 & 0.007296 & 0.953  \\
          &         &           & GRMIS  & -0.05366    & 0.076953 & 1.147114 & 0.07441  & 0.005922 & 0.954  \\
          &         &           & GRMIC  & -0.1297     & 0.076059 & 1.160603 & 0.074555 & 0.005787 & 0.948 
\end{tabular}
}
\end{table}

\begin{table}[H]
\sisetup{
input-decimal-markers={.},
round-mode=places,
detect-all,
round-integer-to-decimal
}
\caption{Simulation results for estimating $\beta_x$ using the data imputation approach for error scenario 2 (errors in event indicator and failure time) with $N=10000$, $n=2000$, and simple random sampling. The $\%$ bias, empirical standard error (ESE), relative efficiency (RE), average standard error (ASE), mean squared error, and coverage probabilities (CP) are presented for 2000 simulated datasets.}
\label{data_imp_large_cohort_err_2}
\centering 
\scalebox{0.80}{
\begin{tabular}{
c
c
c
c
*6{S[table-format=1.3,round-precision=3]}
}
$\beta_z$ & \% Cens & $\beta_x$ & Method & {\% Bias}  & {ESE}      & {RE}       & {ASE}      & {MSE}      & {CP}    \\ \hline
log(0.5)  & 50      & log(1.5)  & True   & -0.19065 & 0.017589 & 2.302689 & 0.017612 & 0.00031  & 0.948 \\
          &         &           & HT     & 0.56596  & 0.040501 & 1        & 0.039375 & 0.001646 & 0.949 \\
          &         &           & GRN    & -0.00021 & 0.033095 & 1.223769 & 0.032559 & 0.001095 & 0.958 \\
          &         &           & GRMIS  & -0.09951 & 0.028698 & 1.411272 & 0.028497 & 0.000824 & 0.952 \\
          &         &           & GRMIC  & 0.057628 & 0.028235 & 1.434443 & 0.028443 & 0.000797 & 0.95  \\
          &         &           & GRFCSMIS & -0.16005 & 0.028514 & 1.420372 & 0.028459 & 0.000813 & 0.954 \\
          &         &           & GRFCSMIC & -0.09481 & 0.028744 & 1.40905  & 0.028368 & 0.000826 & 0.953 \\
          &         &           &        &          &          &          &          &          &       \\
          &         & log(3)    & True   & 0.007371 & 0.020095 & 2.216554 & 0.019705 & 0.000404 & 0.954 \\
          &         &           & HT     & 0.14059  & 0.044541 & 1        & 0.043984 & 0.001986 & 0.958 \\
          &         &           & GRN    & 0.120252 & 0.037165 & 1.198481 & 0.036194 & 0.001383 & 0.953 \\
          &         &           & GRMIS  & -0.00645 & 0.032569 & 1.367607 & 0.031431 & 0.001061 & 0.95  \\
          &         &           & GRMIC  & 0.003242 & 0.032407 & 1.374455 & 0.031323 & 0.00105  & 0.954 \\
          &         &           & GRFCSMIS & -0.10256 & 0.031067 & 1.433718 & 0.031312 & 0.000966 & 0.958 \\
          &         &           & GRFCSMIC & 0.021591 & 0.031842 & 1.398842 & 0.031165 & 0.001014 & 0.95  \\
          &         &           &        &          &          &          &          &          &       \\
          & 75      & log(1.5)  & True   & 0.014001 & 0.025018 & 2.11733  & 0.023801 & 0.000626 & 0.949 \\
          &         &           & HT     & 0.244099 & 0.052971 & 1        & 0.053189 & 0.002807 & 0.951 \\
          &         &           & GRN    & 0.107019 & 0.043513 & 1.217349 & 0.043265 & 0.001894 & 0.951 \\
          &         &           & GRMIS  & 0.342023 & 0.042648 & 1.242038 & 0.041066 & 0.001821 & 0.946 \\
          &         &           & GRMIC  & 0.264359 & 0.041911 & 1.263871 & 0.040974 & 0.001758 & 0.944 \\
          &         &           & GRFCSMIS & 0.310782 & 0.043342 & 1.22217  & 0.041012 & 0.00188  & 0.948 \\
          &         &           & GRFCSMIC & 0.286289 & 0.042452 & 1.24779  & 0.040907 & 0.001803 & 0.944 \\
          &         &           &        &          &          &          &          &          &       \\
          &         & log(3)    & True   & -0.05745 & 0.028097 & 2.082046 & 0.02643  & 0.00079  & 0.946 \\
          &         &           & HT     & 0.080356 & 0.0585   & 1        & 0.059039 & 0.003423 & 0.958 \\
          &         &           & GRN    & -0.01744 & 0.050943 & 1.148345 & 0.051062 & 0.002595 & 0.954 \\
          &         &           & GRMIS  & -0.10403 & 0.047242 & 1.238287 & 0.04617  & 0.002233 & 0.95  \\
          &         &           & GRMIC  & -0.09856 & 0.047631 & 1.228176 & 0.046065 & 0.00227  & 0.954 \\
          &         &           & GRFCSMIS & -0.13389 & 0.046445 & 1.259538 & 0.046111 & 0.002159 & 0.954 \\
          &         &           & GRFCSMIC & -0.11486 & 0.04664  & 1.254269 & 0.045871 & 0.002177 & 0.95  \\
          &         &           &        &          &          &          &          &          &       \\
          & 90      & log(1.5)  & True   & 0.300231 & 0.038368 & 2.093359 & 0.037121 & 0.001474 & 0.946 \\
          &         &           & HT     & 0.614175 & 0.080318 & 1        & 0.082808 & 0.006457 & 0.944 \\
          &         &           & GRN    & 0.516905 & 0.075704 & 1.060956 & 0.074574 & 0.005735 & 0.947 \\
          &         &           & GRMIS  & 0.190219 & 0.071749 & 1.119431 & 0.071582 & 0.005149 & 0.941 \\
          &         &           & GRMIC  & 0.410931 & 0.071522 & 1.122986 & 0.071456 & 0.005118 & 0.942 \\
          &         &           & GRFCSMIS & 0.539946 & 0.070147 & 1.145004 & 0.071652 & 0.004925 & 0.942 \\
          &         &           & GRFCSMIC & 0.64252  & 0.070001 & 1.147396 & 0.071463 & 0.004907 & 0.942 \\
          &         &           &        &          &          &          &          &          &       \\
          &         & log(3)    & True   & -0.10845 & 0.040031 & 2.109783 & 0.039781 & 0.001604 & 0.948 \\
          &         &           & HT     & 0.207349 & 0.084456 & 1        & 0.088877 & 0.007138 & 0.949 \\
          &         &           & GRN    & 0.114892 & 0.079391 & 1.063797 & 0.083919 & 0.006304 & 0.942 \\
          &         &           & GRMIS  & -0.15029 & 0.077382 & 1.091408 & 0.07822  & 0.005991 & 0.948 \\
          &         &           & GRMIC  & -0.2245  & 0.075666 & 1.116166 & 0.077916 & 0.005731 & 0.948 \\
          &         &           & GRFCSMIS & -0.14806 & 0.075271 & 1.122027 & 0.077797 & 0.005668 & 0.948 \\
          &         &           & GRFCSMIC & -0.29095 & 0.074947 & 1.126878 & 0.077503 & 0.005627 & 0.946
\end{tabular}
}
\end{table}

\begin{table}[H]
\sisetup{
input-decimal-markers={.},
round-mode=places,
detect-all,
round-integer-to-decimal
}
\caption{Simulation results for estimating $\beta_x$ using the data imputation approach for error scenario 3 (errors in event indicator, failure time, and X) with $N=10000$, $n=2000$, and simple random sampling. The $\%$ bias, empirical standard error (ESE), relative efficiency (RE), average standard error (ASE), mean squared error, and coverage probabilities (CP) are presented for 2000 simulated datasets.}
\label{data_imp_large_cohort_err_3}
\centering 
\scalebox{0.80}{
\begin{tabular}{
c
c
c
c
*6{S[table-format=1.3,round-precision=3]}
}
$\beta_z$ & \% Cens & $\beta_x$ & Method & {\% Bias}  & {ESE}      & {RE}       & {ASE}      & {MSE}      & {CP}    \\ \hline
log(0.5)  & 50      & log(1.5)  & True   & -0.19065 & 0.017589 & 2.305954 & 0.017612 & 0.00031  & 0.948 \\
          &         &           & HT     & -0.38798 & 0.040559 & 1        & 0.039363 & 0.001647 & 0.947 \\
          &         &           & GRN    & -0.0649  & 0.040211 & 1.008631 & 0.039278 & 0.001617 & 0.949 \\
          &         &           & GRMIS  & -0.40733 & 0.040069 & 1.01222  & 0.039306 & 0.001608 & 0.943 \\
          &         &           & GRMIC  & -0.42348 & 0.040052 & 1.012644 & 0.039306 & 0.001607 & 0.944 \\
          &         &           & GRFCSMIS & -0.19395 & 0.032359 & 1.253383 & 0.033263 & 0.001048 & 0.953 \\
          &         &           & GRFCSMIC & -0.20848 & 0.032963 & 1.230438 & 0.03321  & 0.001087 & 0.95  \\
          &         &           &        &          &          &          &          &          &       \\
          &         & log(3)    & True   & 0.007371 & 0.020095 & 2.176311 & 0.019705 & 0.000404 & 0.954 \\
          &         &           & HT     & 0.152582 & 0.043733 & 1        & 0.043902 & 0.001915 & 0.944 \\
          &         &           & GRN    & 0.182688 & 0.04374  & 0.999833 & 0.043789 & 0.001917 & 0.938 \\
          &         &           & GRMIS  & 0.140377 & 0.043289 & 1.010252 & 0.043781 & 0.001876 & 0.944 \\
          &         &           & GRMIC  & 0.13122  & 0.043248 & 1.01122  & 0.043759 & 0.001872 & 0.945 \\
          &         &           & GRFCSMIS & 0.08376  & 0.039471 & 1.107966 & 0.039375 & 0.001559 & 0.946 \\
          &         &           & GRFCSMIC & 0.082497 & 0.03993  & 1.095225 & 0.03949  & 0.001595 & 0.944 \\
          &         &           &        &          &          &          &          &          &       \\
          & 75      & log(1.5)  & True   & 0.014001 & 0.025018 & 2.070413 & 0.023801 & 0.000626 & 0.949 \\
          &         &           & HT     & -0.42736 & 0.051797 & 1        & 0.053204 & 0.002686 & 0.953 \\
          &         &           & GRN    & -0.08905 & 0.050398 & 1.027756 & 0.052399 & 0.00254  & 0.952 \\
          &         &           & GRMIS  & -0.19243 & 0.051672 & 1.002418 & 0.052889 & 0.002671 & 0.95  \\
          &         &           & GRMIC  & -0.14175 & 0.051877 & 0.998465 & 0.052852 & 0.002692 & 0.95  \\
          &         &           & GRFCSMIS & -0.1983  & 0.046177 & 1.121696 & 0.046052 & 0.002133 & 0.946 \\
          &         &           & GRFCSMIC & 0.056028 & 0.045931 & 1.127705 & 0.045945 & 0.00211  & 0.942 \\
          &         &           &        &          &          &          &          &          &       \\
          &         & log(3)    & True   & -0.05745 & 0.028097 & 2.160778 & 0.02643  & 0.00079  & 0.946 \\
          &         &           & HT     & 0.257044 & 0.060712 & 1        & 0.058916 & 0.003694 & 0.94  \\
          &         &           & GRN    & 0.304253 & 0.060606 & 1.001741 & 0.058098 & 0.003684 & 0.939 \\
          &         &           & GRMIS  & 0.304018 & 0.061782 & 0.982683 & 0.058593 & 0.003828 & 0.937 \\
          &         &           & GRMIC  & 0.279433 & 0.061726 & 0.983575 & 0.058552 & 0.003819 & 0.938 \\
          &         &           & GRFCSMIS & 0.247202 & 0.055262 & 1.098615 & 0.05378  & 0.003061 & 0.947 \\
          &         &           & GRFCSMIC & 0.162435 & 0.054445 & 1.115095 & 0.053735 & 0.002967 & 0.942 \\
          &         &           &        &          &          &          &          &          &       \\
          & 90      & log(1.5)  & True   & 0.300231 & 0.038368 & 2.135321 & 0.037121 & 0.001474 & 0.946 \\
          &         &           & HT     & -0.13418 & 0.081929 & 1        & 0.082664 & 0.006713 & 0.944 \\
          &         &           & GRN    & 0.403334 & 0.081083 & 1.010423 & 0.079681 & 0.006577 & 0.949 \\
          &         &           & GRMIS  & 0.539478 & 0.081062 & 1.010686 & 0.079814 & 0.006576 & 0.948 \\
          &         &           & GRMIC  & 0.397156 & 0.08106  & 1.010708 & 0.079601 & 0.006573 & 0.949 \\
          &         &           & GRFCSMIS & 0.113621 & 0.078127 & 1.048663 & 0.076604 & 0.006104 & 0.948 \\
          &         &           & GRFCSMIC & 0.324641 & 0.076865 & 1.065877 & 0.076394 & 0.00591  & 0.947 \\
          &         &           &        &          &          &          &          &          &       \\
          &         & log(3)    & True   & -0.10845 & 0.040031 & 2.365923 & 0.039781 & 0.001604 & 0.948 \\
          &         &           & HT     & 0.106186 & 0.094709 & 1        & 0.08875  & 0.008971 & 0.944 \\
          &         &           & GRN    & 0.368331 & 0.090655 & 1.044722 & 0.086697 & 0.008235 & 0.946 \\
          &         &           & GRMIS  & 0.397741 & 0.089394 & 1.059461 & 0.086097 & 0.00801  & 0.944 \\
          &         &           & GRMIC  & 0.384319 & 0.091384 & 1.036391 & 0.085945 & 0.008369 & 0.944 \\
          &         &           & GRFCSMIS & 0.205223 & 0.087776 & 1.078985 & 0.084764 & 0.00771  & 0.943 \\
          &         &           & GRFCSMIC & 0.16134  & 0.090074 & 1.051463 & 0.084645 & 0.008116 & 0.946
\end{tabular}
}
\end{table}

\section*{Appendix E: Type 1 error results}

\begin{table}[H]
\sisetup{
input-decimal-markers={.},
round-mode=places,
detect-all,
round-integer-to-decimal
}
\caption{Type 1 error results for $\beta_x=0$ using the data imputation approach for error scenario 3 (errors in event indicator, failure time, and X) with $N=10000$, $n=2000$, and simple random sampling. The absolute bias, empirical standard error (ESE), relative efficiency (RE), average standard error (ASE), mean squared error (MSE), and type 1 error are presented for 2000 simulated datasets.}
\label{data_imp_type1_err}
\centering 
\scalebox{0.80}{
\begin{tabular}{
c
c
c
c
*6{S[table-format=1.3,round-precision=3]}
}
$\beta_z$ & \% Cens & $\beta_x$ & Method & {Bias}  & {ESE}     & {RE}      & {ASE}     & {MSE}     & {Type 1 error} \\ \hline
log(0.5)  & 50      & 0         & True   & 0.00091  & 0.04430 & 2.19292 & 0.04309 & 0.00196 & 0.052        \\
          &         &           & HT     & 0.00171  & 0.09715 & 1.00000 & 0.09609 & 0.00944 & 0.052        \\
          &         &           & GRN    & 0.00282  & 0.09357 & 1.03825 & 0.09300 & 0.00876 & 0.052        \\
          &         &           & GRMIS  & 0.00557  & 0.09364 & 1.03750 & 0.09291 & 0.00880 & 0.052        \\
          &         &           & GRMIC  & 0.00609  & 0.09248 & 1.05050 & 0.09270 & 0.00859 & 0.055        \\
          &         &           & GRFCSMIS & 0.00383  & 0.08964 & 1.08383 & 0.09020 & 0.00805 & 0.052        \\
          &         &           & GRFCSMIC & 0.00468  & 0.09041 & 1.07457 & 0.09000 & 0.00820 & 0.056        \\
          &         &           &        &          &         &         &         &         &              \\
          & 75      & 0         & True   & -0.00268 & 0.06448 & 2.42275 & 0.06675 & 0.00417 & 0.044        \\
          &         &           & HT     & -0.00357 & 0.15623 & 1.00000 & 0.14633 & 0.02442 & 0.057        \\
          &         &           & GRN    & 0.00064  & 0.15408 & 1.01394 & 0.14484 & 0.02374 & 0.06         \\
          &         &           & GRMIS  & -0.00021 & 0.15210 & 1.02713 & 0.14405 & 0.02314 & 0.06         \\
          &         &           & GRMIC  & -0.00194 & 0.15477 & 1.00944 & 0.14364 & 0.02396 & 0.064        \\
          &         &           & GRFCSMIS & -0.00429 & 0.15389 & 1.01521 & 0.14389 & 0.02370 & 0.057        \\
          &         &           & GRFCSMIC & -0.00005 & 0.14833 & 1.05325 & 0.14309 & 0.02200 & 0.059        \\
          &         &           &        &          &         &         &         &         &              \\
          & 90      & 0         & True   & 0.00094  & 0.11292 & 2.24950 & 0.11122 & 0.01275 & 0.057        \\
          &         &           & HT     & 0.00193  & 0.25401 & 1.00000 & 0.23831 & 0.06453 & 0.066        \\
          &         &           & GRN    & -0.00086 & 0.25437 & 0.99859 & 0.23712 & 0.06470 & 0.069        \\
          &         &           & GRMIS  & 0.00193  & 0.25698 & 0.98846 & 0.23662 & 0.06604 & 0.068        \\
          &         &           & GRMIC  & -0.00063 & 0.26162 & 0.97090 & 0.23583 & 0.06845 & 0.07         \\
          &         &           & GRFCSMIS & 0.00212  & 0.25010 & 1.01563 & 0.23629 & 0.06256 & 0.071        \\
          &         &           & GRFCSMIC & -0.00458 & 0.24366 & 1.04246 & 0.23422 & 0.05939 & 0.072      
\end{tabular}
}
\end{table}

\section*{Appendix F: Influence function imputation results}

\begin{table}[H]
\sisetup{
input-decimal-markers={.},
round-mode=places,
detect-all,
round-integer-to-decimal
}
\caption{Simulation results for estimating $\beta_x$ using the IF imputation approach for error scenario 1 (error only in event indicator) with $N=2000$, $n=400$, and simple random sampling. The $\%$ bias, empirical standard error (ESE), relative efficiency (RE), average standard error (ASE), mean squared error, and coverage probabilities (CP) are presented for 2000 simulated datasets.}
\label{IF_imp_err_1}
\centering 
\scalebox{0.80}{
\begin{tabular}{
c
c
c
c
*6{S[table-format=1.3,round-precision=3]}
}
$\beta_z$ & \% Cens & $\beta_x$ & Method & {\% Bias} & {ESE}  & {RE}   & {ASE}  & {MSE}  & {CP} \\ \hline
log(0.5)  & 50      & log(1.5)  & True   & -0.03595    & 0.039644 & 2.289193 & 0.039422 & 0.001572 & 0.956  \\
          &         &           & HT     & 1.228958    & 0.090753 & 1        & 0.087874 & 0.008261 & 0.949  \\
          &         &           & GRN    & 1.40684     & 0.07401  & 1.226214 & 0.072528 & 0.00551  & 0.95   \\
          &         &           & GRMIS  & -0.90195    & 0.064965 & 1.396945 & 0.064525 & 0.004234 & 0.946  \\
          &         &           & GRMIC  & -1.01073    & 0.065105 & 1.393946 & 0.064286 & 0.004255 & 0.95   \\
          &         &           &        &             &          &          &          &          &        \\
          &         & log(3)    & True   & 0.041168    & 0.041582 & 2.453957 & 0.04415  & 0.001729 & 0.948  \\
          &         &           & HT     & 0.631089    & 0.10204  & 1        & 0.097775 & 0.01046  & 0.939  \\
          &         &           & GRN    & 0.282312    & 0.082568 & 1.235824 & 0.080447 & 0.006827 & 0.942  \\
          &         &           & GRMIS  & -0.22609    & 0.07478  & 1.364544 & 0.071255 & 0.005598 & 0.952  \\
          &         &           & GRMIC  & -0.22782    & 0.073727 & 1.384023 & 0.070891 & 0.005442 & 0.954  \\
          &         &           &        &             &          &          &          &          &        \\
          & 75      & log(1.5)  & True   & 0.119394    & 0.051672 & 2.266392 & 0.053276 & 0.00267  & 0.954  \\
          &         &           & HT     & 0.781363    & 0.117109 & 1        & 0.118644 & 0.013725 & 0.952  \\
          &         &           & GRN    & 0.916624    & 0.097339 & 1.203106 & 0.096548 & 0.009489 & 0.945  \\
          &         &           & GRMIS  & -0.96486    & 0.094689 & 1.236776 & 0.090898 & 0.008981 & 0.94   \\
          &         &           & GRMIC  & -0.55219    & 0.095677 & 1.224011 & 0.090511 & 0.009159 & 0.94   \\
          &         &           &        &             &          &          &          &          &        \\
          &         & log(3)    & True   & -0.01311    & 0.06088  & 2.241353 & 0.059211 & 0.003706 & 0.949  \\
          &         &           & HT     & 1.034735    & 0.136454 & 1        & 0.131041 & 0.018749 & 0.938  \\
          &         &           & GRN    & 0.386125    & 0.119288 & 1.143905 & 0.113786 & 0.014248 & 0.934  \\
          &         &           & GRMIS  & -0.24879    & 0.104165 & 1.309981 & 0.102151 & 0.010858 & 0.945  \\
          &         &           & GRMIC  & -0.1159     & 0.102101 & 1.336464 & 0.101521 & 0.010426 & 0.942  \\
          &         &           &        &             &          &          &          &          &        \\
          & 90      & log(1.5)  & True   & 0.0138      & 0.084364 & 2.222885 & 0.083155 & 0.007117 & 0.947  \\
          &         &           & HT     & 1.805251    & 0.187531 & 1        & 0.184444 & 0.035222 & 0.943  \\
          &         &           & GRN    & 0.30929     & 0.167181 & 1.121725 & 0.165789 & 0.027951 & 0.94   \\
          &         &           & GRMIS  & -2.53059    & 0.16078  & 1.166381 & 0.154901 & 0.025956 & 0.942  \\
          &         &           & GRMIC  & -1.37636    & 0.161713 & 1.159658 & 0.152769 & 0.026182 & 0.933  \\
          &         &           &        &             &          &          &          &          &        \\
          &         & log(3)    & True   & -0.04654    & 0.088525 & 2.315872 & 0.089229 & 0.007837 & 0.95   \\
          &         &           & HT     & 1.160558    & 0.205013 & 1        & 0.197598 & 0.042193 & 0.938  \\
          &         &           & GRN    & 0.945284    & 0.194363 & 1.054793 & 0.187058 & 0.037885 & 0.941  \\
          &         &           & GRMIS  & -1.02709    & 0.182758 & 1.121773 & 0.164852 & 0.033528 & 0.931  \\
          &         &           & GRMIC  & -0.94924    & 0.178754 & 1.146899 & 0.163399 & 0.032062 & 0.924 
\end{tabular}
}
\end{table}

\begin{table}[H]
\sisetup{
input-decimal-markers={.},
round-mode=places,
detect-all,
round-integer-to-decimal
}
\caption{Simulation results for estimating $\beta_x$ using the IF imputation approach for error scenario 2 (errors in event indicator and failure time) with $N=2000$, $n=400$, and simple random sampling. The $\%$ bias, empirical standard error (ESE), relative efficiency (RE), average standard error (ASE), mean squared error, and coverage probabilities (CP) are presented for 2000 simulated datasets.}
\label{IF_imp_err_2}
\centering 
\scalebox{0.80}{
\begin{tabular}{
c
c
c
c
*6{S[table-format=1.3,round-precision=3]}
}
$\beta_z$ & \% Cens & $\beta_x$ & Method & {\% Bias}  & {ESE}      & {RE}       & {ASE}      & {MSE}     & {CP}    \\ \hline
log(0.5)  & 50      & log(1.5)  & True   & -0.03595 & 0.039644 & 2.259058 & 0.039422 & 0.001572 & 0.956 \\
          &         &           & HT     & 1.193166 & 0.089558 & 1        & 0.088263 & 0.008044 & 0.944 \\
          &         &           & GRN    & 0.36901  & 0.073756 & 1.21425  & 0.072739 & 0.005442 & 0.94  \\
          &         &           & GRMIS  & -0.66916 & 0.067803 & 1.320857 & 0.064383 & 0.004605 & 0.95  \\
          &         &           & GRMIC  & -1.17444 & 0.067482 & 1.327148 & 0.064126 & 0.004576 & 0.947 \\
          &         &           & GRFCSMIS & -0.6887  & 0.066363 & 1.349513 & 0.064489 & 0.004412 & 0.948 \\
          &         &           & GRFCSMIC & -1.04496 & 0.066929 & 1.338113 & 0.064188 & 0.004497 & 0.948 \\
          &         &           &        &          &          &          &          &          &       \\
          &         & log(3)    & True   & 0.041168 & 0.041582 & 2.387503 & 0.04415  & 0.001729 & 0.948 \\
          &         &           & HT     & 0.159486 & 0.099277 & 1        & 0.097769 & 0.009859 & 0.942 \\
          &         &           & GRN    & -0.05843 & 0.08363  & 1.187094 & 0.080697 & 0.006994 & 0.94  \\
          &         &           & GRMIS  & -0.33902 & 0.074376 & 1.334794 & 0.072157 & 0.005546 & 0.942 \\
          &         &           & GRMIC  & -0.33888 & 0.074533 & 1.331979 & 0.071715 & 0.005569 & 0.942 \\
          &         &           & GRFCSMIS & -0.33134 & 0.074262 & 1.336838 & 0.071762 & 0.005528 & 0.94  \\
          &         &           & GRFCSMIC & -0.41767 & 0.072415 & 1.370951 & 0.071454 & 0.005265 & 0.941 \\
          &         &           &        &          &          &          &          &          &       \\
          & 75      & log(1.5)  & True   & 0.119394 & 0.051672 & 2.277078 & 0.053276 & 0.00267  & 0.954 \\
          &         &           & HT     & 0.106275 & 0.117662 & 1        & 0.118679 & 0.013844 & 0.946 \\
          &         &           & GRN    & 0.073904 & 0.099255 & 1.185442 & 0.096827 & 0.009852 & 0.938 \\
          &         &           & GRMIS  & -1.39746 & 0.095574 & 1.2311   & 0.091069 & 0.009167 & 0.934 \\
          &         &           & GRMIC  & -2.02299 & 0.098117 & 1.199203 & 0.090413 & 0.009694 & 0.928 \\
          &         &           & GRFCSMIS & -1.13488 & 0.095578 & 1.231054 & 0.09105  & 0.009156 & 0.938 \\
          &         &           & GRFCSMIC & -1.8046  & 0.096456 & 1.21985  & 0.090576 & 0.009357 & 0.929 \\
          &         &           &        &          &          &          &          &          &       \\
          &         & log(3)    & True   & -0.01311 & 0.06088  & 2.182885 & 0.059211 & 0.003706 & 0.949 \\
          &         &           & HT     & 0.358656 & 0.132894 & 1        & 0.131964 & 0.017676 & 0.946 \\
          &         &           & GRN    & -0.08903 & 0.115453 & 1.151068 & 0.114087 & 0.01333  & 0.939 \\
          &         &           & GRMIS  & -1.04825 & 0.101828 & 1.305086 & 0.103487 & 0.010502 & 0.944 \\
          &         &           & GRMIC  & -1.14319 & 0.103824 & 1.279992 & 0.103089 & 0.010937 & 0.948 \\
          &         &           & GRFCSMIS & -1.18049 & 0.100957 & 1.31635  & 0.102971 & 0.01036  & 0.943 \\
          &         &           & GRFCSMIC & -1.16145 & 0.101611 & 1.307871 & 0.102171 & 0.010488 & 0.944 \\
          &         &           &        &          &          &          &          &          &       \\
          & 90      & log(1.5)  & True   & 0.0138   & 0.084364 & 2.241325 & 0.083155 & 0.007117 & 0.947 \\
          &         &           & HT     & -0.12238 & 0.189087 & 1        & 0.184527 & 0.035754 & 0.94  \\
          &         &           & GRN    & -0.47182 & 0.167907 & 1.126137 & 0.166491 & 0.028197 & 0.937 \\
          &         &           & GRMIS  & -4.90776 & 0.165713 & 1.141049 & 0.155956 & 0.027857 & 0.925 \\
          &         &           & GRMIC  & -3.43053 & 0.166734 & 1.134061 & 0.153326 & 0.027994 & 0.923 \\
          &         &           & GRFCSMIS & -4.20924 & 0.162172 & 1.165962 & 0.154132 & 0.026591 & 0.931 \\
          &         &           & GRFCSMIC & -2.69717 & 0.166436 & 1.136093 & 0.152002 & 0.027821 & 0.927 \\
          &         &           &        &          &          &          &          &          &       \\
          &         & log(3)    & True   & -0.04654 & 0.088525 & 2.307755 & 0.089229 & 0.007837 & 0.95  \\
          &         &           & HT     & 1.212813 & 0.204295 & 1        & 0.200054 & 0.041914 & 0.946 \\
          &         &           & GRN    & 1.110413 & 0.195368 & 1.045689 & 0.188311 & 0.038318 & 0.942 \\
          &         &           & GRMIS  & -1.13955 & 0.177245 & 1.152611 & 0.171462 & 0.031573 & 0.929 \\
          &         &           & GRMIC  & -1.08415 & 0.179691 & 1.13692  & 0.16897  & 0.032431 & 0.923 \\
          &         &           & GRFCSMIS & -1.55686 & 0.172123 & 1.186911 & 0.168801 & 0.029919 & 0.928 \\
          &         &           & GRFCSMIC & -0.90004 & 0.176818 & 1.155395 & 0.166461 & 0.031362 & 0.926
\end{tabular}
}
\end{table}

\begin{table}[H]
\sisetup{
input-decimal-markers={.},
round-mode=places,
detect-all,
round-integer-to-decimal
}
\caption{Simulation results for estimating $\beta_x$ using the IF imputation approach for error scenario 1 (error only in event indicator) with $N=10000$, $n=2000$, and simple random sampling. The $\%$ bias, empirical standard error (ESE), relative efficiency (RE), average standard error (ASE), mean squared error, and coverage probabilities (CP) are presented for 2000 simulated datasets.}
\label{IF_imp_large_cohort_err_1}
\centering 
\scalebox{0.80}{
\begin{tabular}{
c
c
c
c
*6{S[table-format=1.3,round-precision=3]}
}
$\beta_z$ & \% Cens & $\beta_x$ & Method & {\% Bias} & {ESE}  & {RE}   & {ASE}  & {MSE}  & {CP} \\ \hline
log(0.5)  & 50      & log(1.5)  & True   & 0.054808    & 0.017284 & 2.362416 & 0.017613 & 0.000299 & 0.951  \\
          &         &           & HT     & -0.02749    & 0.040833 & 1        & 0.039395 & 0.001667 & 0.958  \\
          &         &           & GRN    & -0.17062    & 0.032145 & 1.270254 & 0.032467 & 0.001034 & 0.946  \\
          &         &           & GRMIS  & 0.100479    & 0.029726 & 1.373624 & 0.029255 & 0.000884 & 0.949  \\
          &         &           & GRMIC  & 0.088663    & 0.029435 & 1.387224 & 0.02926  & 0.000867 & 0.95   \\
          &         &           &        &             &          &          &          &          &        \\
          &         & log(3)    & True   & -0.05218    & 0.020492 & 2.155264 & 0.019701 & 0.00042  & 0.942  \\
          &         &           & HT     & -0.06065    & 0.044165 & 1        & 0.043853 & 0.001951 & 0.946  \\
          &         &           & GRN    & -0.09902    & 0.038639 & 1.143009 & 0.035966 & 0.001494 & 0.945  \\
          &         &           & GRMIS  & -0.09059    & 0.032648 & 1.352761 & 0.032008 & 0.001067 & 0.952  \\
          &         &           & GRMIC  & -0.0665     & 0.032999 & 1.338383 & 0.032021 & 0.001089 & 0.95   \\
          &         &           &        &             &          &          &          &          &        \\
          & 75      & log(1.5)  & True   & -0.34616    & 0.023745 & 2.238788 & 0.023804 & 0.000566 & 0.956  \\
          &         &           & HT     & -0.29482    & 0.053159 & 1        & 0.053226 & 0.002827 & 0.944  \\
          &         &           & GRN    & -0.23871    & 0.043041 & 1.235098 & 0.043158 & 0.001853 & 0.943  \\
          &         &           & GRMIS  & -0.32534    & 0.043819 & 1.213156 & 0.041613 & 0.001922 & 0.95   \\
          &         &           & GRMIC  & -0.46772    & 0.04413  & 1.204598 & 0.041607 & 0.001951 & 0.948  \\
          &         &           &        &             &          &          &          &          &        \\
          &         & log(3)    & True   & -0.03541    & 0.027097 & 2.085792 & 0.026437 & 0.000734 & 0.942  \\
          &         &           & HT     & 0.006602    & 0.056519 & 1        & 0.058911 & 0.003194 & 0.948  \\
          &         &           & GRN    & -0.03041    & 0.052345 & 1.079745 & 0.050762 & 0.00274  & 0.948  \\
          &         &           & GRMIS  & -0.28792    & 0.047163 & 1.198373 & 0.046367 & 0.002234 & 0.948  \\
          &         &           & GRMIC  & -0.23723    & 0.046623 & 1.212257 & 0.04631  & 0.00218  & 0.948  \\
          &         &           &        &             &          &          &          &          &        \\
          & 90      & log(1.5)  & True   & 0.300231    & 0.038368 & 2.189938 & 0.037121 & 0.001474 & 0.946  \\
          &         &           & HT     & -0.27875    & 0.084024 & 1        & 0.082716 & 0.007061 & 0.95   \\
          &         &           & GRN    & -0.15046    & 0.076347 & 1.100555 & 0.074226 & 0.005829 & 0.951  \\
          &         &           & GRMIS  & -0.38222    & 0.075817 & 1.108242 & 0.072511 & 0.005751 & 0.947  \\
          &         &           & GRMIC  & -0.47893    & 0.074779 & 1.123631 & 0.072106 & 0.005596 & 0.946  \\
          &         &           &        &             &          &          &          &          &        \\
          &         & log(3)    & True   & -0.10845    & 0.040031 & 2.20517  & 0.039781 & 0.001604 & 0.948  \\
          &         &           & HT     & -0.03262    & 0.088274 & 1        & 0.08865  & 0.007792 & 0.952  \\
          &         &           & GRN    & -0.06333    & 0.085414 & 1.033488 & 0.083241 & 0.007296 & 0.953  \\
          &         &           & GRMIS  & -0.1807     & 0.078796 & 1.120288 & 0.07501  & 0.006213 & 0.947  \\
          &         &           & GRMIC  & -0.13735    & 0.078226 & 1.128451 & 0.074792 & 0.006122 & 0.948 
\end{tabular}
}
\end{table}

\begin{table}[H]
\sisetup{
input-decimal-markers={.},
round-mode=places,
detect-all,
round-integer-to-decimal
}
\caption{Simulation results for estimating $\beta_x$ using the IF imputation approach for error scenario 2 (errors in event indicator and failure time) with $N=10000$, $n=2000$, and simple random sampling. The $\%$ bias, empirical standard error (ESE), relative efficiency (RE), average standard error (ASE), mean squared error, and coverage probabilities (CP) are presented for 2000 simulated datasets.}
\label{IF_imp_large_cohort_err_2}
\centering 
\scalebox{0.80}{
\begin{tabular}{
c
c
c
c
*6{S[table-format=1.3,round-precision=3]}
}
$\beta_z$ & \% Cens & $\beta_x$ & Method & {\% Bias}  & {ESE}      & {RE}       & {ASE}      & {MSE}      & {CP}    \\ \hline
log(0.5)  & 50      & log(1.5)  & True   & 0.054808 & 0.017284 & 2.34789  & 0.017613 & 0.000299 & 0.951 \\
          &         &           & HT     & 0.439737 & 0.040582 & 1        & 0.03938  & 0.00165  & 0.953 \\
          &         &           & GRN    & -0.05916 & 0.033695 & 1.204375 & 0.032522 & 0.001135 & 0.955 \\
          &         &           & GRMIS  & 0.002278 & 0.030202 & 1.343668 & 0.029332 & 0.000912 & 0.953 \\
          &         &           & GRMIC  & 0.040389 & 0.030626 & 1.325078 & 0.029336 & 0.000938 & 0.951 \\
          &         &           & GRFCSMIS & 0.028835 & 0.030745 & 1.319925 & 0.029295 & 0.000945 & 0.955 \\
          &         &           & GRFCSMIC & 0.001368 & 0.030748 & 1.319817 & 0.029306 & 0.000945 & 0.956 \\
          &         &           &        &          &          &          &          &          &       \\
          &         & log(3)    & True   & -0.05218 & 0.020492 & 2.161159 & 0.019701 & 0.00042  & 0.942 \\
          &         &           & HT     & 0.061535 & 0.044286 & 1        & 0.043865 & 0.001962 & 0.951 \\
          &         &           & GRN    & -0.14009 & 0.035509 & 1.247183 & 0.036138 & 0.001263 & 0.95  \\
          &         &           & GRMIS  & -0.17072 & 0.031834 & 1.391139 & 0.032509 & 0.001017 & 0.95  \\
          &         &           & GRMIC  & -0.20784 & 0.031965 & 1.385436 & 0.032543 & 0.001027 & 0.95  \\
          &         &           & GRFCSMIS & -0.26753 & 0.031827 & 1.391444 & 0.032351 & 0.001022 & 0.945 \\
          &         &           & GRFCSMIC & -0.24417 & 0.031706 & 1.396765 & 0.032372 & 0.001012 & 0.946 \\
          &         &           &        &          &          &          &          &          &       \\
          & 75      & log(1.5)  & True   & -0.34616 & 0.023745 & 2.312845 & 0.023804 & 0.000566 & 0.956 \\
          &         &           & HT     & 0.621078 & 0.054918 & 1        & 0.053213 & 0.003022 & 0.949 \\
          &         &           & GRN    & -0.17351 & 0.043689 & 1.257005 & 0.043311 & 0.001909 & 0.943 \\
          &         &           & GRMIS  & -0.27204 & 0.041358 & 1.327858 & 0.04161  & 0.001712 & 0.95  \\
          &         &           & GRMIC  & -0.32999 & 0.040478 & 1.356729 & 0.041675 & 0.00164  & 0.949 \\
          &         &           & GRFCSMIS & -0.238   & 0.041502 & 1.323273 & 0.041606 & 0.001723 & 0.952 \\
          &         &           & GRFCSMIC & -0.35271 & 0.040672 & 1.350258 & 0.041595 & 0.001656 & 0.951 \\
          &         &           &        &          &          &          &          &          &       \\
          &         & log(3)    & True   & -0.03541 & 0.027097 & 2.136482 & 0.026437 & 0.000734 & 0.942 \\
          &         &           & HT     & 0.174341 & 0.057892 & 1        & 0.059047 & 0.003355 & 0.948 \\
          &         &           & GRN    & 0.101953 & 0.050843 & 1.138661 & 0.051027 & 0.002586 & 0.936 \\
          &         &           & GRMIS  & -0.26643 & 0.046069 & 1.256639 & 0.047328 & 0.002131 & 0.949 \\
          &         &           & GRMIC  & -0.23906 & 0.046182 & 1.253583 & 0.047276 & 0.00214  & 0.952 \\
          &         &           & GRFCSMIS & -0.38985 & 0.0466   & 1.242337 & 0.047083 & 0.00219  & 0.946 \\
          &         &           & GRFCSMIC & -0.43044 & 0.046508 & 1.244774 & 0.046946 & 0.002185 & 0.95  \\
          &         &           &        &          &          &          &          &          &       \\
          & 90      & log(1.5)  & True   & 0.300231 & 0.038368 & 2.138466 & 0.037121 & 0.001474 & 0.946 \\
          &         &           & HT     & 0.681825 & 0.082049 & 1        & 0.082724 & 0.00674  & 0.945 \\
          &         &           & GRN    & -0.02971 & 0.075265 & 1.090132 & 0.074518 & 0.005665 & 0.942 \\
          &         &           & GRMIS  & -0.51917 & 0.072877 & 1.125852 & 0.073122 & 0.005316 & 0.948 \\
          &         &           & GRMIC  & -0.53355 & 0.073562 & 1.115369 & 0.072863 & 0.005416 & 0.948 \\
          &         &           & GRFCSMIS & -0.5727  & 0.072729 & 1.128148 & 0.07256  & 0.005295 & 0.95  \\
          &         &           & GRFCSMIC & -0.72049 & 0.073653 & 1.114003 & 0.072181 & 0.005433 & 0.948 \\
          &         &           &        &          &          &          &          &          &       \\
          &         & log(3)    & True   & -0.10845 & 0.040031 & 2.21953  & 0.039781 & 0.001604 & 0.948 \\
          &         &           & HT     & 0.172047 & 0.088849 & 1        & 0.088705 & 0.007898 & 0.946 \\
          &         &           & GRN    & -0.0031  & 0.082597 & 1.075695 & 0.08387  & 0.006822 & 0.94  \\
          &         &           & GRMIS  & -0.57558 & 0.078538 & 1.131282 & 0.07818  & 0.006208 & 0.945 \\
          &         &           & GRMIC  & -0.62865 & 0.077182 & 1.151156 & 0.078037 & 0.006005 & 0.945 \\
          &         &           & GRFCSMIS & -0.68889 & 0.077174 & 1.151274 & 0.077335 & 0.006013 & 0.942 \\
          &         &           & GRFCSMIC & -0.7213  & 0.07917  & 1.122251 & 0.077035 & 0.006331 & 0.941
\end{tabular}
}
\end{table}

\begin{table}[H]
\sisetup{
input-decimal-markers={.},
round-mode=places,
detect-all,
round-integer-to-decimal
}
\caption{Simulation results for estimating $\beta_x$ using the IF imputation approach for error scenario 3 (errors in event indicator, failure time, and X) with $N=10000$, $n=2000$, and simple random sampling. The $\%$ bias, empirical standard error (ESE), relative efficiency (RE), average standard error (ASE), mean squared error, and coverage probabilities (CP) are presented for 2000 simulated datasets.}
\label{IF_imp_large_cohort_err_3}
\centering 
\scalebox{0.80}{
\begin{tabular}{
c
c
c
c
*6{S[table-format=1.3,round-precision=3]}
}
$\beta_z$ & \% Cens & $\beta_x$ & Method & {\% Bias}  & {ESE}      & {RE}       & {ASE}      & {MSE}      & {CP}    \\ \hline
log(0.5)  & 50      & log(1.5)  & True   & 0.054808 & 0.017284 & 2.283291 & 0.017613 & 0.000299 & 0.951 \\
          &         &           & HT     & -0.00955 & 0.039465 & 1        & 0.0394   & 0.001557 & 0.952 \\
          &         &           & GRN    & 0.521031 & 0.039278 & 1.004761 & 0.039304 & 0.001547 & 0.95  \\
          &         &           & GRMIS  & -0.23187 & 0.039863 & 0.990027 & 0.038274 & 0.00159  & 0.939 \\
          &         &           & GRMIC  & -0.34967 & 0.03956  & 0.997613 & 0.038145 & 0.001567 & 0.938 \\
          &         &           & GRFCSMIS & 0.158378 & 0.034421 & 1.146544 & 0.033399 & 0.001185 & 0.944 \\
          &         &           & GRFCSMIC & 0.04174  & 0.033688 & 1.171498 & 0.033346 & 0.001135 & 0.94  \\
          &         &           &        &          &          &          &          &          &       \\
          &         & log(3)    & True   & -0.05218 & 0.020492 & 2.194225 & 0.019701 & 0.00042  & 0.942 \\
          &         &           & HT     & 0.033411 & 0.044964 & 1        & 0.043911 & 0.002022 & 0.95  \\
          &         &           & GRN    & 0.10533  & 0.044923 & 1.000899 & 0.043838 & 0.002019 & 0.953 \\
          &         &           & GRMIS  & 0.104264 & 0.043356 & 1.037085 & 0.042279 & 0.001881 & 0.946 \\
          &         &           & GRMIC  & 0.101906 & 0.043544 & 1.032593 & 0.042242 & 0.001897 & 0.946 \\
          &         &           & GRFCSMIS & -0.08375 & 0.041255 & 1.089891 & 0.040115 & 0.001703 & 0.948 \\
          &         &           & GRFCSMIC & -0.09068 & 0.04112  & 1.093472 & 0.040135 & 0.001692 & 0.945 \\
          &         &           &        &          &          &          &          &          &       \\
          & 75      & log(1.5)  & True   & -0.34616 & 0.023745 & 2.223618 & 0.023804 & 0.000566 & 0.956 \\
          &         &           & HT     & -0.14543 & 0.052799 & 1        & 0.053227 & 0.002788 & 0.952 \\
          &         &           & GRN    & 0.48857  & 0.052468 & 1.006304 & 0.052421 & 0.002757 & 0.95  \\
          &         &           & GRMIS  & 1.162651 & 0.051602 & 1.023206 & 0.050093 & 0.002685 & 0.948 \\
          &         &           & GRMIC  & 1.205129 & 0.052288 & 1.009773 & 0.049919 & 0.002758 & 0.946 \\
          &         &           & GRFCSMIS & 0.014211 & 0.044344 & 1.19068  & 0.046387 & 0.001966 & 0.955 \\
          &         &           & GRFCSMIC & -0.05963 & 0.043994 & 1.200136 & 0.046299 & 0.001936 & 0.954 \\
          &         &           &        &          &          &          &          &          &       \\
          &         & log(3)    & True   & -0.03541 & 0.027097 & 2.195267 & 0.026437 & 0.000734 & 0.942 \\
          &         &           & HT     & 0.210299 & 0.059485 & 1        & 0.058999 & 0.003544 & 0.955 \\
          &         &           & GRN    & 0.343256 & 0.058068 & 1.024415 & 0.058181 & 0.003386 & 0.952 \\
          &         &           & GRMIS  & 0.179398 & 0.054387 & 1.093735 & 0.05498  & 0.002962 & 0.948 \\
          &         &           & GRMIC  & 0.178737 & 0.055385 & 1.074034 & 0.05504  & 0.003071 & 0.944 \\
          &         &           & GRFCSMIS & 0.118899 & 0.053781 & 1.106067 & 0.053714 & 0.002894 & 0.947 \\
          &         &           & GRFCSMIC & 0.082925 & 0.054461 & 1.092252 & 0.053663 & 0.002967 & 0.944 \\
          &         &           &        &          &          &          &          &          &       \\
          & 90      & log(1.5)  & True   & 0.300231 & 0.038368 & 2.272879 & 0.037121 & 0.001474 & 0.946 \\
          &         &           & HT     & 0.093613 & 0.087206 & 1        & 0.08286  & 0.007605 & 0.95  \\
          &         &           & GRN    & 0.390952 & 0.083979 & 1.038428 & 0.079908 & 0.007055 & 0.95  \\
          &         &           & GRMIS  & 1.612847 & 0.083257 & 1.047438 & 0.077403 & 0.006974 & 0.94  \\
          &         &           & GRMIC  & 1.605483 & 0.083066 & 1.049849 & 0.077219 & 0.006942 & 0.938 \\
          &         &           & GRFCSMIS & 1.232738 & 0.078398 & 1.11235  & 0.076839 & 0.006171 & 0.946 \\
          &         &           & GRFCSMIC & 0.715285 & 0.08002  & 1.089812 & 0.076573 & 0.006412 & 0.944 \\
          &         &           &        &          &          &          &          &          &       \\
          &         & log(3)    & True   & -0.10845 & 0.040031 & 2.176706 & 0.039781 & 0.001604 & 0.948 \\
          &         &           & HT     & -0.21345 & 0.087135 & 1        & 0.088947 & 0.007598 & 0.956 \\
          &         &           & GRN    & -0.02096 & 0.086341 & 1.009188 & 0.086725 & 0.007455 & 0.954 \\
          &         &           & GRMIS  & 0.446222 & 0.087491 & 0.995923 & 0.085366 & 0.007679 & 0.948 \\
          &         &           & GRMIC  & 0.344767 & 0.088426 & 0.985395 & 0.085189 & 0.007834 & 0.948 \\
          &         &           & GRFCSMIS & -0.18669 & 0.08665  & 1.005597 & 0.083511 & 0.007512 & 0.952 \\
          &         &           & GRFCSMIC & -0.12586 & 0.087487 & 0.995971 & 0.083299 & 0.007656 & 0.945
\end{tabular}
}
\end{table}

\section*{Appendix G: Sampling design comparison results}

\begin{table}[H]
\sisetup{
input-decimal-markers={.},
round-mode=places,
detect-all,
round-integer-to-decimal
}
\caption{Simulation results for estimating $\beta_x$ using the data imputation approach for error scenario 2 (errors in event indicator and failure time) with $N=4000$, $n=800$ comparing simple random sampling (SRS), case-control sampling (CC), and stratified case-control sampling (SCC). The $\%$ bias, empirical standard error (ESE), relative efficiency (RE), average standard error (ASE), mean squared error, and coverage probabilities (CP) are presented for 2000 simulated datasets.}
\label{data_imp_design_compare_err_2}
\centering 
\scalebox{0.80}{
\begin{tabular}{
c
c
c
c
c
*6{S[table-format=1.3,round-precision=3]}
}
$\beta_z$ & \% Cens & $\beta_x$ & Design & Method & {\% Bias}  & {ESE}      & {RE}       & {ASE}      & {MSE}      & {CP}    \\ \hline
log(0.5)  & 90      & log(1.5)  & SRS    & True   & -0.19575 & 0.056825 & 2.316178 & 0.058701 & 0.00323  & 0.953 \\
          &         &           &        & HT     & -0.09687 & 0.131616 & 1        & 0.130847 & 0.017323 & 0.946 \\
          &         &           &        & GRN    & -0.21918 & 0.104644 & 1.257759 & 0.104506 & 0.010951 & 0.948 \\
          &         &           &        & GRMIS  & -0.45879 & 0.105389 & 1.248862 & 0.103657 & 0.01111  & 0.946 \\
          &         &           &        & GRMIC  & -0.09587 & 0.104913 & 1.254526 & 0.102638 & 0.011007 & 0.94  \\
          &         &           &        & GRFCSMIS & -0.28308 & 0.104808 & 1.255791 & 0.10352  & 0.010986 & 0.946 \\
          &         &           &        & GRFCSMIC & 0.056619 & 0.106483 & 1.236032 & 0.102959 & 0.011339 & 0.944 \\
          &         &           &        &        &          &          &          &          &          &       \\
          &         &           & CC     & True   & -0.19575 & 0.056825 & 2.173331 & 0.058701 & 0.00323  & 0.953 \\
          &         &           &        & HT     & 1.443152 & 0.123499 & 1        & 0.120076 & 0.015286 & 0.936 \\
          &         &           &        & GRN    & 1.055416 & 0.108371 & 1.139601 & 0.105822 & 0.011763 & 0.934 \\
          &         &           &        & GRMIS  & -0.3051  & 0.110035 & 1.122364 & 0.106001 & 0.012109 & 0.935 \\
          &         &           &        & GRMIC  & 1.012974 & 0.107239 & 1.151629 & 0.105819 & 0.011517 & 0.928 \\
          &         &           &        & GRFCSMIS & 0.350989 & 0.107726 & 1.146416 & 0.105584 & 0.011607 & 0.939 \\
          &         &           &        & GRFCSMIC & -0.05063 & 0.110085 & 1.12185  & 0.105693 & 0.012119 & 0.93  \\
          &         &           &        &        &          &          &          &          &          &       \\
          &         &           & SCC    & True   & -0.19575 & 0.056825 & 2.024597 & 0.058701 & 0.00323  & 0.953 \\
          &         &           &        & HT     & 1.541675 & 0.115047 & 1        & 0.109119 & 0.013275 & 0.942 \\
          &         &           &        & GRN    & 1.3337   & 0.103861 & 1.107711 & 0.099885 & 0.010816 & 0.934 \\
          &         &           &        & GRMIS  & 1.34774  & 0.104582 & 1.100072 & 0.099689 & 0.010967 & 0.936 \\
          &         &           &        & GRMIC  & 1.120339 & 0.104766 & 1.098134 & 0.099614 & 0.010997 & 0.935 \\
          &         &           &        & GRFCSMIS & 1.454201 & 0.105636 & 1.089092 & 0.099647 & 0.011194 & 0.939 \\
          &         &           &        & GRFCSMIC & 0.826524 & 0.104132 & 1.104821 & 0.099525 & 0.010855 & 0.934 \\
          &         &           &        &        &          &          &          &          &          &       \\
          &         & log(3)    & SRS    & True   & 0.1293   & 0.064842 & 2.198686 & 0.06303  & 0.004206 & 0.954 \\
          &         &           &        & HT     & 0.1639   & 0.142567 & 1        & 0.139971 & 0.020328 & 0.946 \\
          &         &           &        & GRN    & 0.046035 & 0.120143 & 1.186638 & 0.115264 & 0.014435 & 0.943 \\
          &         &           &        & GRMIS  & -0.33881 & 0.116575 & 1.222958 & 0.114212 & 0.013604 & 0.951 \\
          &         &           &        & GRMIC  & -0.36136 & 0.117864 & 1.209587 & 0.113062 & 0.013908 & 0.941 \\
          &         &           &        & GRFCSMIS & -0.35298 & 0.117358 & 1.214796 & 0.113453 & 0.013788 & 0.946 \\
          &         &           &        & GRFCSMIC & -0.29133 & 0.117659 & 1.211692 & 0.11276  & 0.013854 & 0.945 \\
          &         &           &        &        &          &          &          &          &          &       \\
          &         &           & CC     & True   & 0.1293   & 0.064842 & 2.034957 & 0.06303  & 0.004206 & 0.954 \\
          &         &           &        & HT     & 0.863599 & 0.13195  & 1        & 0.130468 & 0.017501 & 0.93  \\
          &         &           &        & GRN    & 0.906118 & 0.118367 & 1.114757 & 0.113327 & 0.01411  & 0.93  \\
          &         &           &        & GRMIS  & 0.354587 & 0.116178 & 1.135762 & 0.113018 & 0.013512 & 0.934 \\
          &         &           &        & GRMIC  & 0.490041 & 0.115837 & 1.1391   & 0.112904 & 0.013447 & 0.93  \\
          &         &           &        & GRFCSMIS & 0.275773 & 0.115996 & 1.137543 & 0.112389 & 0.013464 & 0.931 \\
          &         &           &        & GRFCSMIC & 0.28242  & 0.116929 & 1.128468 & 0.112369 & 0.013682 & 0.93  \\
          &         &           &        &        &          &          &          &          &          &       \\
          &         &           & SCC    & True   & 0.1293   & 0.064842 & 1.918313 & 0.06303  & 0.004206 & 0.954 \\
          &         &           &        & HT     & 0.744288 & 0.124387 & 1        & 0.119663 & 0.015539 & 0.938 \\
          &         &           &        & GRN    & 0.856953 & 0.111821 & 1.112379 & 0.108713 & 0.012592 & 0.94  \\
          &         &           &        & GRMIS  & 0.448073 & 0.11011  & 1.129659 & 0.108819 & 0.012148 & 0.944 \\
          &         &           &        & GRMIC  & 0.589966 & 0.111043 & 1.120168 & 0.108831 & 0.012373 & 0.944 \\
          &         &           &        & GRFCSMIS & 0.529959 & 0.109927 & 1.131544 & 0.108549 & 0.012118 & 0.942 \\
          &         &           &        & GRFCSMIC & 0.279283 & 0.110541 & 1.125258 & 0.108364 & 0.012229 & 0.94 
\end{tabular}
}
\end{table}

\begin{table}[H]
\sisetup{
input-decimal-markers={.},
round-mode=places,
detect-all,
round-integer-to-decimal
}
\caption{Simulation results for estimating $\beta_x$ using the IF imputation approach for error scenario 2 (errors in event indicator and failure time) with $N=4000$, $n=800$ comparing simple random sampling (SRS), case-control sampling (CC), and stratified case-control sampling (SCC). The $\%$ bias, empirical standard error (ESE), relative efficiency (RE), average standard error (ASE), mean squared error, and coverage probabilities (CP) are presented for 2000 simulated datasets.}
\label{IF_imp_design_compare_err_2}
\centering 
\scalebox{0.80}{
\begin{tabular}{
c
c
c
c
c
*6{S[table-format=1.3,round-precision=3]}
}
$\beta_z$ & \% Cens & $\beta_x$ & Design & Method & {\% Bias}  & {ESE}      & {RE}       & {ASE}      & {MSE}      & {CP}    \\ \hline
log(0.5)  & 90      & log(1.5)  & SRS    & True   & 0.013181 & 0.059384 & 2.257952 & 0.05879  & 0.003526 & 0.951 \\
          &         &           &        & HT     & 0.590898 & 0.134085 & 1        & 0.130847 & 0.017985 & 0.94  \\
          &         &           &        & GRN    & 0.764177 & 0.109177 & 1.228152 & 0.104129 & 0.011929 & 0.94  \\
          &         &           &        & GRMIS  & -1.38011 & 0.108987 & 1.230285 & 0.103116 & 0.01191  & 0.93  \\
          &         &           &        & GRMIC  & -1.16293 & 0.109731 & 1.221942 & 0.102465 & 0.012063 & 0.924 \\
          &         &           &        & GRFCSMIS & -1.4325  & 0.110198 & 1.216764 & 0.1025   & 0.012177 & 0.929 \\
          &         &           &        & GRFCSMIC & -0.97002 & 0.112221 & 1.194832 & 0.102177 & 0.012609 & 0.926 \\
          &         &           &        &        &          &          &          &          &          &       \\
          &         &           & CC     & True   & 0.013181 & 0.059384 & 2.085539 & 0.05879  & 0.003526 & 0.951 \\
          &         &           &        & HT     & 3.005687 & 0.123847 & 1        & 0.121097 & 0.015487 & 0.94  \\
          &         &           &        & GRN    & 2.412196 & 0.111991 & 1.105862 & 0.106782 & 0.012638 & 0.934 \\
          &         &           &        & GRMIS  & -0.87831 & 0.114389 & 1.082684 & 0.10729  & 0.013097 & 0.935 \\
          &         &           &        & GRMIC  & -0.09236 & 0.12005  & 1.031628 & 0.10668  & 0.014412 & 0.922 \\
          &         &           &        & GRFCSMIS & -1.99286 & 0.112352 & 1.102314 & 0.106793 & 0.012688 & 0.931 \\
          &         &           &        & GRFCSMIC & -0.97418 & 0.116849 & 1.059884 & 0.106057 & 0.013669 & 0.916 \\
          &         &           &        &        &          &          &          &          &          &       \\
          &         &           & SCC    & True   & 0.013181 & 0.059384 & 1.908976 & 0.05879  & 0.003526 & 0.951 \\
          &         &           &        & HT     & 0.351249 & 0.113362 & 1        & 0.109603 & 0.012853 & 0.946 \\
          &         &           &        & GRN    & 0.059211 & 0.102666 & 1.104187 & 0.10025  & 0.01054  & 0.945 \\
          &         &           &        & GRMIS  & -2.49652 & 0.104083 & 1.089149 & 0.099592 & 0.010936 & 0.934 \\
          &         &           &        & GRMIC  & -2.39435 & 0.102572 & 1.105198 & 0.099262 & 0.010615 & 0.931 \\
          &         &           &        & GRFCSMIS & -2.94422 & 0.102822 & 1.102511 & 0.099375 & 0.010715 & 0.937 \\
          &         &           &        & GRFCSMIC & -2.92328 & 0.10429  & 1.086985 & 0.099046 & 0.011017 & 0.933 \\
          &         &           &        &        &          &          &          &          &          &       \\
          &         & log(3)    & SRS    & True   & 0.090194 & 0.065292 & 2.228051 & 0.06311  & 0.004264 & 0.948 \\
          &         &           &        & HT     & 0.98782  & 0.145474 & 1        & 0.140929 & 0.02128  & 0.942 \\
          &         &           &        & GRN    & 0.462722 & 0.114252 & 1.273267 & 0.115252 & 0.013079 & 0.947 \\
          &         &           &        & GRMIS  & -0.88636 & 0.113281 & 1.284186 & 0.112866 & 0.012927 & 0.94  \\
          &         &           &        & GRMIC  & -1.00917 & 0.116412 & 1.249647 & 0.111686 & 0.013675 & 0.933 \\
          &         &           &        & GRFCSMIS & -1.29521 & 0.11259  & 1.292064 & 0.111025 & 0.012879 & 0.94  \\
          &         &           &        & GRFCSMIC & -1.21127 & 0.113696 & 1.279502 & 0.109949 & 0.013104 & 0.936 \\
          &         &           &        &        &          &          &          &          &          &       \\
          &         &           & CC     & True   & 0.090194 & 0.065292 & 2.178578 & 0.06311  & 0.004264 & 0.948 \\
          &         &           &        & HT     & 1.522176 & 0.142244 & 1        & 0.129912 & 0.020513 & 0.918 \\
          &         &           &        & GRN    & 1.018669 & 0.121551 & 1.170234 & 0.113392 & 0.0149   & 0.922 \\
          &         &           &        & GRMIS  & -0.35563 & 0.125685 & 1.13175  & 0.111195 & 0.015812 & 0.911 \\
          &         &           &        & GRMIC  & 0.023545 & 0.129196 & 1.100987 & 0.110825 & 0.016692 & 0.907 \\
          &         &           &        & GRFCSMIS & -0.50886 & 0.120788 & 1.177634 & 0.110295 & 0.014621 & 0.919 \\
          &         &           &        & GRFCSMIC & -0.13024 & 0.128134 & 1.110117 & 0.109476 & 0.01642  & 0.898 \\
          &         &           &        &        &          &          &          &          &          &       \\
          &         &           & SCC    & True   & 0.090194 & 0.065292 & 2.036797 & 0.06311  & 0.004264 & 0.948 \\
          &         &           &        & HT     & 0.601277 & 0.132986 & 1        & 0.119273 & 0.017729 & 0.934 \\
          &         &           &        & GRN    & 0.594733 & 0.115309 & 1.153306 & 0.108376 & 0.013339 & 0.938 \\
          &         &           &        & GRMIS  & -0.58776 & 0.11268  & 1.180213 & 0.106481 & 0.012738 & 0.942 \\
          &         &           &        & GRMIC  & -0.44739 & 0.115677 & 1.14964  & 0.106537 & 0.013405 & 0.935 \\
          &         &           &        & GRFCSMIS & -0.97012 & 0.113241 & 1.174364 & 0.105788 & 0.012937 & 0.934 \\
          &         &           &        & GRFCSMIC & -0.87356 & 0.110695 & 1.201377 & 0.105449 & 0.012345 & 0.927
\end{tabular}
}
\end{table}

\begin{table}[H]
\sisetup{
input-decimal-markers={.},
round-mode=places,
detect-all,
round-integer-to-decimal
}
\caption{Simulation results for estimating $\beta_x$ using the IF imputation approach for error scenario 3 (errors in event indicator, failure time, and X) with $N=4000$, $n=800$ comparing simple random sampling (SRS), case-control sampling (CC), and stratified case-control sampling (SCC). The $\%$ bias, empirical standard error (ESE), relative efficiency (RE), average standard error (ASE), mean squared error, and coverage probabilities (CP) are presented for 2000 simulated datasets.}
\label{IF_imp_design_compare_err_3}
\centering 
\scalebox{0.80}{
\begin{tabular}{
c
c
c
c
c
*6{S[table-format=1.3,round-precision=3]}
}
$\beta_z$ & \% Cens & $\beta_x$ & Design & Method & {\% Bias}  & {ESE}      & {RE}       & {ASE}      & {MSE}      & {CP}    \\ \hline
log(0.5)  & 90      & log(1.5)  & SRS    & True   & 0.013181 & 0.059384 & 2.333618 & 0.05879  & 0.003526 & 0.951 \\
          &         &           &        & HT     & 0.644516 & 0.138579 & 1        & 0.130362 & 0.019211 & 0.948 \\
          &         &           &        & GRN    & 1.081391 & 0.125327 & 1.10574  & 0.120056 & 0.015726 & 0.951 \\
          &         &           &        & GRMIS  & 3.093863 & 0.126174 & 1.098315 & 0.115137 & 0.016077 & 0.928 \\
          &         &           &        & GRMIC  & 2.868721 & 0.124744 & 1.110902 & 0.115237 & 0.015696 & 0.926 \\
          &         &           &        & GRFCSMIS & 0.401896 & 0.119263 & 1.161954 & 0.111873 & 0.014226 & 0.94  \\
          &         &           &        & GRFCSMIC & 0.594765 & 0.120779 & 1.147378 & 0.111389 & 0.014593 & 0.939 \\
          &         &           &        &        &          &          &          &          &          &       \\
          &         &           & CC     & True   & 0.013181 & 0.059384 & 2.112828 & 0.05879  & 0.003526 & 0.951 \\
          &         &           &        & HT     & 1.421846 & 0.125467 & 1        & 0.120629 & 0.015775 & 0.944 \\
          &         &           &        & GRN    & 1.616807 & 0.124611 & 1.00687  & 0.120092 & 0.015571 & 0.942 \\
          &         &           &        & GRMIS  & 2.025107 & 0.125633 & 0.998682 & 0.115642 & 0.015851 & 0.926 \\
          &         &           &        & GRMIC  & 2.742054 & 0.130046 & 0.964796 & 0.115097 & 0.017035 & 0.919 \\
          &         &           &        & GRFCSMIS & -0.48853 & 0.114874 & 1.092217 & 0.111443 & 0.0132   & 0.936 \\
          &         &           &        & GRFCSMIC & -0.44037 & 0.121669 & 1.031218 & 0.110755 & 0.014807 & 0.93  \\
          &         &           &        &        &          &          &          &          &          &       \\
          &         &           & SCC    & True   & 0.013181 & 0.059384 & 1.850378 & 0.05879  & 0.003526 & 0.951 \\
          &         &           &        & HT     & 0.544971 & 0.109882 & 1        & 0.110417 & 0.012079 & 0.944 \\
          &         &           &        & GRN    & 1.084536 & 0.110315 & 0.996079 & 0.110318 & 0.012189 & 0.946 \\
          &         &           &        & GRMIS  & 1.545407 & 0.112827 & 0.973897 & 0.107852 & 0.012769 & 0.93  \\
          &         &           &        & GRMIC  & 1.325522 & 0.114064 & 0.963336 & 0.107612 & 0.01304  & 0.93  \\
          &         &           &        & GRFCSMIS & -1.44068 & 0.105472 & 1.041814 & 0.104069 & 0.011158 & 0.947 \\
          &         &           &        & GRFCSMIC & -0.42746 & 0.107909 & 1.018281 & 0.103619 & 0.011647 & 0.94  \\
          &         &           &        &        &          &          &          &          &          &       \\
          &         & log(3)    & SRS    & True   & 0.090194 & 0.065292 & 2.211227 & 0.06311  & 0.004264 & 0.948 \\
          &         &           &        & HT     & 0.407488 & 0.144375 & 1        & 0.14084  & 0.020864 & 0.94  \\
          &         &           &        & GRN    & 0.385225 & 0.136955 & 1.054183 & 0.130108 & 0.018774 & 0.942 \\
          &         &           &        & GRMIS  & 1.24374  & 0.138733 & 1.040671 & 0.127417 & 0.019434 & 0.93  \\
          &         &           &        & GRMIC  & 1.30716  & 0.140311 & 1.028966 & 0.127054 & 0.019893 & 0.93  \\
          &         &           &        & GRFCSMIS & -0.73771 & 0.135317 & 1.066941 & 0.122208 & 0.018376 & 0.934 \\
          &         &           &        & GRFCSMIC & -0.66206 & 0.135598 & 1.064733 & 0.121757 & 0.01844  & 0.925 \\
          &         &           &        &        &          &          &          &          &          &       \\
          &         &           & CC     & True   & 0.090194 & 0.065292 & 2.015433 & 0.06311  & 0.004264 & 0.948 \\
          &         &           &        & HT     & 1.264558 & 0.131591 & 1        & 0.130024 & 0.017509 & 0.935 \\
          &         &           &        & GRN    & 1.1846   & 0.134597 & 0.977673 & 0.128859 & 0.018286 & 0.932 \\
          &         &           &        & GRMIS  & 1.728262 & 0.134406 & 0.979058 & 0.126532 & 0.018426 & 0.924 \\
          &         &           &        & GRMIC  & 1.827018 & 0.136783 & 0.962044 & 0.126234 & 0.019113 & 0.919 \\
          &         &           &        & GRFCSMIS & -0.35803 & 0.131977 & 0.997079 & 0.120769 & 0.017433 & 0.921 \\
          &         &           &        & GRFCSMIC & 0.105282 & 0.129613 & 1.015268 & 0.120185 & 0.016801 & 0.915 \\
          &         &           &        &        &          &          &          &          &          &       \\
          &         &           & SCC    & True   & 0.090194 & 0.065292 & 1.900692 & 0.06311  & 0.004264 & 0.948 \\
          &         &           &        & HT     & 1.19953  & 0.1241   & 1        & 0.122704 & 0.015574 & 0.934 \\
          &         &           &        & GRN    & 1.281997 & 0.12254  & 1.01273  & 0.121992 & 0.015214 & 0.934 \\
          &         &           &        & GRMIS  & 1.168752 & 0.125257 & 0.99076  & 0.120536 & 0.015854 & 0.932 \\
          &         &           &        & GRMIC  & 1.34881  & 0.124024 & 1.000608 & 0.120272 & 0.015602 & 0.928 \\
          &         &           &        & GRFCSMIS & -0.73769 & 0.122563 & 1.012538 & 0.116614 & 0.015087 & 0.93  \\
          &         &           &        & GRFCSMIC & -0.36084 & 0.1256   & 0.988055 & 0.116154 & 0.015791 & 0.928
\end{tabular}
}
\end{table}

\section*{Appendix H: Complex misclassification results}

\begin{table}[H]
\caption{Misclassification generation process for the simulations testing misclassification generation with interactions. The sensitivity (Sens), specificity (Spec), positive predictive value (PPV), and negative predictive value (NPV) for the event indicator are presented.}
\label{chpt3:misclass_interact_metrics}
\centering 
\begin{tabular}{cccccccc}
$\Delta^{\star}$   & $\%$ Cens & $\beta_x$ & $\beta_z$                                                  & Sens  & Spec  & PPV   & NPV   \\ \hline
\begin{tabular}[c]{@{}c@{}}$\textrm{Bernoulli}(\textrm{expit}(-1.1+0.5*\Delta-$\\ $0.25*X-0.1*U+0.2*Z+$\\$0.85*\Delta*X+0.2*\Delta*U+$\\$0.8*\Delta*z))$\end{tabular} & 50 & log(1.5)  & log(0.5)   & 0.833 & 0.889 & 0.860 & 0.867 \\
            & & log(3) & log(0.5) & 0.874 & 0.892 & 0.880 & 0.887 \\
            & & & & & & & \\
& 75 & log(1.5)  & log(0.5)   & 0.768 & 0.818 & 0.573 & 0.917 \\
            & & log(3) & log(0.5) & 0.826 & 0.797 & 0.553 & 0.938 \\         & & & & & & & \\
& 90 & log(1.5)  & log(0.5)   & 0.709 & 0.734 & 0.224 & 0.959 \\
            & & log(3) & log(0.5) & 0.797 & 0.717 & 0.226 & 0.972 \\ 
\end{tabular}
\end{table}

\begin{table}[H]
\sisetup{
input-decimal-markers={.},
round-mode=places,
detect-all,
round-integer-to-decimal
}
\caption{Simulation results for estimating $\beta_x$ using the data imputation approach for error scenario 3 (errors in event indicator, failure time, and X) with interaction terms in the misclassification generation, $N=2000$, $n=400$, and simple random sampling. The $\%$ bias, empirical standard error (ESE), relative efficiency (RE), average standard error (ASE), mean squared error, and coverage probabilities (CP) are presented for 2000 simulated datasets.}
\label{data_imp_misclass_interact_err_3}
\centering 
\scalebox{0.80}{
\begin{tabular}{
c
c
c
c
*6{S[table-format=1.3,round-precision=3]}
}
$\beta_z$ & \% Cens & $\beta_x$ & Method   & {\% Bias}  & {ESE}      & {RE}       & {ASE}      & {MSE}      & {CP}    \\ \hline
log(0.5)  & 50      & log(1.5)  & True     & -0.03595 & 0.039644 & 2.357942 & 0.039422 & 0.001572 & 0.956 \\
          &         &           & HT       & 0.968647 & 0.093478 & 1        & 0.087968 & 0.008754 & 0.944 \\
          &         &           & GRN      & 2.065881 & 0.093214 & 1.002836 & 0.087707 & 0.008759 & 0.942 \\
          &         &           & GRMIS    & 2.037858 & 0.092802 & 1.007285 & 0.087701 & 0.00868  & 0.944 \\
          &         &           & GRMIC    & 2.068534 & 0.092468 & 1.010926 & 0.087639 & 0.008621 & 0.942 \\
          &         &           & GRFCSMIS & 1.111458 & 0.06979  & 1.339419 & 0.07076  & 0.004891 & 0.954 \\
          &         &           & GRFCSMIC & 0.806618 & 0.069849 & 1.338291 & 0.069445 & 0.00489  & 0.947 \\
          &         &           &          &          &          &          &          &          &       \\
          &         & log(3)    & True     & 0.041168 & 0.041582 & 2.490834 & 0.04415  & 0.001729 & 0.948 \\
          &         &           & HT       & 0.313211 & 0.103573 & 1        & 0.097851 & 0.010739 & 0.942 \\
          &         &           & GRN      & 0.533937 & 0.1046   & 0.990187 & 0.097608 & 0.010976 & 0.946 \\
          &         &           & GRMIS    & 0.450849 & 0.10535  & 0.983136 & 0.097619 & 0.011123 & 0.945 \\
          &         &           & GRMIC    & 0.524909 & 0.104816 & 0.988144 & 0.097605 & 0.01102  & 0.947 \\
          &         &           & GRFCSMIS & 0.28924  & 0.088922 & 1.164762 & 0.085733 & 0.007917 & 0.943 \\
          &         &           & GRFCSMIC & 0.231675 & 0.088009 & 1.17685  & 0.08482  & 0.007752 & 0.938 \\
          &         &           &          &          &          &          &          &          &       \\
          & 75      & log(1.5)  & True     & 0.119394 & 0.051672 & 2.316876 & 0.053276 & 0.00267  & 0.954 \\
          &         &           & HT       & 1.004049 & 0.119718 & 1        & 0.118566 & 0.014349 & 0.948 \\
          &         &           & GRN      & 1.895293 & 0.11939  & 1.002747 & 0.11707  & 0.014313 & 0.949 \\
          &         &           & GRMIS    & 2.027832 & 0.120512 & 0.993414 & 0.117998 & 0.014591 & 0.95  \\
          &         &           & GRMIC    & 2.267477 & 0.121239 & 0.987452 & 0.117415 & 0.014784 & 0.949 \\
          &         &           & GRFCSMIS & 0.287842 & 0.099554 & 1.202546 & 0.104128 & 0.009912 & 0.952 \\
          &         &           & GRFCSMIC & 0.811668 & 0.099056 & 1.208593 & 0.102347 & 0.009823 & 0.946 \\
          &         &           &          &          &          &          &          &          &       \\
          &         & log(3)    & True     & -0.01311 & 0.06088  & 2.250031 & 0.059211 & 0.003706 & 0.949 \\
          &         &           & HT       & 0.836351 & 0.136982 & 1        & 0.131293 & 0.018849 & 0.952 \\
          &         &           & GRN      & 1.165105 & 0.134278 & 1.020136 & 0.130112 & 0.018195 & 0.95  \\
          &         &           & GRMIS    & 1.064519 & 0.136716 & 1.001951 & 0.130604 & 0.018828 & 0.95  \\
          &         &           & GRMIC    & 1.028011 & 0.135967 & 1.007466 & 0.130518 & 0.018615 & 0.952 \\
          &         &           & GRFCSMIS & 0.639607 & 0.123603 & 1.108244 & 0.121525 & 0.015327 & 0.949 \\
          &         &           & GRFCSMIC & 0.474697 & 0.121686 & 1.125707 & 0.12072  & 0.014835 & 0.944 \\
          &         &           &          &          &          &          &          &          &       \\
          & 90      & log(1.5)  & True     & 0.0138   & 0.084364 & 2.251745 & 0.083155 & 0.007117 & 0.947 \\
          &         &           & HT       & 1.897751 & 0.189966 & 1        & 0.183082 & 0.036146 & 0.94  \\
          &         &           & GRN      & 1.971642 & 0.18939  & 1.00304  & 0.179804 & 0.035933 & 0.94  \\
          &         &           & GRMIS    & 2.311279 & 0.189845 & 1.000635 & 0.179986 & 0.036129 & 0.948 \\
          &         &           & GRMIC    & 2.802321 & 0.186701 & 1.017489 & 0.177676 & 0.034986 & 0.94  \\
          &         &           & GRFCSMIS & -0.05938 & 0.184643 & 1.02883  & 0.174435 & 0.034093 & 0.94  \\
          &         &           & GRFCSMIC & -0.06001 & 0.18401  & 1.032369 & 0.171773 & 0.03386  & 0.934 \\
          &         &           &          &          &          &          &          &          &       \\
          &         & log(3)    & True     & -0.04654 & 0.088525 & 2.348938 & 0.089229 & 0.007837 & 0.95  \\
          &         &           & HT       & 0.928622 & 0.207941 & 1        & 0.196985 & 0.043343 & 0.939 \\
          &         &           & GRN      & 0.855951 & 0.204852 & 1.015079 & 0.19409  & 0.042053 & 0.938 \\
          &         &           & GRMIS    & 1.023834 & 0.205516 & 1.011799 & 0.193344 & 0.042363 & 0.939 \\
          &         &           & GRMIC    & 1.051391 & 0.203195 & 1.023355 & 0.190818 & 0.041422 & 0.937 \\
          &         &           & GRFCSMIS & 0.819666 & 0.20135  & 1.032732 & 0.190641 & 0.040623 & 0.933 \\
          &         &           & GRFCSMIC & 0.471738 & 0.197965 & 1.050389 & 0.1891   & 0.039217 & 0.935
\end{tabular}
}
\end{table}

\begin{table}[H]
\sisetup{
input-decimal-markers={.},
round-mode=places,
detect-all,
round-integer-to-decimal
}
\caption{Simulation results for estimating $\beta_x$ using the IF imputation approach for error scenario 3 (errors in event indicator, failure time, and X) with interaction terms in the misclassification generation, $N=2000$, $n=400$, and simple random sampling. The $\%$ bias, empirical standard error (ESE), relative efficiency (RE), average standard error (ASE), mean squared error, and coverage probabilities (CP) are presented for 2000 simulated datasets.}
\label{IF_imp_misclass_interact_err_3}
\centering 
\scalebox{0.80}{
\begin{tabular}{
c
c
c
c
*6{S[table-format=1.3,round-precision=3]}
}
$\beta_z$ & \% Cens & $\beta_x$ & Method   & {\% Bias}  & {ESE}      & {RE}       & {ASE}      & {MSE}      & {CP}    \\ \hline
log(0.5)  & 50      & log(1.5)  & True     & -0.03595 & 0.039644 & 2.357942 & 0.039422 & 0.001572 & 0.956 \\
          &         &           & HT       & 0.968647 & 0.093478 & 1        & 0.087968 & 0.008754 & 0.944 \\
          &         &           & GRN      & 2.065881 & 0.093214 & 1.002836 & 0.087707 & 0.008759 & 0.942 \\
          &         &           & GRMIS    & 1.469612 & 0.093884 & 0.995677 & 0.08226  & 0.00885  & 0.914 \\
          &         &           & GRMIC    & 0.960839 & 0.093115 & 1.003903 & 0.081983 & 0.008686 & 0.92  \\
          &         &           & GRFCSMIS & -0.01847 & 0.07019  & 1.331791 & 0.06906  & 0.004927 & 0.944 \\
          &         &           & GRFCSMIC & -0.27818 & 0.069723 & 1.340707 & 0.068681 & 0.004863 & 0.942 \\
          &         &           &          &          &          &          &          &          &       \\
          &         & log(3)    & True     & 0.041168 & 0.041582 & 2.490834 & 0.04415  & 0.001729 & 0.948 \\
          &         &           & HT       & 0.313211 & 0.103573 & 1        & 0.097851 & 0.010739 & 0.942 \\
          &         &           & GRN      & 0.533937 & 0.1046   & 0.990187 & 0.097608 & 0.010976 & 0.946 \\
          &         &           & GRMIS    & 1.79989  & 0.103708 & 0.998707 & 0.092364 & 0.011146 & 0.927 \\
          &         &           & GRMIC    & 1.777451 & 0.102679 & 1.008709 & 0.09228  & 0.010924 & 0.925 \\
          &         &           & GRFCSMIS & -0.13418 & 0.095758 & 1.081613 & 0.084204 & 0.009172 & 0.934 \\
          &         &           & GRFCSMIC & -0.28799 & 0.094592 & 1.094954 & 0.083523 & 0.008958 & 0.932 \\
          &         &           &          &          &          &          &          &          &       \\
          & 75      & log(1.5)  & True     & 0.119394 & 0.051672 & 2.316876 & 0.053276 & 0.00267  & 0.954 \\
          &         &           & HT       & 1.004049 & 0.119718 & 1        & 0.118566 & 0.014349 & 0.948 \\
          &         &           & GRN      & 1.895293 & 0.11939  & 1.002747 & 0.11707  & 0.014313 & 0.949 \\
          &         &           & GRMIS    & 3.457688 & 0.120251 & 0.99557  & 0.107183 & 0.014657 & 0.926 \\
          &         &           & GRMIC    & 3.698102 & 0.119798 & 0.999333 & 0.106444 & 0.014576 & 0.924 \\
          &         &           & GRFCSMIS & 0.700943 & 0.104505 & 1.145569 & 0.101122 & 0.010929 & 0.947 \\
          &         &           & GRFCSMIC & 0.953667 & 0.103681 & 1.154675 & 0.100249 & 0.010765 & 0.943 \\
          &         &           &          &          &          &          &          &          &       \\
          &         & log(3)    & True     & -0.01311 & 0.06088  & 2.250031 & 0.059211 & 0.003706 & 0.949 \\
          &         &           & HT       & 0.836351 & 0.136982 & 1        & 0.131293 & 0.018849 & 0.952 \\
          &         &           & GRN      & 1.165105 & 0.134278 & 1.020136 & 0.130112 & 0.018195 & 0.95  \\
          &         &           & GRMIS    & 1.141388 & 0.133559 & 1.025635 & 0.121304 & 0.017995 & 0.931 \\
          &         &           & GRMIC    & 1.081814 & 0.134695 & 1.016982 & 0.121179 & 0.018284 & 0.933 \\
          &         &           & GRFCSMIS & -0.38405 & 0.127498 & 1.074387 & 0.11702  & 0.016274 & 0.934 \\
          &         &           & GRFCSMIC & -0.25338 & 0.125074 & 1.095207 & 0.11659  & 0.015651 & 0.93  \\
          &         &           &          &          &          &          &          &          &       \\
          & 90      & log(1.5)  & True     & 0.0138   & 0.084364 & 2.251745 & 0.083155 & 0.007117 & 0.947 \\
          &         &           & HT       & 1.897751 & 0.189966 & 1        & 0.183082 & 0.036146 & 0.94  \\
          &         &           & GRN      & 1.971642 & 0.18939  & 1.00304  & 0.179804 & 0.035933 & 0.94  \\
          &         &           & GRMIS    & 8.575347 & 0.207643 & 0.914869 & 0.168291 & 0.044324 & 0.902 \\
          &         &           & GRMIC    & 8.465425 & 0.199277 & 0.953278 & 0.165431 & 0.040889 & 0.892 \\
          &         &           & GRFCSMIS & 4.650762 & 0.183287 & 1.036438 & 0.165352 & 0.03395  & 0.925 \\
          &         &           & GRFCSMIC & 4.944207 & 0.18214  & 1.042967 & 0.161798 & 0.033577 & 0.91  \\
          &         &           &          &          &          &          &          &          &       \\
          &         & log(3)    & True     & -0.04654 & 0.088525 & 2.348938 & 0.089229 & 0.007837 & 0.95  \\
          &         &           & HT       & 0.928622 & 0.207941 & 1        & 0.196985 & 0.043343 & 0.939 \\
          &         &           & GRN      & 0.855951 & 0.204852 & 1.015079 & 0.19409  & 0.042053 & 0.938 \\
          &         &           & GRMIS    & 4.226444 & 0.204751 & 1.015579 & 0.183868 & 0.044079 & 0.916 \\
          &         &           & GRMIC    & 4.045888 & 0.205878 & 1.010016 & 0.181946 & 0.044362 & 0.915 \\
          &         &           & GRFCSMIS & 1.592917 & 0.195551 & 1.063359 & 0.178513 & 0.038546 & 0.911 \\
          &         &           & GRFCSMIC & 1.238013 & 0.201628 & 1.031308 & 0.176939 & 0.040839 & 0.908
\end{tabular}
}
\end{table}

\section*{Appendix I: VCCC analysis details}

For this study, we analyzed data on 4797 HIV-positive patients that had been fully validated and applied some common inclusion/exclusion criteria used in HIV studies to obtain the final analysis dataset. Specifically, any patients that had an indeterminate ART start date, no CD4 count measurement between 180 days before or 30 days after starting ART, no follow-up visits in the clinic after starting ART, an ADE before starting ART, or an indeterminate ADE date were excluded. In addition, patients must have been at least 18 years of age at ART start and not started ART prior to enrollment. Lastly, any ADE within 6 months of starting ART were not considered a true failure due to the time required for ART to be efficacious. After application of these criteria, the unvalidated and validated data contained 1995 and 1595 patients, respectively. The 1595 patients that met the criteria in the validated dataset were used for the analysis of the ADE outcome. 

The censoring rate among the 1595 patients was very high at $93.8\%$, suggesting that an outcome-dependent sampling design that oversamples cases would be necessary. Of the 1595 patients, $11\%$ had a misclassified ADE; specifically, $161$ were incorrectly classified as having an ADE and $12$ were incorrectly classified as having been censored. For the failure times, $34.5\%$ were incorrect, with the errors having mean and standard deviation of $-0.75$ and $2.89$ years, respectively. There were errors in the CD4 count at ART start for only $6.7\%$ of the patients; however, the errors were right skewed, having mean and standard deviation of $10$ and $154$ $\textrm{cell}/\textrm{mm}^3$, respectively. In addition, the errors in the failure times and CD4 count at ART start had a correlation of $-0.10$.

\begin{table}[H]
\sisetup{
input-decimal-markers={.},
round-mode=places,
detect-all,
round-integer-to-decimal
}
\caption{The median hazard ratios (HR) and their corresponding $95\%$ confidence interval widths calculated using the IF imputation method from 100 different sampled validation subsets for a 100 cell/$\textrm{mm}^3$ increase in CD4 count at ART initiation and 10-year increase in age at CD4 count measurement.}
\label{IF_imp_vccc_table}
\centering 
\scalebox{0.75}{
\begin{tabular}{
c
c
c
c
*4{S[table-format=1.3,round-precision=3]}
}
Subset size & Sampling & Method   & {CD4 HR} & {CD4 CI width} & {Age HR} & {Age CI width} \\ \hline
340         & CC       & True     & 0.693  & 0.19         & 0.829  & 0.361        \\
            &          & Naive    & 0.91   & 0.125        & 1.087  & 0.275        \\
            &          & HT       & 0.677  & 0.323        & 0.805  & 0.576        \\
            &          & GRN      & 0.68   & 0.284        & 0.821  & 0.477        \\
            &          & GRMIS    & 0.704  & 0.323        & 0.807  & 0.526        \\
            &          & GRMIC    & 0.695  & 0.296        & 0.804  & 0.492        \\
            &          & GRFCSMIS & 0.69   & 0.307        & 0.813  & 0.488        \\
            &          & GRFCSMIC & 0.684  & 0.299        & 0.813  & 0.463        \\
            &          &          &        &              &        &              \\
            & SCCB     & True     & 0.693  & 0.19         & 0.829  & 0.361        \\
            &          & Naive    & 0.91   & 0.125        & 1.087  & 0.275        \\
            &          & HT       & 0.682  & 0.283        & 0.855  & 0.571        \\
            &          & GRN      & 0.682  & 0.278        & 0.835  & 0.497        \\
            &          & GRMIS    & 0.691  & 0.284        & 0.851  & 0.515        \\
            &          & GRMIC    & 0.691  & 0.277        & 0.861  & 0.499        \\
            &          & GRFCSMIS & 0.7    & 0.289        & 0.846  & 0.506        \\
            &          & GRFCSMIC & 0.702  & 0.282        & 0.848  & 0.49         \\
            &          &          &        &              &        &              \\
            & SCCN     & True     & 0.693  & 0.19         & 0.829  & 0.361        \\
            &          & Naive    & 0.91   & 0.125        & 1.087  & 0.275        \\
            &          & HT       & 0.694  & 0.31         & 0.829  & 0.702        \\
            &          & GRN      & 0.69   & 0.304        & 0.813  & 0.609        \\
            &          & GRMIS    & 0.711  & 0.303        & 0.82   & 0.583        \\
            &          & GRMIC    & 0.715  & 0.298        & 0.824  & 0.57         \\
            &          & GRFCSMIS & 0.708  & 0.301        & 0.838  & 0.566        \\
            &          & GRFCSMIC & 0.723  & 0.298        & 0.826  & 0.561        \\
            &          &          &        &              &        &              \\
680         & CC       & True     & 0.693  & 0.19         & 0.829  & 0.361        \\
            &          & Naive    & 0.91   & 0.125        & 1.087  & 0.275        \\
            &          & HT       & 0.691  & 0.237        & 0.839  & 0.411        \\
            &          & GRN      & 0.69   & 0.227        & 0.83   & 0.386        \\
            &          & GRMIS    & 0.696  & 0.234        & 0.829  & 0.391        \\
            &          & GRMIC    & 0.7    & 0.232        & 0.834  & 0.385        \\
            &          & GRFCSMIS & 0.696  & 0.232        & 0.832  & 0.388        \\
            &          & GRFCSMIC & 0.702  & 0.23         & 0.83   & 0.386        \\
            &          &          &        &              &        &              \\
            & SCCB     & True     & 0.693  & 0.19         & 0.829  & 0.361        \\
            &          & Naive    & 0.91   & 0.125        & 1.087  & 0.275        \\
            &          & HT       & 0.688  & 0.228        & 0.828  & 0.413        \\
            &          & GRN      & 0.69   & 0.227        & 0.821  & 0.387        \\
            &          & GRMIS    & 0.694  & 0.23         & 0.831  & 0.398        \\
            &          & GRMIC    & 0.697  & 0.229        & 0.83   & 0.39         \\
            &          & GRFCSMIS & 0.698  & 0.23         & 0.826  & 0.393        \\
            &          & GRFCSMIC & 0.7    & 0.231        & 0.824  & 0.388        \\
            &          &          &        &              &        &              \\
            & SCCN     & True     & 0.693  & 0.19         & 0.829  & 0.361        \\
            &          & Naive    & 0.91   & 0.125        & 1.087  & 0.275        \\
            &          & HT       & 0.688  & 0.231        & 0.832  & 0.438        \\
            &          & GRN      & 0.687  & 0.231        & 0.832  & 0.409        \\
            &          & GRMIS    & 0.693  & 0.232        & 0.825  & 0.407        \\
            &          & GRMIC    & 0.694  & 0.231        & 0.825  & 0.402        \\
            &          & GRFCSMIS & 0.694  & 0.233        & 0.823  & 0.402        \\
            &          & GRFCSMIC & 0.698  & 0.233        & 0.828  & 0.4         
\end{tabular}
}
\end{table}

\end{document}